\documentclass[twocolumn]{article}
\usepackage{amsmath}
\usepackage{icqproc-1}
\usepackage{rawfonts}
\usepackage{colordvi}
\usepackage{color}                     
\usepackage{graphicx}
\usepackage[]{rotating}

\usepackage{figsize}

\pdfoutput=1
\input prepictex
\input pictex
\input postpictex

\def\Label[#1]{\label{#1}}
\def\Ref[#1]{(\ref{#1})}

\def\bm{\boldmath}

\definecolor{Red1}{rgb}{1.0,0.2,0.3}
\definecolor{Sgreen}{rgb}{0.1,0.6,0.2}
\definecolor{Yel1}{rgb}{0.95,0.85,0.1}
\definecolor{Yel2}{rgb}{1.0,0.7,0.0}
\definecolor{Yel3}{rgb}{0.9,0.8,0.0}
\definecolor{Yel4}{rgb}{0.4,0.7,0.0}
\definecolor{Yel5}{rgb}{0.32,0.6,0.0}
\definecolor{Red1}{rgb}{1.0,0.2,0.3}
\definecolor{Red2}{rgb}{0.7,0.2,0.1}
\definecolor{Grn1}{rgb}{0.1,0.7,0.1}
\definecolor{Grn2}{rgb}{0.1,0.8,0.1}
\definecolor{Blue1}{rgb}{0.2,0.2,1.0}
\definecolor{Mag2}{rgb}{0.7,0.0,0.8}
\definecolor{Mix1}{rgb}{0.2,0.5,0.3}
\definecolor{Mix2}{rgb}{0.7,0.4,0.3}
\definecolor{Mix3}{rgb}{0.95,0.75,0.15}
\definecolor{Mix4}{rgb}{0.6,0.0,0.6}
\definecolor{Mix5}{rgb}{0.8,0.0,0.3}
\definecolor{Mix6}{rgb}{0.0,0.0,1.0}
\definecolor{Mix7}{rgb}{0.95,0.25,0.0}
\definecolor{Whit}{rgb}{1.0,1.0,1.0}
\definecolor{White}{rgb}{1.0,1.0,1.0}
\definecolor{Cy1}{rgb}{0.1,0.7,0.7}
\definecolor{Cy2}{rgb}{0.1,0.6,0.6}
\definecolor{Blue1}{rgb}{0.2,0.2,1.0}
\definecolor{Blue2}{rgb}{0.5,0.2,0.9}
\definecolor{Blue3}{rgb}{0.4,0.4,0.9}
\definecolor{Grey}{rgb}{0.5,0.5,0.8}
\def\ROT[#1]{%
\makebox[0.8cm]{%
\begin{sideways}
\makebox[0.8cm][l]{#1}\end{sideways}}}

\begin{document}
\thispagestyle{empty}
\fontnew

\twocolumn[
%\vspace*{30mm}
\begin{LARGE} 
\begin{center}
%
%  TITLE
%
On structure-stabilizing electronic interferences in bcc-related phases. \\
{\Large A research report}

%
%  END TITLE
%
\end{center}
\end{LARGE}
\begin{normalsize}
\begin{center} 
%
%  AUTHORS
%
H. Solbrig$^0$
%
%  END AUTHORS
%
\end{center}
\end{normalsize}
\begin{footnotesize}
\begin{it}
\begin{center}
%
%  ADDRESS
%
Chemnitz University of Technology, Institute of Physics,
D-09107 Chemnitz, Germany
%
%  END ADDRESS
%
\end{center}
\end{it}
\end{footnotesize}
\begin{footnotesize}
\begin{center}
%
%  DATE
%
\today
%
%  END DATE
%
\end{center}
\end{footnotesize}
\vspace{4ex}
\begin{small}
\hrule\vspace{3ex}
\begin{minipage}{\textwidth}
{\bf Abstract}\vspace{2ex}\\
\hp 
%
%  ABSTRACT
%
This study deals with cubic crystals where the contents of the simple cubic unit cells are close to
n$\times$n$\times$n-bcc sublattices ($n$ = 2: diamond- and zinc-blende type, $n$ = 3: $\gamma$-brasses).
First-principle results on the electronic structure are obtained from augmented LMTO-ASA calculations
and interpreted within a VEC-based Hume-Rothery concept which employs joined planar-radial interferences to
treat interference and hybridization on the same footing.
We show that the charge redistribution supports enhanced electronic interference which causes the band energy to decrease.
Several topics are included such as stabilizing networks, 
hardness and $s$-to-$p$ transfer,
cooperation of interferences, 
interplay between local radial order and global planar order,
electron-per-atom ratio, 
and the comparison with recent FLAPW-based results.
%
%  END ABSTRACT
%
\vspace{2.5ex} \\
%
%{\it Keywords:}\/ 
%
%  KEYWORDS
%
%
%  END KEYWORDS
%
\end{minipage}\vspace{3ex}
\hrule
\end{small}\vspace{6ex}
]
\footnotetext{solbrig@physik.tu-chemnitz.de}
\noindent
{\large \bf Contents} \\[2ex]
\begin{tabular}{ l  l  l  l }
%\hspace*{0.3cm} & \hspace*{0.3cm} & 				&\\
1.    & {\bf Introduction}					&1\\
2.    & {\bf Structure stability and electronic} 		&3\\
      & {\bf interference} 					&\\
  2.1 & $\;$ Momentum sphere					&3\\
  2.2 & $\;$ The general interference condition			&4\\
  2.3 & $\;$ Planar interference				&4\\
  2.4 & $\;$ Joined planar-radial interference 			&5\\
      & $\;$ and hybridization					&\\
  2.5 & $\;$ Virtual valence				 	&7\\
3.    & {\bf Diamond and zinc blende phases} 			&8\\
  3.1 & $\;$ Stability of essential length scales		&10\\
  3.2 & $\;$ Stabilizing $p$-networks				&12\\
  3.3 & $\;$ Summarizing 					&15\\
4.    & {\bf Electronic interference in $\gamma$-brasses} 	&15\\
  4.1 & $\;$ $\gamma$-$\rm Ag_5Li_8$				&15\\
  4.2 & $\;$ $\gamma$-$\rm V_5Al_8$				&24\\
  4.3 & $\;$ $\gamma$-$\rm Cu_5Zn_8$				&29\\
5.    & {\bf Conclusions} 					&35\\
6.    & {\bf Appendices} 					&37\\
\end{tabular}
\vspace{0.1ex}
%
%
%  MAIN TEXT
%
\section{Introduction}

Since the 1920th materials science has developed from purely empirical roots 
to an applied branch of natural sciences.
It was William Hume-Rothery \cite{Hume26} who did a decisive step in 1926.
He announced a tight connection between the crystal structures of certain alloys and the 
estimated numbers of valence electrons per atom (the $e/a$-ratio).
This finding defines the origin of the Hume-Rothery rules which predict the
crystal structures depending on the $e/a$-ratios.
Even today the rules are useful for two reasons: 
(i) Simple estimates of the $e/a$-ratios can be obtained from 
the $sp$-electron configurations of the component free atoms.
(ii) The Hume-Rothery rules provide a valuable classification tool for the wealth of 
crystalline alloys \cite{Massalski10a}.

However, going into detail, the $e/a$-ratio thus estimated checks the involved free atoms 
regarding the itinerant quantum weight they could possibly acquire up to the Fermi energy.
Hence, the bonding effect is supposed around the Fermi energy and nearly-free electron like.
This must cause difficulties if atomic $d$-states are involved or if
the bonding contributions arise deep in the valence band.
Following Raynor \cite{Raynor49} amended empirical rules are available which
employ even negative valences for transition metals \cite{TdL05a}.

In 1936, Mott and Jones \cite{Mott58a} have contributed the basic ideas
for more realistic treatments of the electron-structure interrelations.
They attributed structure stability to so-called 
''Fermi-surface Brillouin-zone interactions``.
This means that structures are considered stable if they allow for electron transitions on 
the surface of a momentum sphere (MS) in the extended $k$-space 
which are supported by Bragg-reflections.
The MS encloses the itinerant weight of the valence electrons up to the Fermi energy,
i.e. the contributions $e/a$ of all atoms in the unit cell.
However, applications have clarified that the proper MS 
differs from the real Fermi body \cite{Hume62a,Asahi05a,Paxton97}.
Since the 1930th the community is busy with the determination of appropriate $e/a$-values.
Recently, the activities of the group around Mizutani have been reviewed \cite{Mizutani17a}.

Among the great amount of experimental and theoretical contributions 
to the Hume-Rothery stabilization,
a paper by Watanabe and Ishii \cite{Watanabe08a} reveals an interesting aspect of the 
cooperation between interference and hybridization.
The paper deals with simple Al-Mn alloys where $sp$-$d$ hybridization and 
$d$-band splitting dominate around the Fermi energy,
clearly not a standard case of Hume-Rothery stabilization.
The authors emphasize: The bonding effect at the Fermi-energy is hybridization-based.
However, unhybridized band states allow for interference around the lower bound 
of the $d$Mn-band which configures the partial weights for hybridization 
close to the Fermi energy.
Hence, interference clearly below the Fermi energy supports the hybridization
close to the Fermi energy.
We will show that this kind of cooperations happens in $\gamma$-brasses, too.

The present study examines cubic crystals where the contents of the cubic unit cells 
are close to $n$$\times$$n$$\times$$n$-bcc sublattices. 
There belong
with $n$ = 1 the CsCl-type phases (number 221), 
with $n$ = 2 the diamond- (number 227), zinc-blende- (number 216), 
and Heusler-phases (number 225), 
with $n$ = 3 the $\gamma$-brasses (number 217), and 
with $n$ = 4 certain quasicrystalline approximants such as of Al-Cu-Fe with 128 atoms 
in the simple cubic (SC) unit cell (number 198).
This study deals only with cases $n$ = 2 and $n$ = 3.

Using published structure models in augmented LMTO-ASA calculations 
we obtain various projected densities of states together with the
 compositions of the band states in the atomic-site angular-momentum representation.
The main purpose of the study is the subsequent interpretation of the 
first-principle results within a Hume-Rothery concept based on the 
calculated valence-electron concentrations (VEC).

For diamond and related phases with pronounced covalent bonding character 
optimized quantum-chemical approaches are available.
The present attempt within the atomic-sphere approximation with local XC-potential
may thus appear quite strange.
Admittedly, the resulting gaps are not wide enough.
However, we put emphasis on treating different phases on equal footing regarding
the stabilizing aspects of electronic interference.
It will turn out that the mentioned drawback does not devalue the subsequent analysis
 of the first-principle results.

Three $\gamma$-phases will be examined which were formerly proved quite special,
namely the exceptions to the Hume-Rothery rule $\gamma$-$\rm Ag_5Li_8$ \cite{Mizutani08a}
and $\gamma$-$\rm V_5Al_8$ \cite{Mizutani06a} and the prototype alloy 
$\gamma$-$\rm Cu_5Zn_8$ of the group-one $\gamma$-brasses \cite{Mizutani17a}
which matches the Hume-Rothery rule perfectly.
In each case, we put emphasis on the consequences of the improved hybridization 
after the transition from 3$\times$3$\times$3-bcc to the $\gamma$-phase.
Both systems differ only slightly in view of the planar order 
and the assigned planar interferences.

We focus on those properties of the valence electrons which promote the 
electronic influence and ensure this way structure stability.
Such properties are low band energy, enhanced planar interference, 
and efficient hybridization on the atomic scale.
However, despite of many successful applications 
of each separate principle in the literature,
more insight into the mutual cooperation is required.
The following two issues will be tackled in this study: \\
(i) For planar interferences a quantitative description is available 
in terms of interference conditions in the extended $k$-space.
For hybridization, on the contrary, this level is not yet achieved.
Generally, only evidence for the influence of hybridization is announced \cite{TdL05a}.
We show that efficient hybridization at an atom goes along 
with efficient radial interference of inward backscattered waves 
by the sequence of neighbor shells.
Radial interference conditions can be found which reveal possible cooperation with 
nearby planar interferences via the common momentum pool.
We use such joined planar-radial interferences at specific energies 
to analyze the results of the first-principle calculations. \\
(ii) Low band energy and enhanced joined planar-radial interference 
should be seen as equivalent orientations on the way to structure stability.
They are both indicators of electronic reinforcement.
We show, in particular for $\gamma$-brasses,
 that structure relaxation which improves the interference status 
does also reduce the electronic band energy.
Improved interference concentrates spectral weight at fulfilled interference conditions
which provides a practicable tool to evaluate the achieved interference status.

We deal with interferences in transitions on the surface of a MS in the extended $k$-space.
The size of the MS is derived from the integrated DOS projected to the 
respective active part of the electron quantum space (EQS),
e.g. the $sp$-subspace or the $d$-subspace of certain atomic components.
To identify active subspaces we inspect the fluctuating decompositions of the norm One
 of band states and search for preferences of the fluctuation patterns. 
This procedure is a generalization of the reduction to unhybridized band
states in the approach by Watanabe and Ishii \cite{Watanabe08a}.  

Joined planar-radial interferences link the local radial order 
with the extended planar order.
The crystalline translation symmetry thus acts on the content of the unit cell,
because certain interatomic distances get access to strong planar reflections 
$G^2 \equiv h^2+k^2+l^2$ of the crystal.
The shape and the size of the equilibrium unit cell must depend on the
inside-outside tuning by guiding joined interferences.

This concept opens a wide field of possible applications.
There belong also cases, such as quasicrystals,
where this equilibrium is not achieved with lattice periodicity in three dimensions.
The long-range planar partners of the local radial processes
reveal rather Fibonacci-like arrangements.
One may conjecture that the formation of icosahedral clusters drives the system to develop this kind of 
long-range order.

In amorphous phases the long-range co-ordination is suppressed despite 
of well developed local radial interference and hybridization.
In wide composition fields of thin-film amorphous phases,
the experiments by P. H\"aussler and coworkers \cite{Haussler83c,Haussler07,Stiehler07} 
have verified the concept of structure stability due to resonance between the electrons and 
the average neighbor-shell order.

Even gaps in the sequence of planar interferences at certain values of $G^2$
prove important.
In that case, the planar interference $G^2 - 1$ can join 
a radial interference somewhat above which operates inside a 
larger spectral range for optimized hybridization 
without being hampered by a second planar interference.
In particular the absent simple-cubic interferences $G^2$ = 7 and 15 
ensure freedom for local radial interferences and hybridizations in the examined phases.

Section 2 gives a detailed introduction to the concept of 
joined planar-radial interferences.
A few examples of the planar-radial interplay support the confidence into the
reality of this unification.
Interference conditions for the planar and radial components 
of the joined interferences are supplied for the subsequent applications in the 
Sections 3 and 4. 
Section 3 shows that diamond- and zinc-blende-type phases distribute the charges  
according to interference aspects.
In particular the role of voids (empty spheres) will be examined.
Moreover, $p$-dominated networks are shown to be highly important for phase stability.
Section 4 goes the step to three $\gamma$-brasses which give rise to remember
the intrinsic physical content of the Hume-Rothery rules.
 
\section{Structure stability and electronic interference}

\subsection{Momentum sphere}   

Suppose a cubic crystal with $N_{as}$ effective atoms (atomic spheres, AS) per SC unit cell (side length $a$)
and $\nu_{as}$ spin-orbitals per AS.
This defines a space-filling atomic-site angular-momentum basis of the total electron quantum space, 
dimension $D_{tot}$=$N_{as}\nu_{as}$.
Each electron band state is specified by a normalized set of orbital amplitudes
which are subjected to fermionic restrictions with respect to other band states at lower energies.
Various integrated subspace-projected densities of states, $I_{sub}(\epsilon)$, describe the resulting utilization of the EQS
up to given energies, $\epsilon$.

The present study obtains the $I_{sub}(\epsilon)$ by means of  
the LMTO-ASA \cite{Andersen75} with the special feature \cite{Arnold97b}
that the atomic spheres contain muffin-tin potentials (non-overlapping).
This LMTO-ASMT ensures some proximity to both the muffin-tin scattered-wave (MT-SW) concept and the FLAPW method
(full-potential linearized augmented plane wave \cite{Wimmer81,Weinert82}).
Above all, we intend to describe various systems on equal footing, knowing that the AS-approximation has drawbacks (e.g. reduced gap widths).
However, charges can definitely be assigned to atomic sites.
On this basis we will examine the role of electronic interference in the stabilization of structures.

Following Hume-Rothery, Mott, and Jones \cite{Hume26,Mott58a} (HMJ-concept)
we adopt that structure stability on a given length scale, $\Lambda$, can arise from electronic interference
in an active space (total EQS or subspace) around a specific energy, e.g. an interplanar distance, 
the $sp$-subspace, and the Fermi energy, $\epsilon_F$. 
An active space is spanned by all AS-orbitals which are clearly involved in the interference process.

For estimating electronic interference in valence band states we skip to the extended momentum space ($k$-space). 
Band states at the energy $\epsilon$ must be subjected 
to equivalent conditions in the momentum representation as in the atomic-site angular-momentum representation.
Since the very beginning, the adequate characterization of these conditions turned out to be the crucial point.
Generally, the HMJ-concept has proved successful if interference is estimated 
at a $k$-space sphere instead of the real Fermi surface \cite{Hume62a, Asahi05a,Paxton97}.
At the energy $\epsilon$, we adopt a $k$-space sphere of the 
volume $(I_{\rm act}(\epsilon)/2) (2\pi/a)^3$,
hereafter referred to as the active momentum sphere (MS).
This ensures that the band states in both representations are subjected to fermionic restrictions 
with respect to the really active quantum volume.

Checking interference conditions with the MS refers to the self-consistently determined 
valence-electron concentration (VEC) in the active space.
At specific VEC values, the size of the MS allows for constructive interference,
and the band states are notably affected by this interference.
Consequently, spectral weight is accumulated in the active space.
Pronounced spectral features of the DOS projected to the active space can thus be labelled by the generating interferences. 

Note that the influence of electronic interference is two-fold.
On the one hand, depending on the size of the MS,
fulfilled interference conditions determine the composition of each band state.
On the other hand, the number of band states per unit energy (total DOS, $t$DOS)
is affected by fulfilled interference conditions, too.
The DOS projected to the active space reveals the average common effect.

Besides the MS we will apply the $k$-space sphere of the corresponding free-electron space, 
counted from the bottom of the valence band.
The deposition of spectral weight in the active space acts bonding if the MS indicates fulfilled interference conditions
at lower energies than this free-electron $k$-space sphere. 

\subsection{The general interference condition}   

Interference in the backscattering along the diameter of the MS (true backscattering, Figure \ref{fig:fig01s2}) is most important because there are many 
related cases nearby.
True backscattering by a stack of $\Lambda$-spaced obstacles (lattice planes, shells of neighbor atoms) 
is supported by constructive interference 
once the phase gain, $2k\Lambda$, on going forth and back between successive obstacles is 2$\pi$.
The diameter of the MS should thus equal 2$\pi/\Lambda$,
and the $k$-space volume enclosed is $(\pi/6)(2\pi/\Lambda)^3 \equiv (M_{\rm act}(\Lambda)/2)(2\pi/a)^3$.
We define $M_{\rm act}(\Lambda)$ which is twice the number of cells $(2\pi/a)^3$ inside the MS or, in other terms, the quantum volume in the active space 
to be enclosed by the MS at interference on the length scale $\Lambda$.
This provides the general interference condition,

\begin{equation}
           2\left(3\pi^2\frac{M_{\rm act}(\Lambda)}{a^3}\right)^{\frac{1}{3}}=\frac{2\pi}{\Lambda}.
\Label[INTL]
\end{equation}

Suppose that $D_{\rm act}$=$N_{\rm act}$$\nu_{\rm act}$ is the dimension of the active space where $N_{\rm act}$ sites contribute each $\nu_{\rm act}$ spin-orbitals.
Hence, the required quantum volume $M_{\rm act}(\Lambda)$ per SC unit cell can be decomposed into contributions $Z(\Lambda)$ per participating site,

\begin{equation}
           M_{\rm act}(\Lambda) = Z(\Lambda) N_{\rm act}.
\Label[VVL]
\end{equation}

\noindent
$Z(\Lambda)$ will be referred to as the virtual valence for interference on the length scale $\Lambda$.
This is just the classical result of Mott and Jones \cite{Mott58a}. 

We ask moreover for the energy $\epsilon(\Lambda)$ where interference on the scale $\Lambda$ affects the band states which
implies $I_{\rm act}(\epsilon(\Lambda))=M_{\rm act}(\Lambda)$.
After replacing $I_{\rm act}(\epsilon(\Lambda))=i_{\rm act}(\epsilon(\Lambda)) N_{\rm act}$ by the integrated DOS projected to the active space
 per participating site we obtain
\begin{equation}
           i_{\rm act}(\epsilon(\Lambda))=Z(\Lambda).
\Label[INTL1]
\end{equation}
\noindent
We insert (\ref{VVL}) with (\ref{INTL1}) into (\ref{INTL}) and arrive at the implicit condition for the energy $\epsilon(\Lambda)$,

\begin{equation}
	2\left(3\pi^2\frac{N_{\rm act} i_{\rm act}(\epsilon(\Lambda))}{a^3}\right)^{\frac{1}{3}}=\frac{2\pi}{\Lambda}.
\Label[INTL2]
\end{equation}

This shows that the present study follows a rather classical concept:
The required momentum transfer in the extended $k$-space results from an empty-lattice consideration whereas
the size of the MS accounts for both the actual deposition of quantum weight and the fermionic restrictions.

In the following, we will ever start with the search for the a possible active space.
Thereafter we calculate the DOS projected to the active space by the LMTO-ASMT 
and apply (\ref{INTL2}) with the right-hand side specified to the sequence 
of planar respectively radial length scales of the employed structure model.

\subsection{Planar interference}   

In the real space, the SC crystal is composed of equidistant lattice planes
with interplanar distances,

\begin{equation}
           d(G^2)= \frac{a}{\sqrt{h^2+k^2+l^2}}=\frac{a}{\sqrt{G^2}}.
\Label[HKL]
\end{equation}

\noindent
The Miller indices $(hkl)$ are integers without a common measure. 
Lattice planes with the same interplanar distances form a group 
labelled by $G^2$=$h^2$+$k^2$+$l^2$.

The sequence of $G^2$ of the SC lattice reveals gaps for two reasons: 
(i) There are values such as 7 which cannot arise from three squared integers. 
(ii) There are other values such as 8 which cannot arise from three squared integers without a common measure. 
Below we will deal with cases where sublattices are inscribed into the SC unit cell, 
and the SC lattice itself serves as a reference.
Some $(hkl)$ triples with common measures may thus represent occupied lattice planes, too,
such as (220) with half the planar spacing of the (110) planes in the case of 
an inscribed 2$\times$2$\times$2-bcc sublattice. 

\begin{figure}
\centering
\includegraphics[width=5cm]{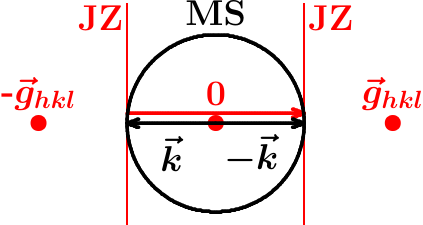}
\caption{True backscattering with fulfilled Bragg condition.}
\label{fig:fig01s2}
\end{figure}

The SC reciprocal lattice vectors including a factor of 2$\pi$,
${\vec{g}}_{hkl}$=$(2\pi/a)(h,k,l)$, 
point along the normal directions of the lattice planes $(hkl)$,
and the interplanar distance is specified by $|{\vec{g}}_{hkl}|$=(2$\pi/a)\sqrt{G^2}$=2$\pi/d(G^2)$.
Jones zones (JZ) are assigned to the vectors {$\vec{g}_{hkl}$}
as the perpendicular bisecting planes (Figure \ref{fig:fig01s2}). 
Once the MS touches a pair of JZ
the plane waves $\vec{k}$ and -$\vec{k}$ at the contact points are related by the fulfilled Bragg condition in the true backscattering,

\begin{equation}
\vec{k}+\vec{g}_{hkl}=-\vec{k}, \hspace*{0.5cm} 2|\vec{k}|=|\vec{g}_{hkl}|={\rm \frac{2\pi}{d(G^2)}}.
\Label[BRAGG]
\end{equation}

\noindent
For such planar interferences $G^2$ the general interference condition (\ref{INTL}) with (\ref{VVL}) reads as

\begin{equation}
        N_{\rm act} Z(d(G^2)) = \frac{\pi}{3} \left(\frac{a}{d(G^2)}\right)^3 = \frac{\pi}{3} (G^2)^{\frac{3}{2}}.
\label{PLZ}
\end{equation}

Electron-structure interrelations via $(hkl)$ Bragg effects emerge in the true backscattering with high statistical weights
and continue acting at rising energy as specular reflections with lower statistical weights.
This causes some broadening of spectral features.

Note that only the utilized quantum volume per SC unit cell is determined by (\ref{PLZ}) at the planar interference $G^2$.
The assigned virtual valence, $Z(d(G^2))$, depends on the number of contributing sites, $N_{\rm act}$. 

Note moreover that structure-specific selection rules
of the diffraction theory, such as $G^2$=3,8,11,16,19,\dots for diamond structures, are due to single scattering and distant fields.
The band electrons contribute close fields and are subjected to multiple scattering.
One may rather expect the whole sequence of the Bragg effects in the SC reference crystal
with some preference for allowed diffraction events.

Even gaps in the sequence of $G^2$ mentioned above may be significant due to the 
absence of planar interference which allows for undistorted local processes.

\subsection{Joined planar-radial interference and hybridization}   

One may suppose that the electronic stabilization of structures is finally
due to the charge redistribution on the shortest interatomic length scales,
 i.e. due to hybridization on these scales.
Inside the band states, hybridization adjusts the orbital amplitudes at single sites (such as $sp^3$) or at sites in contact (molecule-like  bonding/anti-bonding superpositions).
Regarding the HMJ-concept, however, the term hybridization is often restricted to the
integration of resonant $d$-orbitals into band states besides $sp$-orbitals.
There is confident evidence for close interrelation between this kind of hybridization and the HMJ-stabilization of structures.
For various systems, it has been demonstrated that the stabilizing pseudogaps are weakened or even disappear after removing hybridizing links in
LMTO calculations \cite{TdL95,TdL05a,Mizutani06a,Mizutani11b}.

The HMJ-concept rests on interference conditions for plane waves.
Interference aspects are thus exclusively attributed to the extended planar order
whereas hybridization is treated as a competing local effect.
However, the discovery of amorphous Hume-Rothery phases \cite{Haussler83a} has proved that interference-based stabilization occurs due to the medium-range radial order around reference atoms even without long-range planar order.
The stability of many amorphous phases has been experimentally investigated by H\"aussler $et\;al.$ \cite{Haussler07,Stiehler07}
within the concept of the resonance between the electrons and the spherically periodic sequence of neighbor-shells.

Following Watanabe and Ishii \cite{Watanabe08a}, the ranking of the elementary processes in the stabilization 
of structures should be reconsidered with some preference for hybridization.
However, on-site hybridization at a given AS requires backscattering by a non-spherical environment,
and this includes the radial interference of the reflected waves from the neighbor shells.
A planar interference at a related momentum transfer can link the local radial interferences
throughout the crystal to become an extended phenomenon.
Hence, at given translation symmetry of the crystal (size and shape of the unit cell)
certain hybridizations inside the unit cells (interatomic distances) must be favored.

In the following we focus on two issues, (i) the momentum transfers where radial interferences and the assigned hybridizations 
occur and (ii) the formation of joined planar-radial interferences. \\[1ex]
{\bm $Radial$ $order$ $is$ $linked$ $with$ $planar$ $order$}.
Radial order appears in the average radial density of neighbor atoms around reference atoms, 4$\pi r^2{\cal{N}}g(r)$, where
$\cal{N}$ is the average atom number density of the system.
The pair correlation function, $g(r)$, reveals the sequence of neighbor shells around reference atoms by ($g(r)$ - 1),
i.e. by the deviation from the homogeneous distribution of neighbors.

Suppose a strong spectral component, $\vec{k}$, of a band state close to a fulfilled planar interference condition in the true backscattering 
by a stack of lattice planes orthogonal to $\vec{k}$.
How does it notice the existing radial order?
The situation resembles a diffraction experiment.
For estimation purposes we are only interested in general structure-related properties and 

\begin{figure}
\centering
\includegraphics[width=6.3cm]{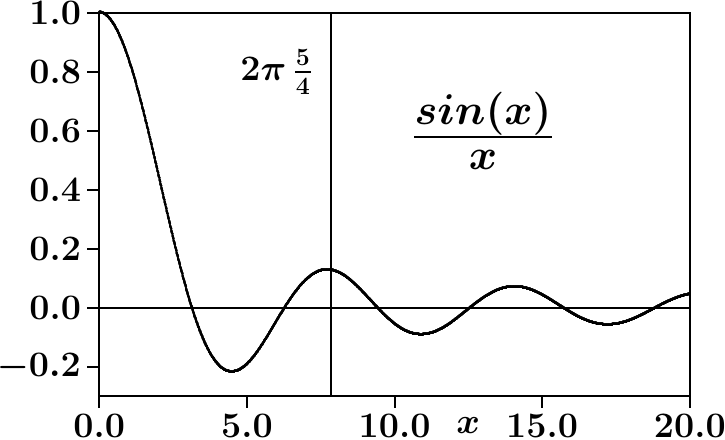}
\caption{The interference function ($x=2kr$).}
\label{fig:fig02s2}
\end{figure}

\noindent
confine thus to identical $s$-scatterers at all lattice sites.
Calculating the intensity of the true backscattering of this spectral component, $\vec{k}$, now accounting for the average radial order,
one arrives at the structure factor $S(q)$ with the momentum transfer $q=2k$, 
\begin{equation}
           S(2k)-1 = \int dr \; 4\pi r^2 {\cal{N}} \; (g(r)-1) \; \frac{sin(2kr)}{2kr}.
\Label[SF]
\end{equation}
\noindent
The structure factor (\ref{SF}) gets large at the main diffraction peak (MDP), $2k=K_1$,
where the sequence of the neighbor-shell radii, $R_s$, and $2k$ are tuned 
as to encounter nearly the maxima of the interference function, $sin(2kr)/2kr$ (Figure \ref{fig:fig02s2}),
\begin{equation}
	K_1 R_s=x_s = 2\pi(1/4 + s), \; s=1,2,3,\dots \;\; .
\Label[K1RS]
\end{equation}
Scattered waves of all neighbor shells interfere constructively with each other.
We conclude that the average neighbor-shell sequence at maximum coupling to an assigned set of lattice planes
is restricted to spherical periodicity,
\begin{equation}
        R_{\rm s}=\frac{2\pi}{K_1}(1/4 + s), \;  s=1,2,3,\dots \;\; .
\Label[RS]
\end{equation}
Similar to the planar case (\ref{BRAGG}), the momentum transfer $K_1$ in (\ref{K1RS}) is related 
to the neighbor shell spacing, $\Lambda$, by $K_1$=2$\pi$/$\Lambda$. 
A second relation can be drawn from (\ref{K1RS}), $R_1$=(5/4)$\Lambda$, which has no analog in the planar case.
It indicates that the spherically periodic neighbor-shell sequence (\ref{RS}) is pinned at the central atom.
Hence, the radial interference condition is twofold,

\begin{equation}
        K_1=\frac{2\pi}{\Lambda}=\frac{2\pi}{(4/5)R_1},
\Label[RADI]
\end{equation}

\noindent
and the general interference condition (\ref{INTL}) with (\ref{VVL}) can be specified to radial interferences $R$ by

\begin{equation}
        N_{\rm act} Z(R) = \frac{125\pi}{192} \left(\frac{a}{R}\right)^3.
\label{RAZ}
\end{equation}

The strongest planar-radial coupling via mutual true backscatterings requires
the momentum transfer at the MDP (\ref{RADI}) close to the momentum transfer at the assigned planar interference (\ref{BRAGG}).
Hence, we define the equivalent interatomic distances, $R(G^2)$, of an interplanar distances, $d(G^2)$, by
\begin{eqnarray}
\Label[RHKL]
      	\frac{2\pi}{(4/5)R(G^2)}&=&\frac{2\pi}{d(G^2)}=\frac{2\pi}{a}\sqrt{G^2} \\
	R(G^2)&=&\frac{5}{4}d(G^2)=\frac{5}{4}\frac{a}{\sqrt{G^2}}
\Label[RHKL1]
\end{eqnarray}
\noindent
In the following, the term ''joined planar-radial interferences and hybridizations`` (short ''joined interferences``)
will be applied to interferences which fulfil approximately the conditions (\ref{RHKL}). \\[1ex]
\noindent
{\bm $Radial$ $interference$, $DOS$, $hybridization$}.
The periodic sequence of neighbor-shells (\ref{RS}) around central atoms must cause interferences of the inward backscattered spherical waves
which are due to a singular outgoing spherical wave emerging from the central atom.
This kind of radial interference is part of the one-particle Green function, and
hence it must affect both the AS-DOS and the hybridization at the central atom.
We will show that just the interference conditions (\ref{RADI}) specify such situations.

Searching for a diffraction-related representation of the AS-DOS in a finite atom cluster 
\cite{Solbrig87a,Solbrig92a,Arnold95} we have used
the Green function in terms of a scattering-path matrix,
\begin{equation}
      T_{sL,s^{\prime}L^{\prime}} = e^{i\eta_{sl}} \langle sL| \left(I - P F\right)^{-1}P|s^{\prime}L^{\prime}\rangle \, e^{i\eta_{s^{\prime}l^{\prime}}},
\label{SPMAT}
\end{equation}
\noindent
which is a special adaption of the scattering-path operator (Gyorffy and Stott \cite{Gyorffy71}).
The $\eta_{sl}$ denote the real scattering phase shifts of the AS number $s$ (AS$s$), 
and the matrices $P$ and $F$ contain the interatomic vacuum-wave propagators
respectively the scattering amplitudes of the AS (cf. Appendix 6.2).
$T_{sL,s^{\prime}L^{\prime}}$ is the amplitude of the regular orbital in the AS$s$, angular momentum $L$=$(l,m)$,
which appears in response to a singular outgoing wave with unit amplitude in the AS$s^{\prime}$, angular momentum $L^{\prime}$.

Let $n_{sl}^o(\epsilon)$ be the partial DOS of the bare AS$s$ (without environment, a function of $\eta_{sl}$).
The diagonal elements, $T_{sL,sL}(\epsilon)$, specify the relative environment contribution in the partial DOS, $n_{sl}(\epsilon)$, 
of the embedded AS$s$ (with environment) \cite{Solbrig87a,Solbrig92a,Arnold95},
\begin{equation}
	\frac{n_{sl}(\epsilon) - n_{sl}^o(\epsilon)}{n_{sl}^o(\epsilon)} = \frac{1}{2\,l + 1} \Re\sum\limits_{m=-l}\limits^l T_{sL,sL}(\epsilon).
\label{DELD}
\end{equation}
\noindent
For estimation purposes we confine (i) to single scattering in the environment, $(I - P F)^{-1}P \approx (P + PFP)$, 
suppose (ii) identical $s$-scatterers at all atomic sites ($L$ = (0,0) $\equiv$ 0, phase shift $\eta_0$, scattering amplitude $f_0$),
and take (iii) the average, $\overline{T}_0$, of all $N$ individual $T_{s0,s0}$ in the large cluster.
\begin{equation}
      \overline{T}_0 = e^{i\eta_0} \frac{1}{N}\sum\limits_s \langle s0| PFP |s0\rangle \; e^{i\eta_0} 
\label{SPMAT1}
\end{equation}
Each element $\langle s0| PFP |s0 \rangle$ implies the sum of all neighbors at $\vec{R}_{s^\prime}$ around the reference atom at $\vec{R}_s$,
but only $|\vec{R}_{s^\prime} -\vec{R}_s|$ matters, not the orientation of $(\vec{R}_{s^\prime} -\vec{R}_s)$.
Hence, we replace the explicite summations in (\ref{SPMAT1}) by an integral with the pair-correlation function, $g(r)$.
\begin{equation}
      \overline{T}_0 = e^{i\eta_0} \frac{2\pi {\cal{N}} f_0}{k^2}\left(1 + (2k)^2 \int dr r (g(r)-1) \frac{e^{i2kr}}{i2kr}\right) \; e^{i\eta_0} 
\label{SPMAT2}
\end{equation}
\noindent
The term between the exponential factors $e^{i\eta_0}$ in (\ref{SPMAT2}) describes the average inward reflection of vacuum waves
whereas the factors $e^{i\eta_0}$ are due to the transitions from inside the AS$s$ to the vacuum and vice versa.

As a result of neutron optics, the prefactor $2\pi{\cal{N}}f_0/k^2$ is the inward-reflection coefficient, $n_{\rm Fermi} - 1$,
of an infinitesimal vacuum sphere immersed into a homogeneous medium with the Fermi refraction index, $n_{\rm Fermi} = 1 + 2\pi{\cal{N}}f_0/k^2$.
The vacuum part in (\ref{SPMAT2}) is thus composed of a contribution due to a homogeneous medium and another contribution due to the radial
neighbor-shell order which is represented by $(g(r)-1)$.

Similar to the structure factor $S(2k)$ (\ref{SF}), the latter contribution is reinforced on matching the shell sequence to the 
interference function, $e^{i2kr}/i2kr$, in (\ref{SPMAT2}).
The interference function offers two extreme alignments: On the one hand, the shell radii can follow 
the imaginary part, $-cos(2kr)/2kr$, which implies that the real part, $sin(2kr)/2kr$, is suppressed in the integral (\ref{SPMAT2}).
This applies to a Friedel-like sequence, $R_s = s \Lambda$ ($s = 1, 2, 3, \dots$).
On the other hand, the shell radii can follow the real part, $sin(2kr)/2kr$, with the corresponding suppression of the imaginary part
which applies to the sequence (\ref{RS}).

Kroha $et \; al.$ \cite{Kroha95a} have analyzed this $\pi/2$-shift between the alternative shell sequences by field-theoretical 
means.
It turned out that the sequence (\ref{RS}) should be favored even in common amorphous metals where the quasiparticle life time is large
in comparison with the transport relaxation time.
In the present approach, due to non-zero phase shifts, a small admixture of the imaginary part besides the dominating real part 
may support the relaxation towards the neighbor-shell sequence with the lowest electronic band energy.

We conclude that the radial interference conditions (\ref{RADI}) favor neighbor-shell sequences around (\ref{RS})
which enable joining radial order to planar order with notable consequences for the AS-DOS at the center of the shell sequence 
and hence for the electronic band energy.

Hybridization in electron eigenstates, $|\Psi_n\rangle$, of a large atom cluster 
appears already in the incident vacuum fields at the AS, $|\Phi_n\rangle$.
They are self-sustaining because it takes no initiating source: 
$|\Phi_n\rangle = P F |\Phi_n\rangle$. 
Hybridization results from the structure-dependent mixing of the atomic-site angular-momentum components of $|\Phi_n\rangle$
by powers of $PF$.
Hence, enhanced radial interference is generally accompanied by enhanced on-site hybridization.
The conditions (\ref{RADI}) for radial interferences indicate on-site hybridization of active orbitals, too.
The first part of (\ref{RADI}) ensures that all inward-backscattered waves from the $\Lambda$-spaced neighbor shells interfere constructively.
It depends on $R_1$ how this interference acts on the AS-DOS and hybridization 
at the central atom of the shell sequence. \\[1ex]
\noindent
{\bm $Lateral$ $stability$ $of$ $stacks$ $of$ $lattice$ $planes$}.
A stack of (hkl) planes is barely stabilized against lateral shifts
of single planes if the interplanar links have the length $R=d(G^2)$.
This is evident for simple geometrical reasons and moreover from the electronic point of view, 
because the planar and the radial interferences would occur at the notably different momentum transfers
$2\pi/d(G^2)$ (\ref{BRAGG}) respectively (5/4)$2\pi/d(G^2)$ (\ref{RADI}).
If, on the contrary, both interferences occur at the same momentum transfer, such as proposed by $R(G^2)$ (\ref{RHKL1}),
the planar-radial coupling must be strong.

\begin{figure}
\centering
\includegraphics[width=8.0cm]{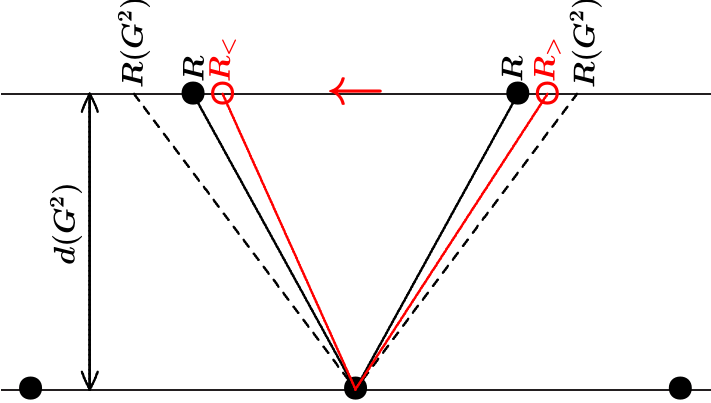}
\caption{Radial links, $R$, between lattice planes (red after
lateral shift). $R<R(G^2)$ ensures restoring forces (arrow).}
\label{fig:fig03s2}
\end{figure}

The stabilizing interplanar links, $R$, of existing systems should thus obey the relation 
\begin{equation}
 	d(G^2) < R < R(G^2) 
\Label[RIPL]
\end{equation}
\noindent
because small lateral shifts around this equilibrium must cause restoring forces.
Suppose $R<R(G^2)$ and a small shift which changes the links $R$ into $R_{>}$ and $R_{<}$ (Figure \ref{fig:fig03s2}).
The attractive coupling will be enhanced along $R_{>}$ respectively weakened along $R_{<}$
which drives back to the equilibrium distances $R$.
The key argument is that the local radial interference $R_>$ increases upon approaching $R(G^2)$ 
due to joining the planar interference $G^2$.
Starting with $R>R(G^2)$, on the contrary, causes the opposite assignment and hence destabilization.

In bcc systems, e.g., where the strongest x-ray peak is due to the planar interference (110),
the stacks of (110)-planes are stabilized against shifts along a cubic axis
by joined interferences [$d(G^2)$,$R$] = [$d(2)$,$a\sqrt{3}/2$].
The condition (\ref{RIPL}) can be reduced to read as $\sqrt{2}<\sqrt{3}<(5/4)\sqrt{2}$, i.e. $1.4142<1.7321<1.7678$.
Lateral stability is achieved at nearly the highest level of planar-radial coupling.

Summarizing this planar-radial cooperation, the Bragg effects in the true 
backscattering support stable interplanar distances, $d(G^2)$.
Radial interference and hybridization in pairs of atoms which link successive lattice planes
can support the missing lateral stability. 

\begin{table}
\small
\centering
\caption{$\Lambda/a$, $Z(\Lambda)$, $N_{act}$ (\ref{PLZ},\ref{RAZ}) of planar and radial interferences, cf. Appendix 6.1}
\vspace{0.2cm}
        \begin{tabular}{ l | l | l | l | l }
        \hline
         		 	&	& bcc 			& fcc			& dia 			\\
        $\Lambda/a$ 		& (hkl)	& $Z(\Lambda)$, 2   	& $Z(\Lambda)$, 4       & $Z(\Lambda)$, 8 	\\
        \hline
        $R3/a=\sqrt{3}/2$         &	& {\bf 1.574}		&      			&			\\
        $R2/a=\sqrt{2}/2$         &	&          		& {\bf 1.446} 		& {\bf 0.723}   	\\
        $R1/a=\sqrt{3}/4$         &	&               	&                    	& {\bf 3.149}   	\\
        $d(2)/a=\sqrt{2}/2$    	& (110)	& {\bf 1.481}       	&                      	&                      	\\
        $d(3)/a=\sqrt{3}/3$     & (111)	&                   	& {\bf 1.360}          	& {\bf 0.680}   	\\
        $d(4)/a=1/2$          	& (200)	& {\bf 4.189}     	& {\bf 2.094}          	&                      	\\
        $d(6)/a=\sqrt{6}/6$    	& (211)	& {\bf 7.695}       	&                    	&                      	\\
        $d(8)/a=\sqrt{2}/4$    	& (220)	&      		  	& {\bf 5.924}          	& {\bf 2.962}   	\\
        \hline
        \end{tabular}
\label{tab:tab01s2}
\end{table}

\subsection{Virtual valence}

Table \ref{tab:tab01s2} presents planar and radial length scales of bcc, fcc, and diamond lattices together with the
virtual valences according to (\ref{PLZ}) and (\ref{RAZ}).
The indicated interplanar distances $d(G^2)$ correspond to strong diffraction peaks.
For the short interatomic distances $R_1, R_2, R_3$ we refer to Appendix 6.1.
Joined interferences, e.g. of the diamond lattice, are [$d(3)$, $R$2] and [$d(8)$, $R$1].
The atoms in the SC unit cell form a tetrahedral arrangement 
(scale $R2 = a\sqrt{2}/2$) of centered tetrahedra 
(internal scale $R1 = a\sqrt{3}/4$) (Appendix 6.1). 
Hence, both joined interferences have specific stabilization missions. 

We deal with systems of SC translation symmetry where  
sublattices close to $n$$\times$$n$$\times$$n$-bcc are inscribed into the SC unit cells.
Table \ref{tab:tab02s2} characterizes the reference systems with true $n$$\times$$n$$\times$$n$-bcc substructure
where the $n^3$ bcc subcells per SC unit cell are stabilized by joined interferences  $[d(2n^2),R(n)] = [a\sqrt{2}/2n,a\sqrt{3}/2n]$
around the planar interferences $(nn0)$ with $G^2 = 2n^2$.

The indicated virtual valences $Z$ (\ref{PLZ},\ref{RAZ}) do not depend on $n$,
provided that all the $2 n^3$ sublattice sites contribute to the MS ($N_{\rm act} = 2 n^3$). 
If part of them is inactive, the active sites must compensate for the missing contributions.

\begin{table}
\small
\centering
\caption{SC unit cells with $n$$\times$$n$$\times$$n$-bcc sublattices inscribed,
planar interferences (nn0) ($d(2 n^2) = a\sqrt{2}/2n$), radial interferences ($R(n) = a\sqrt{3}/2n$),
virtual valences $Z$ (\ref{PLZ},\ref{RAZ}) ($N_{\rm act} = 2n^3$).}
\vspace{0.2cm}
        \begin{tabular}{ l | l | l | l  l | l }
        \hline
         	& sub-	& sites  & planar    	& 		   & radial    			\\
         	& cells	& 	 & (nn0)        & (nn1)		   &	 			\\
        $n$ 	& $n^3$ & 2$n^3$ & $Z(d(2n^2))$ & $Z$($d$(2$n^2$+1))   & $Z(R(n))$        	\\
        \hline
         1	& 1	& 2  	 & 1.481 	& 2.721 	   & 1.574           		\\
         2	& 8	& 16   	 & 1.481	& 1.767 	   & 1.574           		\\
         3	& 27	& 54   	 & 1.481 	& 1.606  	   & 1.574   			\\
         4	& 64	& 128  	 & 1.481 	& 1.551  	   & 1.574     			\\
         5	& 125	& 250  	 & 1.481	& 1.526  	   & 1.574     			\\
        \hline
        \end{tabular}
\label{tab:tab02s2}
\end{table}

\vspace{0.3cm}

Joined interferences $[d(2n^2),R(n)]$ act without perturbation 
by the next higher planar interferences $G^2 = 2n^2+1$ if the virtual valence $Z(d(2n^2+1))$  
is well above $Z(R(n))$.
Table \ref{tab:tab02s2} reveals that the structure types $n$ = 1 (CsCl) and $n$ = 2 (diamond, zinc blende) 
can easily meet this demand.
The $\gamma$-brasses ($n$ = 3), however, have $Z(d(19)) = 1.606$ only slightly above $Z(R(3))$ = 1.574.
Planar interferences $G^2$ = 19 may thus affect the radial interference in the joined interferences $[d(18),R(3)]$.
Hence, $\gamma$-brasses exist at the transition to structure stabilization by blocks of interacting planar interferences
which generally prevails in approximant phases ($n > 3$).

Electronic stabilization of the examined systems relies mainly on joined interferences $[d(2n^2),R(n)]$
because they drive towards bcc-like subcells on the scale $a/n$.
To ensure just these joined interferences the active MS 
should enclose between 2$n^3$ * 1.481 and 2$n^3$ * 1.574 states, 
i.e for $n$ = 3 between 80 and 85 states.

The empirical Hume-Rothery rule for $sp$-type alloys rests on the assumption that the
allocation of partial charges by the effective atoms of the crystal is not too much different
from the free atoms.
For estimation purposes, the effective atoms are thus replaced by free atoms.
This way, supposing only $N_{\rm act}$ = 52 sites per SC unit cell, the empirical Hume-Rothery rule
predicts alloys to form $\gamma$-brasses ($n$ = 3) if the free atoms contribute on the average  
e/a = 21/13 $\approx$ 1.615 $sp$-electrons per site.
At the Fermi energy, the MS should thus enclose 52 * 21/13 = 84 states,  
just in the expected range between 80 and 85 states.
Perturbing planar interferences $G^2$ = 19 occur somewhat above where 54 * 1.606 $\approx$ 86.7 states 
are enclosed by the MS.

Generalizing to arbitrary $\gamma$-brasses ($n$ = 3, $N_{\rm act}$ = 52, $d$-electrons possible), 
we suggest that the MS of any active subspace at any energy gives rise to stabilizing joined interferences $[d(18),R(3)]$
provided that 21/13 states are contributed per site.
This applies to the MS, calculated from first principles and regardless of the $spd$-affiliation of the contributions.
In the following, the label HRR will indicate the energies where the active MS achieves this critical size. 
\section{Diamond and zinc blende phases}

This section examines a group of 18 phases with $Fd\overline{3}m$ structure (number 227, C, Si, Ge, and Sn)
respectively $F\overline{4}3m$ structure (number 216, zinc blende phases). 
They have in common that a 2$\times$2$\times$2-bcc sublattice is inscribed into the SC unit cell 
and that the effective atoms arise from $sp$-type free atoms.

The whole crystal can be seen as a superposition of four interpenetrating 
fcc-sublattices or of two diamond-sublattices (Appendix 6.1).
Only one diamond-sublattice is occupied with atoms. 
In the LMTO-ASMT calculation, we put empty spheres (ES, no nuclear charge) to the other one.
Studies on the role of vacancies in stable structures deserve high interest since a long time.
For just this purpose we let all the 16 AS in the SC unit cell carry only $s$- and $p$-orbitals.
As a consequence, besides the redistribution of the $sp$-charges in the AS with atoms,
charging the ES is the only additional freedom for adjusting to the interference conditions.
Up to 45760 special $k$-points \cite{Chadi73,Monkhorst76} are included in the irreducible wedge of the SC reciprocal lattice.
Besides several projections of the electronic density of states we gather information on individual band states. 

The diamond phases of carbon (dia-C, $a$ = 3.567 \AA
\cite{Riley44}) and cubic boron nitride (cub-BN, $a$ = 3.615 \AA \cite{Wentorf57a}) can be regarded as prototypes of group
IV diamond phases respectively of III-V zinc blende phases.
The atomic spheres have nearly equal diameters,
and the average Pauling electronegativity of B(2.04) and N(3.04) is close to C(2.55) \cite{Pauling32,Alfred61}.

\begin{figure}
\vspace{0.3cm}
\centering
\includegraphics[width=8.5cm]{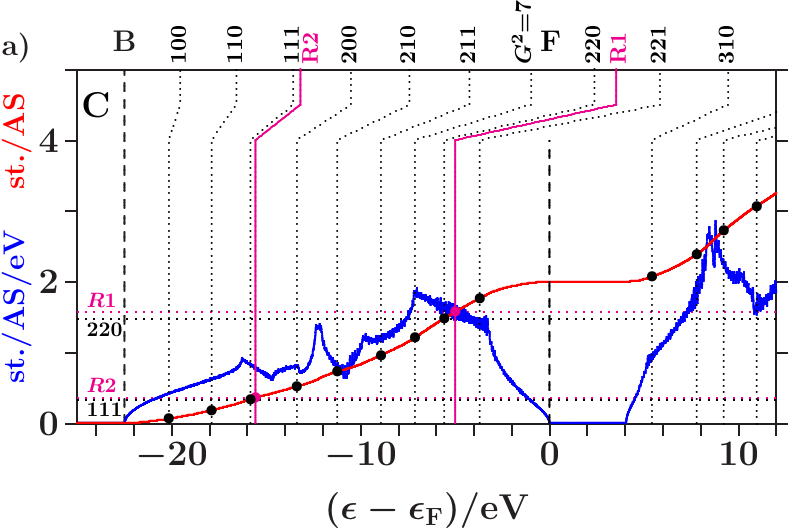}

\vspace{0.5cm}

\includegraphics[width=8.5cm]{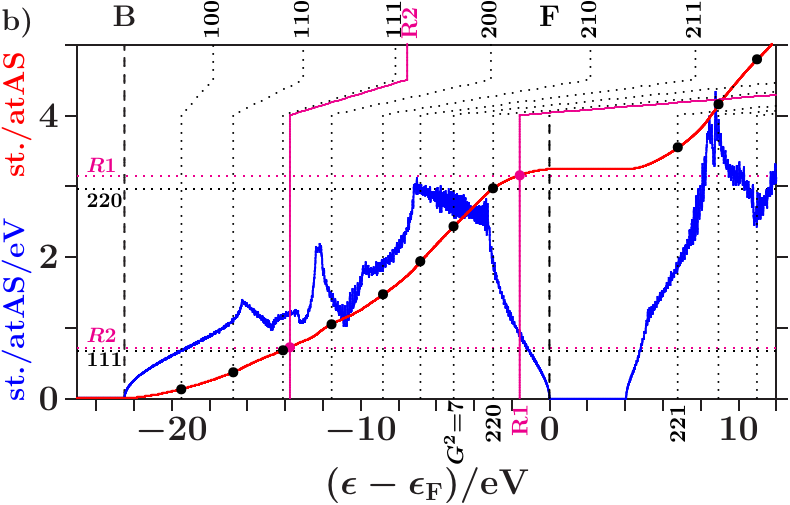}

\vspace{0.5cm}

\includegraphics[width=8.5cm]{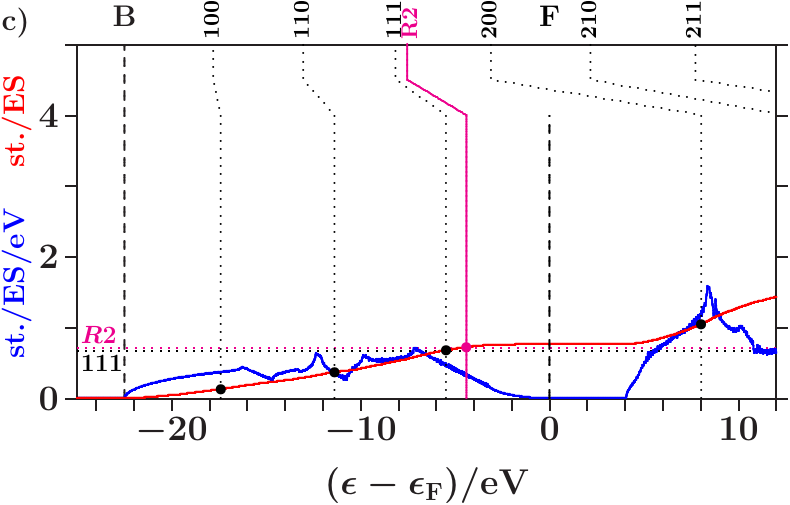}
\caption{Dia-C: Each $sp$DOS$\times$10 (blue), integrated $sp$DOS (red), projections to the AS with atoms and to empty spheres (ES),
a) average $sp$DOS(C,ES), b) $sp$DOS(C), and c) $sp$DOS(ES).
The joined interferences [$d(8)$,$R$1] and [$d(3)$,$R$2] open the gap above the Fermi energy,
see text.}
\label{fig:fig04s3}
\end{figure}

\begin{figure}
\centering
\vspace{0.3cm}
\includegraphics[width=8.5cm]{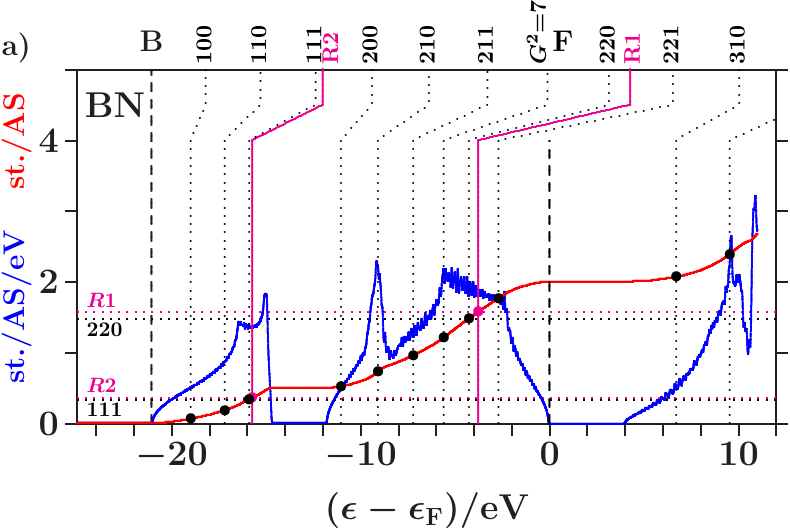}

\vspace{0.5cm}

\includegraphics[width=8.5cm]{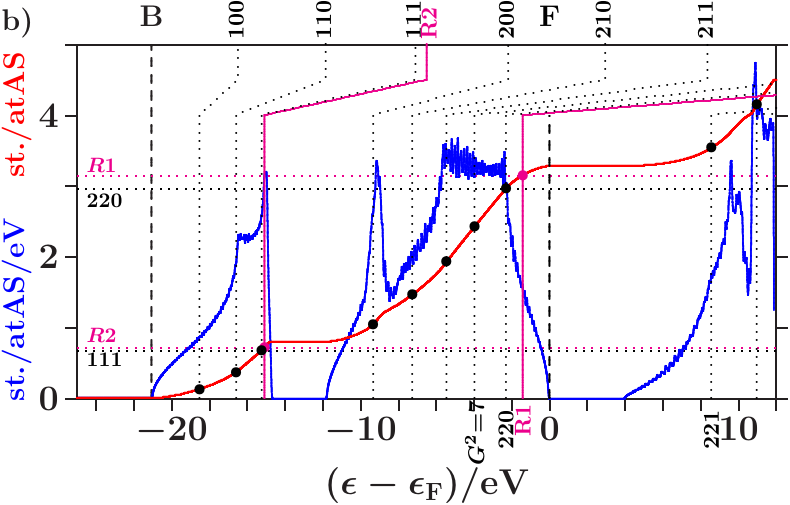}

\vspace{0.5cm}

\includegraphics[width=8.5cm]{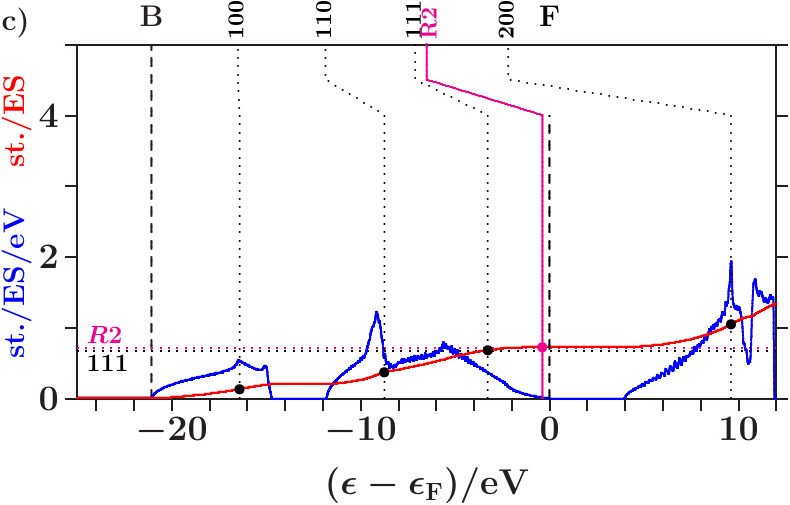}

\caption{Cub-BN: Each $sp$DOS$\times$10 (blue), integrated $sp$DOS (red),
projections to the AS with atoms and to empty spheres (ES),
a) average $sp$DOS(B,N,ES), b) average $sp$DOS(B.N), and c) $sp$DOS(ES).
The joined interferences [$d(3)$,$R$2] open the gap down in the valence band, see text.}
\label{fig:fig05s3}
\end{figure}

\newpage

\noindent
One may expect comparable trends towards charging the ES on the expense
of the AS with atoms.
However, the electronegativities of B and  N are notably different which
suggests a quite complex behavior of the average effective atom.

\subsection{Stability of the essential length scales}

Essential length scales of crystals are the large interplanar and the short interatomic distances.
The corresponding lattice planes are densely occupied which results in strong Bragg reflections,
and close pairs of atoms allow for pronounced electronic multiple scattering and hybridization. 
As shown in Table \ref{tab:tab01s2}, the 
essential length scales of alloys with diamond structure 
give rise to joined interferences [$d(8)$,$R$1] and [$d$(3),$R$2]. \\[1ex]
\noindent
{\bm $Interference$ $energies$}. 
The Figures \ref{fig:fig04s3} and \ref{fig:fig05s3} show the calculated densities of states ($sp$DOS) together with site-related projections 
and the corresponding integrated $sp$DOS.
Note that the $sp$-space is the total EQS of the present approach.
Each $sp$DOS is upscaled by a factor of 10 for reasonable appearance besides the integrated $sp$DOS.

The panels a) present the total $sp$DOS where no
distinction is made between AS with atoms and ES 
($N_{act}$ = 16, the total EQS).
The integrated $sp$DOS at $\epsilon_F$ reaches 2 states/AS.
The panels b) and c) show the projections to the AS with atoms ($N_{act}$ = 8) respectively to the ES ($N_{act}$ = 8).
Averaging b) and c) provides a).

Vertical dotted lines (labels $hkl$, increment 1 of $G^2$) denote
the energies $\epsilon(d(G^2))$ (\ref{INTL1}) of planar interferences.
The lower parts of the dotted lines show where the interference conditions are fulfilled with the MS,
i.e. the integrated $sp$DOS passes the virtual valences $Z(d(G^2))$ (\ref{PLZ}) 
(small bullets, for $\gamma$-brasses cf. Table \ref{tab:tab14s6}).
After kinks the upper parts continue at those energies where the interference
conditions are fulfilled with the integrated free-electron $sp$DOS.
These energies are somewhat larger than the equally-spaced energies, $\epsilon_{\rm fe}(G^2)$,
due to the full free-electron $k$-space sphere,

\begin{equation}
         \epsilon_{\rm fe}(G^2) = \epsilon_b + \left(\frac{\pi}{a}\right)^2 G^2.
\Label[FEDISP]
\end{equation}

\noindent
The symbol $\epsilon_b$ denotes the energy at the bottom of the valence band with reference to $\epsilon_F$.
(\ref{FEDISP}) results from $2(3\pi^2I_{\rm act}(\epsilon)/a^3)^{1/3} = (2\pi/a)\sqrt{G^2}$
replacing $(3\pi^2I_{\rm act}(\epsilon)/a^3)^{1/3} = (\epsilon - \epsilon_b)^{1/2}$.

Radial interferences on the scales $R1$ = $a\sqrt{3}/4$ and $R2$ = $a\sqrt{2}/2$
(Table \ref{tab:tab01s2}) are indicated by magenta lines and bullets in the Figures \ref{fig:fig04s3} and \ref{fig:fig05s3}. \\[1ex]
\noindent
{\bm $Interference$ $in$ $the$ $total$ $EQS$}.
Both panels a) suggest that the wide gaps above $\epsilon_F$ are associated with the joined interferences [d(8),R1]
which control the internal stability of the centered tetrahedra (length scale $R$1).
Depending on the applied $k$-space spheres we obtain different energies of the planar interferences (220).
The free-electron $sp$-spheres predict them inside the gaps where no electron is affected because
spectral weight is fully removed starting with the first scattering event (Figure \ref{fig:fig68s6}).
The self-consistently determined MS, on the contrary, predict them at lower energies in the hat-shaped 
features where spectral weight is accumulated.
Enhanced fluctuations of the $sp$DOS reveal that certain waves are amplified by constructive interference whereas a
great amount of smoothing weak waves is missing 
due to the intentionally sparse $k$-space sampling.
Hence, the assignment (\ref{INTL1}) of interferences to band states is realistic.

Joined interferences [$d(3)$,$R$2] in the lower valence band support the tetrahedral arrangements of the two
types of centered tetrahedra, one with atoms at the center - the other with ES (Figure \ref{fig:fig66s6}).
Cub-BN (Figure \ref{fig:fig05s3}a) generates a wide gap whereas dia-C shows only a weak signature (Figure \ref{fig:fig04s3}a),
despite of the higher bulk modulus (C(442 GPa), BN(369 GPa), \cite{BoXu13}, Table \ref{tab:tab10s6}).
Both systems differ at least regarding the stabilization of the arrangement
of the centered tetrahedra in the SC unit cell. 

Note that the joined interferences [$d(8)$,$R1$] in the hat-shaped features (panels a) 
act fairly deep below the Fermi energy.
Hence, it seems reasonable to ask whether the stabilization processes so far assigned to interferences in the total EQS
do really emerge in subspaces. \\[1ex]
\noindent
{\bm $Are$ $there$ $active$ $subspaces?$}.
We decompose the total EQS into two subspaces, namely that of the
AS with atoms ($N_{act}$ = 8) and that of the ES ($N_{act}$ = 8).
Suppose, on approaching $\epsilon_F$ the band states are more and more confined to the 
subspace of the atoms because the joined interferences [$d(3)$,$R$2]
generate a separate occupation edge in the subspace of the ES fairly below $\epsilon_F$.
Hence, for the interference in band states at $\epsilon_F$, the whole spectral weight in the ES is hidden,
interference conditions include only the spectral weight on the sublattice of the atoms.

In that case  essential length scales are stabilized in separate subspaces on approaching $\epsilon_F$.
No stabilizing gap is required down in the low valence band.
Vice versa, such a gap due to the interferences [$d(3)$,$R$2] is required for a stable arrangement
of the centered tetrahedra if the complete separation of both subspaces below $\epsilon_F$ is not achieved.

\begin{figure}
\vspace{0.15cm}
\centering
\includegraphics[width=8.5cm]{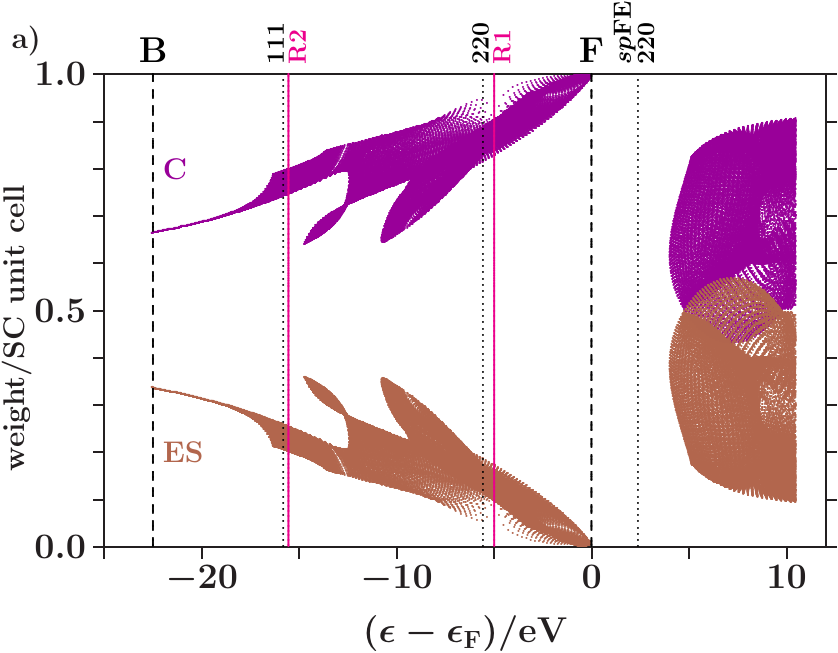}

\vspace{0.3cm}

\includegraphics[width=8.5cm]{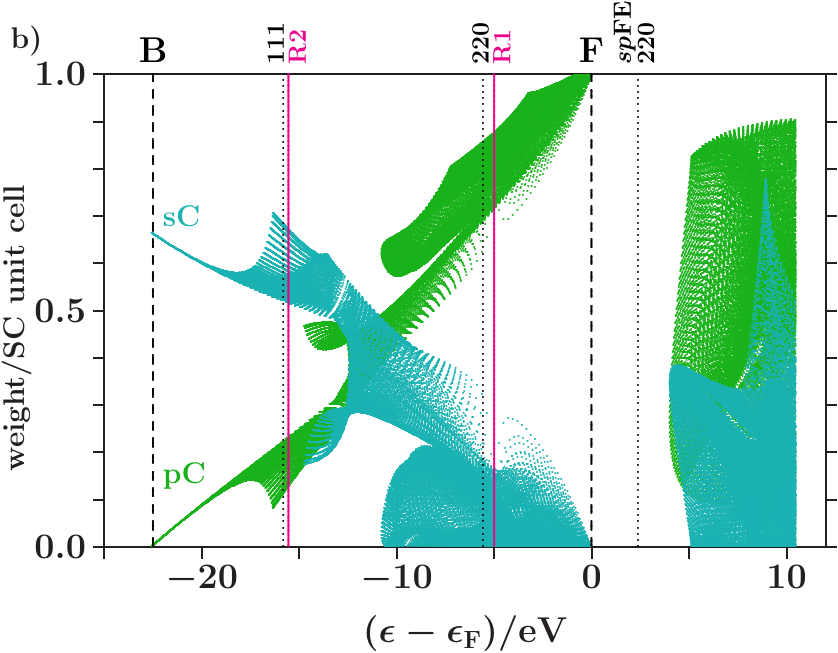}
\caption{Dia-C, the band states: Decomposition of the norm One into partial weights, a) $sp$C-weight versus $sp$ES-weight,
b) $p$C-weight versus $s$C-weight.}
\label{fig:fig06s3}
\end{figure}

It turns out that this separation towards $\epsilon_F$ is almost
perfect in dia-C but incomplete in cub-BN and the other 16 examined phases.
For proof we analyze the compositions of the band states.
In Figure \ref{fig:fig06s3}a each band state of dia-C contributes
two dots which represent the parts of the norm One on
the sublattices of C respectively of the ES.
Approaching $\epsilon_F$, the band states are more and more confined to the 

\vspace{0.1cm}

\begin{figure}
\centering
\includegraphics[width=5.5cm]{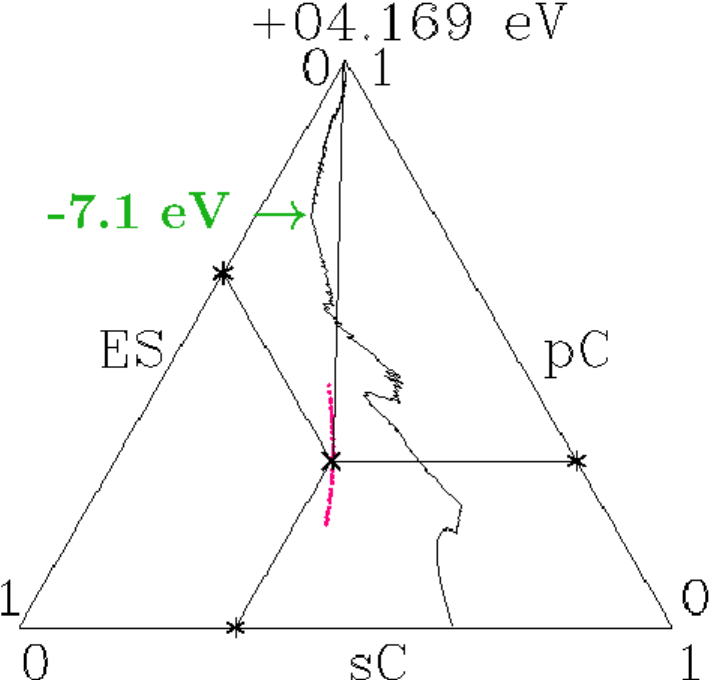}
\caption{Dia-C: The norm One of each band state decomposed into partial 
weights $s$C, $p$C, and ES (red dots)
and the tracking curve of the average in each energy interval. 
A $p$-dominated network forms above -7.1 eV up to the 
topmost point ($s$C,$p$C,ES)=(0.0,0.995,0.005) at the Fermi energy.
A straight line bridges over the gap.}
\label{fig:fig07s3}
\end{figure}

\begin{figure}
\centering
\includegraphics[width=8.5cm]{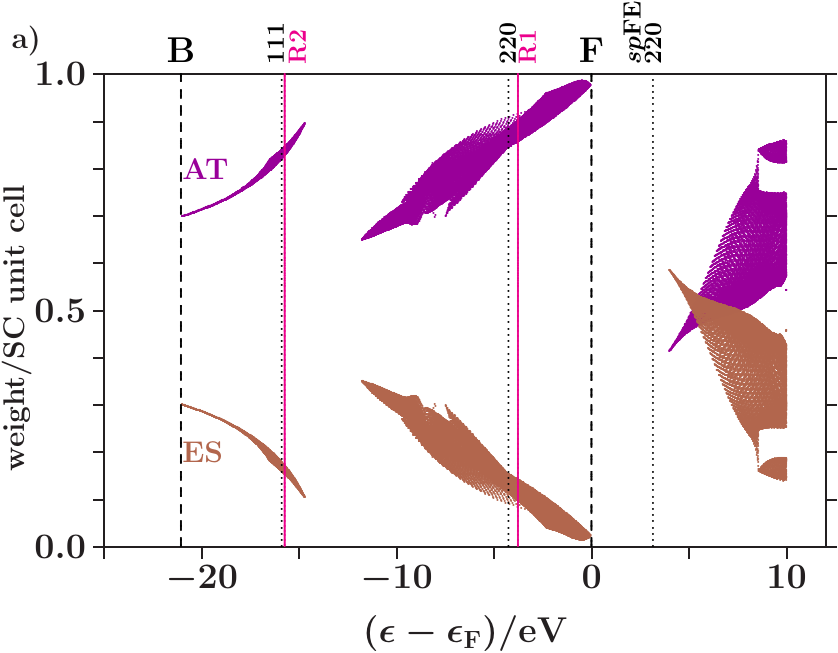}

\vspace{0.4cm}

\includegraphics[width=8.5cm]{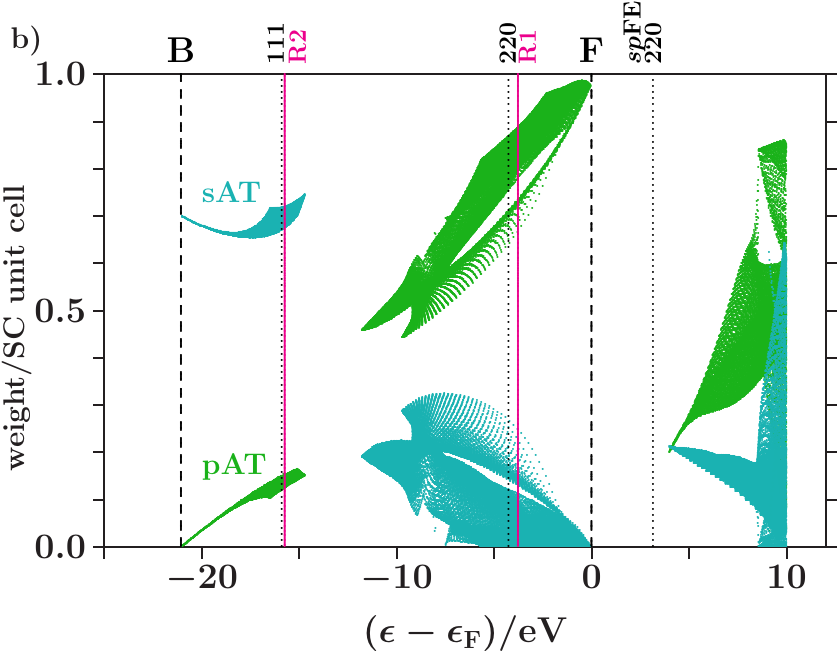}
\caption{Cub-BN, the band states: AT collects the weights on B and N.
Note the incomplete separation of AT from ES at $\epsilon_F$ in a) and the gap in the lower valence band due to the
joined interferences [$d(3)$,$R$2]}
\label{fig:fig08s3}
\end{figure}

\begin{figure}
\centering
\includegraphics[width=5.5cm]{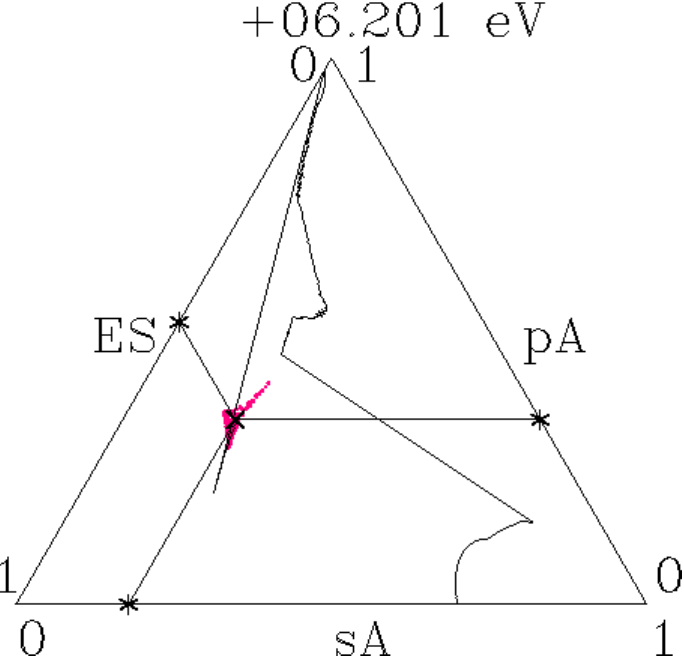}
\caption{Cub-BN: The same specifications as in Figure \ref{fig:fig07s3}.
The symbol A denotes all atoms (B+N).
Two straight lines bridge over two gaps.
The tracking curve reaches only the topmost point ($s$A,$p$A,ES) = (0.0,0.976,0.024) at the Fermi energy, see text.}
\label{fig:fig09s3}
\end{figure}

\noindent
sublattice of C and, moreover, to the $p$C orbitals (Figure \ref{fig:fig06s3}b).
Hence, the states towards $\epsilon_F$ form almost pure $p$C-networks.

Figure \ref{fig:fig08s3}a applies to cub-BN.
The symbol AT designates both atoms, B and N.
Contrary to Figure \ref{fig:fig06s3}a the band states towards $\epsilon_F$ are less clearly confined to the atoms,
a small coupling to the ES remains.
The $s$-orbitals are again excluded (Figure \ref{fig:fig08s3}b).
In both phases, certain branches of band states are thinned out by the joined interferences [$d(8)$,$R1$].
Building $p$-dominated electron networks in band states towards $\epsilon_F$ must be an integral part of stabilization.

The Figures \ref{fig:fig06s3} and \ref{fig:fig08s3} show the areas which are occupied by band states, it is not intended to characterize individual band states.
Just the latter is aimed at by the visualization suggested in Appendix 6.6.  
Employing the data set of Figure \ref{fig:fig06s3} this method is now specified to the
decomposition of the EQS into the three subspaces of $s$C, $p$C, and ES.
By means of the PGPLOT routines \cite{Pearson} an animation is obtained which displays, with stepwise 
rising energy, red dots for all band states in an energy 
interval around the actual energy and the black tracking curve of the average (Figure \ref{fig:fig07s3}).
Note that the tracking curve can also be derived from the relative contributions made by the projected densities 
of states to the total DOS (cf. Appendix 6.6).

Approaching $\epsilon_F$ in Figure \ref{fig:fig07s3}, the fluctuations of the red dots cease down gradually, 
and the tracking curve reaches its topmost point with the 
coordinates ($s$C,$p$C,ES) = (0,1-$\delta$,$\delta$) where $\delta$ $\approx$ 0.005.
The norm One is almost fully deposited to the subspace $p$C, and the band states form nearly pure $p$C-networks.
A green arrow in Figure \ref{fig:fig07s3} indicates the onset of the plateau of the hat-shaped feature in Figure \ref{fig:fig04s3}a.
At this energy the tracking curve turns clearly towards the point
(0,1,0) due to enhanced hybridization in absence of perturbing planar interferences $G^2$ = 7  (Figure \ref{fig:fig04s3}a).

Above the Fermi energy a straight line bridges over the gap.
At the upper bound of the gap, the fluctuation patterns, $\Delta$$p$C = -2$\Delta$ES = -2$\Delta$$s$C, 
reveal that equal losses of ES and $s$C enable gains of $p$C twice as large and vice versa (cf. Appendix 6.6).

Figure \ref{fig:fig09s3} shows the corresponding for cub-BN.
There are two straight lines across two gaps. 
Different to Figure \ref{fig:fig07s3}, the tracking curve reaches only the point 
(0,0.976,0.024) at $\epsilon_F$, 
i.e. the band states are not completely confined to the atoms. \\[1ex]
\noindent
{\bm $Interference$ $in$ $subspaces$}.
At this point some information is available on the composition of band states which can support the 
analysis of the site-projected DOS curves in the Figures \ref{fig:fig04s3} and \ref{fig:fig05s3}.

The panels b) reveal interference in the subspace of the atoms. 
Hybridization along the hat-shaped features takes advantage of missing planar interferences $G^2$ = 7.
In both phases, an interesting transition appears between the upper edge of the plateau and the 
the point of inflection below $\epsilon_F$.
The dominating interference changes from the planar (220)-type to the radial $R$1-type with the assigned hybridization. 
Obviously, the latter processes in the subspace of the atoms ensures the internal stability of the atom-centered tetrahedra.

The panels c) indicate interference in the subspace of the ES.
Contrary to the panels b) the kinks of the vertical lines point towards lower energies. 
This means that the deposition of electrons in the charged ES acts anti-bonding.
Charge is moved to the ES for a resulting bonding effect in the subspace of the atoms.

Joined interferences [$d(3)$,$R2$] in the subspace of the ES control both the size 
of the ES-centered tetrahedra and the arrangement in the SC unit cell.
In dia-C (Figure \ref{fig:fig04s3}c), fairly below $\epsilon_F$, the joined interferences [$d(3)$,$R2$] 
generate a soft occupation edge and enable this way increasingly C-confined band states 
towards $\epsilon_F$ (Figure \ref{fig:fig07s3}).
Approaching $\epsilon_F$, joined interferences [$d(8)$,$R1$] are confined to the C-sublattice,
and the interference conditions refer only to the occupation of the C-sublattice ($N_{\rm act}$ = 8).
Hence, in the panels b) and c), the required contributions per site are twice as large as in panel a).

Different cub-BN (Figure \ref{fig:fig05s3}c) with the same structure but two inequivalent effective atoms.
In the subspace of the ES, radial interference and hybridization on 
the scale $R2$ occur at very low spectral density immediately below $\epsilon_F$.
No separate occupation edge is generated in the ES-subspace.
The arrangement of the ES-centered tetrahedra may thus be less efficiently stabilized than in dia-C.
However, down in the valence band opens a wide gap which can be assigned 
to the joined interferences [$d(3)$,$R2$] in the subspace of the atoms.

\subsection{Stabilizing $p$-networks}
\begin{figure}
\centering
\includegraphics[width=6.0cm]{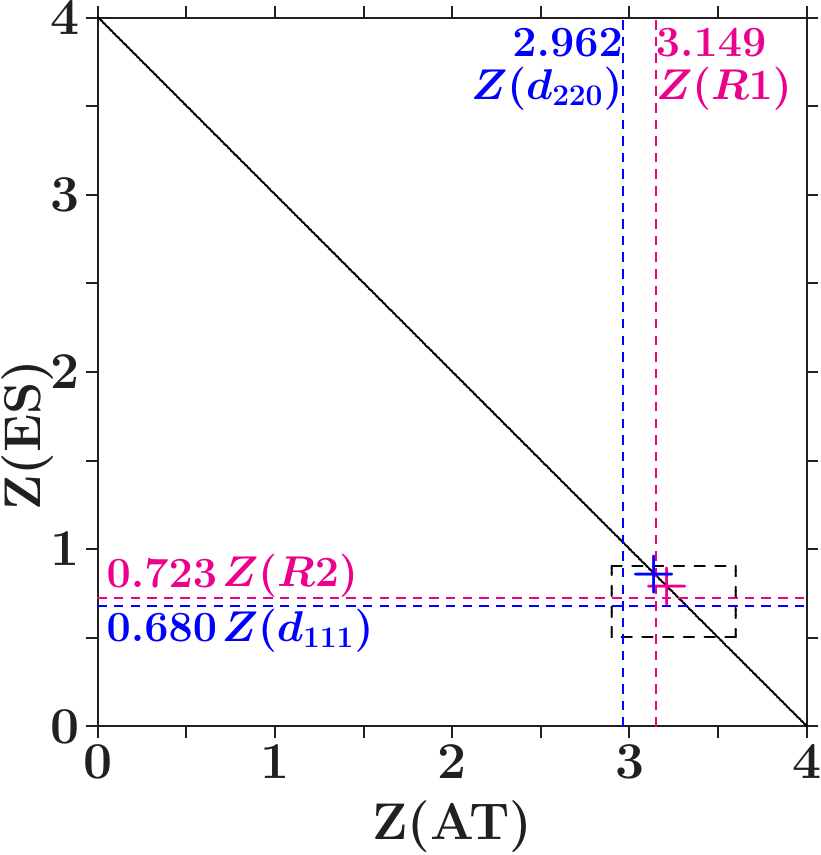}
\caption{Charges of the atoms, $Z$(AT), of the empty spheres, $Z$(ES),
crosses at equal distances to both radial (magenta) and to both planar (blue) virtual valences (dashed lines).
The box encloses the 18 phases of this study.}
\label{fig:fig10s3}
\end{figure}

Essential radial length scales of dia-C are stabilized below $\epsilon_F$
by interferences and hybridizations in separate subspaces.
In that case the 32 valence electrons per SC unit cell which are contributed by the 8 free C-atoms 
should be allocated to these sublattices as to ensure at $\epsilon_F$ equal deviations from the
respective virtual valences.  
In the group of the 18 examined IV-IV, III-V, and II-VI 
phases this expectation may be violated the more the wider a gap opens down in the valence band.
A wide gap indicates preference for the stabilization of the scales $d(3)$ and $R2$ 
in the subspace of the atoms deep in the valence band. \\[1ex]
\noindent
{\bm $Charge \;  balance \; between \; atoms \; and \; ES$}.
For proof we display the charges $Z$(ES) of the ES versus the charges $Z$(AT) of the AS with atoms 
in a figure like Figure \ref{fig:fig10s3}.
The dashed lines indicate the virtual
valences after (\ref{PLZ}) and (\ref{RAZ}) which are significant 
to the respective AS in view of the dual stabilization close to the Fermi energy. 

The diamond phases and the average effective atoms of the zinc blende phases 
must occupy the line $Z$(ES) + $Z$(AT) = 4.
Colored symbols (+) on this line highlight the positions of equal distances to 
both planar respectively to both radial virtual valences.
Regarding the radial interferences the conditions read as $|Z(AT) - 3.149| = |Z(ES) - 0.723|$ (cf. Table \ref{tab:tab01s2})
together with $Z$(ES) + $Z$(AT) = 4 which is solved by $Z(AT)$ = 3.213 and $Z(ES)$ = 0.787. 

We expect that the charge partition is guided by the radial interferences because
they open the door to $sp$-hybridization.
The dashed box encloses the positions of the 18 average effective atoms along
the line $Z$(ES) + $Z$(AT) = 4, obviously well correlated with interference aspects.

\begin{figure}
\centering
\includegraphics[width=8.0cm]{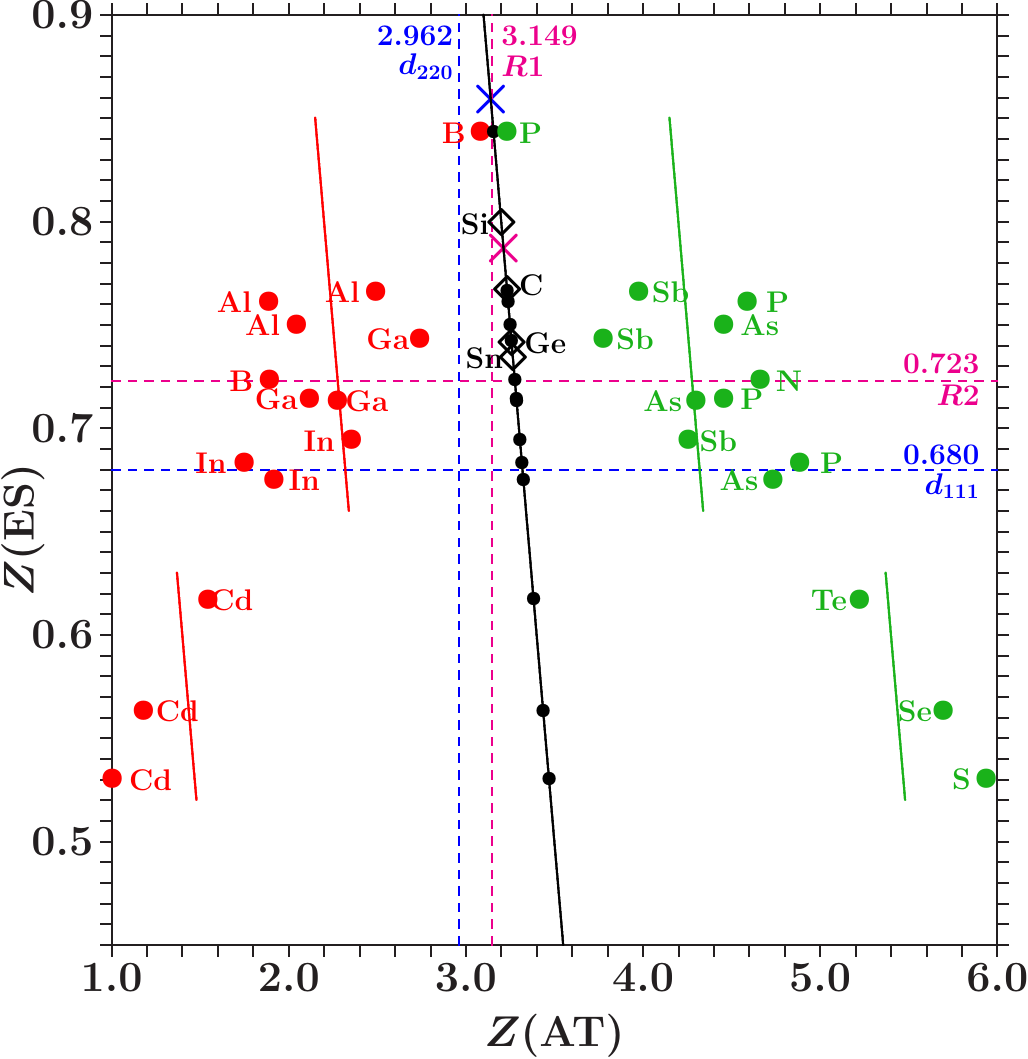}
\caption{Diamond and zinc blende phases.
The content of the box in Figure \ref{fig:fig10s3}, the average atoms (black) and both components (red, green).
Separately for III-V and II-VI phases, the colored lines indicate where the ES are charged on equal expenses of both components.}
\label{fig:fig11s3}
\end{figure}

Figure \ref{fig:fig11s3} shows the content of the box together with the effective atoms of the components.
Dia-C and dia-Si are  very close to the optimum charge partition with respect to both radial interferences.

\begin{figure}
\centering
\includegraphics[width=7.5cm]{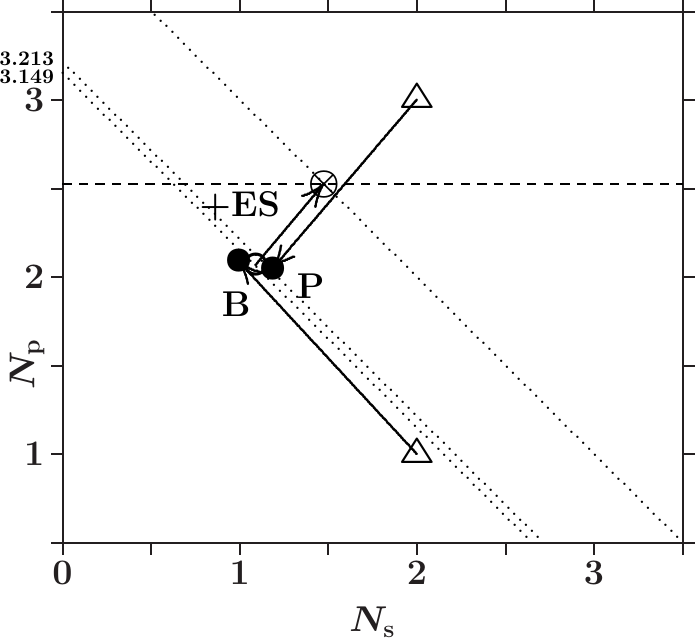}
\caption{BP: The partial charges of the average dimer (AT+ES), {\normalsize\bm$\Delta$} free atoms,
{\Large\bm$\bullet$} effective atoms, {\Large\bm$\circ$} average effective atom,
{\normalsize\bm$\otimes$} dimer(average effective atom + charged ES).}
\label{fig:fig12s3}
\end{figure}

Two systems in Figure \ref{fig:fig11s3} are tuned to only one radial interference immediately below $\epsilon_F$.
Cubic BN has $Z$(ES) $\approx$ $Z(R2)$ = 0.723 and BP $Z$(AT) $\approx$ $Z(R1)$ = 3.149.
In both alloys, the subspace-projected DOS at the radial interference is very low
(BN: Figure \ref{fig:fig05s3}c, BP: similar, not shown).
In the case of cub-BN the size of the ES-centered tetrahedra is concerned (scale $R$2, cf. discussion on Figure \ref{fig:fig05s3}c)
and in BP the internal stability of the atom-centered tetrahedra (length scale $R$1).
With P at the corner and B at the center (distance $R$1) this link is tuned to perfect
radial interference and hybridization.
As a result, the ES and, to minor extent, B are charged on the expense of P until the effective atoms of
B and P reach almost the same $sp$-configurations of the valence electrons (Figures \ref{fig:fig11s3} and \ref{fig:fig12s3}).
The final state may be described by 8 dimers in the SC unit cell which have
almost identical $sp$-configurations but different total charges.
Four dimers (B+ES) carry 0.925 excessive electrons and the other four (P+ES) the neutralizing positive charges.

\begin{figure}
\centering
\includegraphics[width=7.5cm]{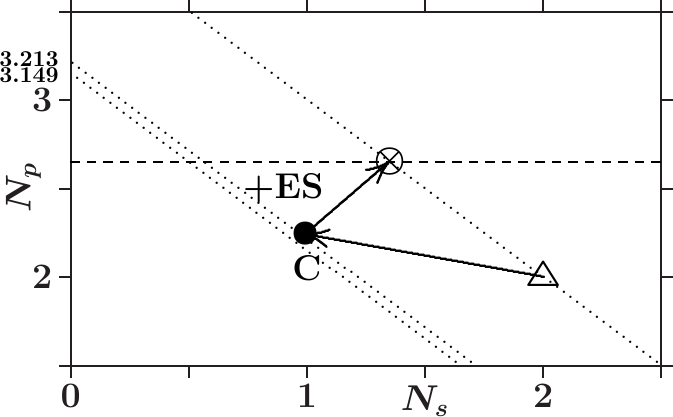}
\caption{Dia-C: The partial charges of the dimer (C+ES),
{\normalsize\bm$\Delta$} free atom C in the initial state,
{\Large\bm$\bullet$} effective atom C,
{\normalsize\bm$\otimes$} effective atom C + charged ES in the final state.}
\label{fig:fig13s3}
\end{figure}

\begin{figure}
\centering
\includegraphics[width=7.5cm]{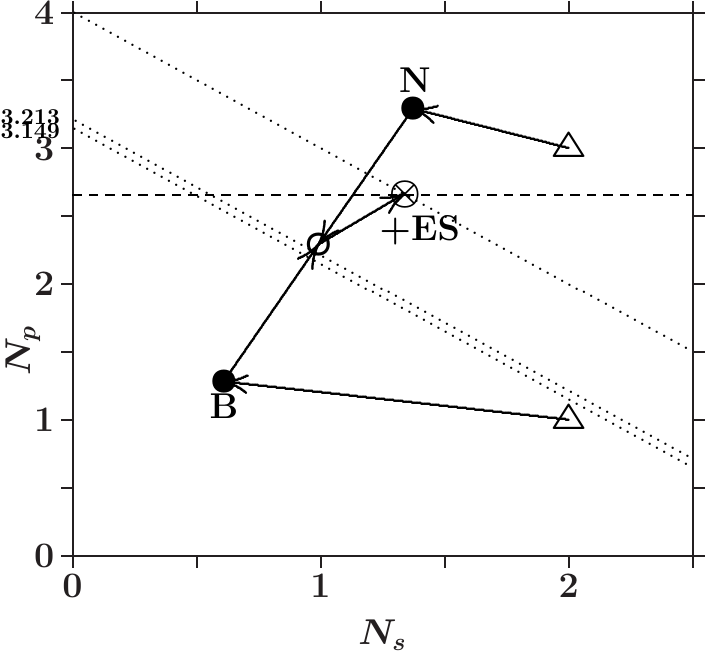}
\caption{Cub-BN: The partial charges of the average dimer (AT+ES), {\normalsize\bm$\Delta$} free atoms,
{\Large\bm$\bullet$} effective atoms, {\Large\bm$\circ$} average effective atom,
{\normalsize\bm$\otimes$} dimer(average effective atom + charged ES).}
\label{fig:fig14s3}
\end{figure}

\begin{figure}
\centering
\includegraphics[width=8.0cm]{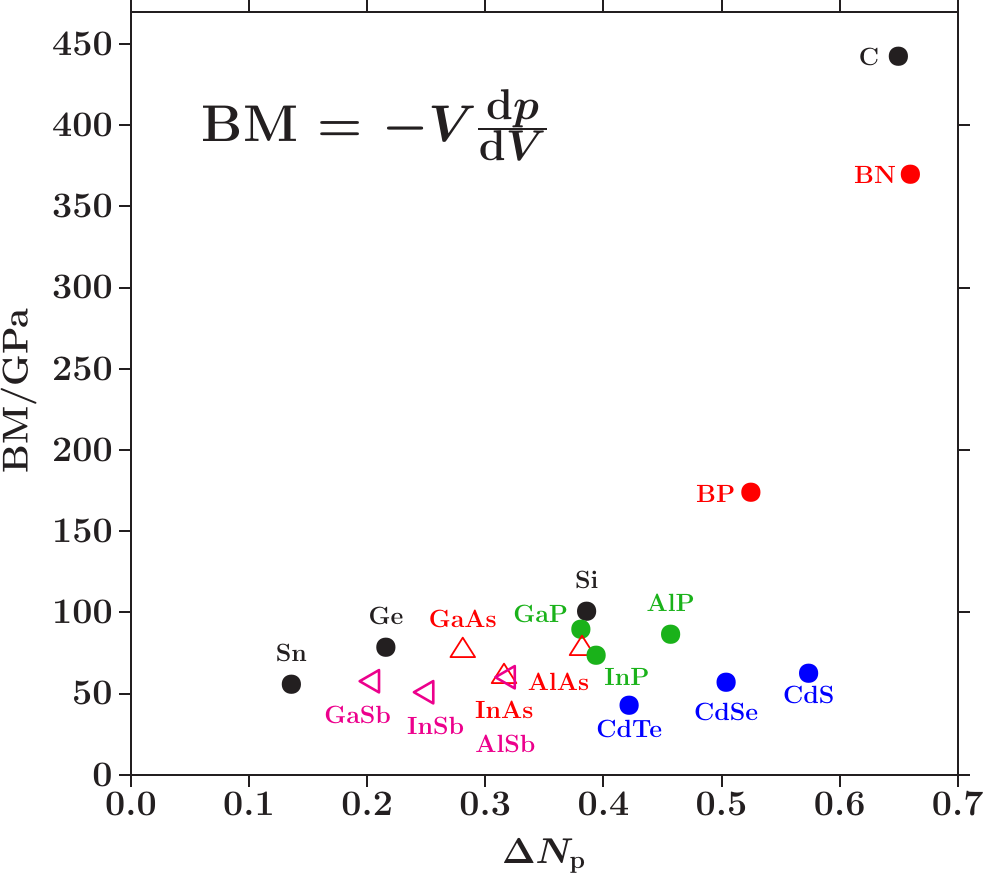}
\caption{The bulk moduli \cite{BoXu13} of the 18 analyzed phases versus the s-to-p transfers of the dimers, $\Delta N_p$.}
\label{fig:fig15s3}
\end{figure}

All systems have the red component with lower electronegativity 
(\cite{Pauling32,Alfred61}, Table \ref{tab:tab10s6}) than the green component.
One expects that the ES are charged mainly on the expense of the component with the lower electronegativity.
Separately for III-V and II-VI the colored full lines indicate where both components have equal charge losses.
Outside the strips between these lines the less electronegative component contributes more and inside vice versa.
Besides BP only AlSb, GaSb, InSb, and CdTe do not meet the expectation. \\[1ex]
\noindent
{\bm $Hume$-$Rothery$ $rule?$}. 
The valence charge of the effective C-atoms in dia-C is 3.233.
The close-fitting value 3.213 ensures equal distances to the fulfilled interference conditions of both
essential radial interferences (red cross in Figure \ref{fig:fig11s3}).
Furthermore, 3.233 is very close to 2 * 21/13 $\approx$ 3.231 which invokes 
a link to the Hume-Rothery condition for $\gamma$-brasses.
For explanation we remember that the electron states at $\epsilon_F$ are confined to the sublattice of the C-atoms.
The size of the MS at $\epsilon_F$ is due to the contributions 3.233 of the 8 C-atoms per SC unit cell.
The same size of the MS must arise if all the 16 sites of the SC unit cell would contribute each 3.233/2 = 1.6165
which is close to 21/13 $\approx$ 1.6154. 
In dia-C, the interference at $\epsilon_F$ fulfils the Hume-Rothery condition for $\gamma$-brasses.

This  guides back to the Table \ref{tab:tab02s2}.
For an arbitrary value of $n$, contributions around 1.6 by each of the $2n^3$ sites per SC unit cell
must allow for joined interferences $[d(2n^2),R(n)]$
which drive towards bcc-like structure on the length scale $a/n$.
Diamond (dia-C) is the case $n$ = 2.
Upon creating empty sites (ES) which are hidden to the electrons at $\epsilon_F$
the polyvalent carbon atoms build stable structure elements 
(the centered tetrahedra) which are reserved to lower valence. \\[1ex]
\noindent
{\bm $The$ $s$-$to$-$p$ $transfer$}.
The analysis above suggests that phase stabilization attempts to transform $s$-weight into $p$-weight.
The former stands for ductility, the latter for hardness.
We follow this idea and describe the SC unit cell again as composed of 8 dimers, each ES joins an AS with atom.
$N_s$ and $N_p$ in Figure \ref{fig:fig13s3} (dia-C) denote the cumulated weights on the
respective AS-orbitals in the dimer (symbols {\small\bm $\triangle$, $\otimes$}).

We start from the hypothetical situation (symbol {\small\bm $\triangle$}, dimer $N_s$ = $N_p$ = 2)
where the ES are really empty and the AS with atoms contain free atoms.
In the final state (symbol {\small\bm $\otimes$}, dimer $N_s$ + $N_p$ = 4) the contribution of the ES (arrow) 
has been added to the contribution of the effective C-atom ($\bullet$, $N_s$(C), $N_p$(C)).
The effective C-atom ends up close to the dotted line $N_s$ + $N_p$ = 3.213 where the sublattice of the atoms 
is equally far away from $Z(R1)$ = 3.149 as the sublattice of the ES is away from $Z(R2)$ = 0.723.
The line $N_s$ + $N_p$ = 3.149 implies perfect adaption only to $Z(R1)$.

The dimer has increased its $p$-weight by
$\Delta N_p$ = 0.650 on the expense of former $s$-weight.
Charging the ES creates new centers for electrons on directed orbitals
which makes the $p$-network more space-filling,
and the mechanical stability should be enhanced.
Other diamond phases of elements are less successful in this respect: 
Si($\Delta N_p$ = 0.386), Ge($\Delta N_p$ = 0.216), and Sn($\Delta N_p$ = 0.136).

Cubic BN ($\Delta N_p$ = 0.660, Figure \ref{fig:fig14s3}) compares with dia-C.
However, $\Delta N_p$ estimates only the average enhancement of the $p$-occupation per dimer.
The $p$-occupation in the
sublattice of B is much lower (cf. Figure \ref{fig:fig14s3}) which acts as a weak chain link.
We conclude that the mechanical stability of cub-BN must be less than that of dia-C. \\[1ex]
\noindent
{\bm $Bulk$ $modulus$ $and$ $s$-$to$-$p$ $transfer \: \Delta N_p$}.
Bulk moduli, $BM = -V(dp/dV)$, characterize the mechanical stability of materials.
Note that dia-C with the highest bulk modulus reveals two peculiarities in the employed approximation:
(i) Band states at $\epsilon_F$ put almost the whole weight to $p$-orbitals.
(ii) Among the diamond phases of elements, dia-C has the highest $s$-to-$p$ transfer starting from free atoms.
One may presume that a general correlation exists between the bulk modulus and the $s$-to-$p$ transfer. 

Table \ref{tab:tab10s6} shows the measured bulk moduli \cite{BoXu13} of the 18 phases.
The expected correlation proves valid for diamond phases (Figure \ref{fig:fig15s3}) where
the bulk moduli rise steeply with increasing $\Delta N_p$.
Zinc blende phases, are not completely characterized by $\Delta N_p$,
because one of both AS with atoms has the weight on $p$-orbitals less than the average.
Hence, at equal $\Delta N_p$ the bulk moduli of zinc blende phases are generally smaller 
than those of the diamond phases.

BP looks like an exception which fits rather into the sequence of the diamond phases.
This is proved valid by Figure \ref{fig:fig12s3}
where the effective atoms of B and P have almost equal $sp$-configurations as a result of 
the strong radial interference on the scale $R$1.
Starting from the free atoms, B even acquires some charge (Figures \ref{fig:fig08s3} and \ref{fig:fig12s3}) whereas P looses 
electrons, despite of the higher electronegativity. 

\subsection{Summarizing}

Structure stability of 18 crystalline phases has been analyzed 
within the concept of joined planar-radial interference and hybridization
where the tight connection is emphasized between the medium-range neighbor-shell order and the
extended planar order.
Hybridization is controlled by radial interferences which are in momentum contact with certain planar interferences.

Stabilizing interferences can be confined to subspaces of the total electron quantum space.
In particular dia-C reveals somewhat below $\epsilon_F$ 
that the internal stability of the C-centered tetrahedra arises in the C-subspace 
from radial interference and hybridization along the links $R$1 between the vertices and the center. 
At slightly lower energy, joined interferences [$d$(3),$R$2] in the ES-subspace 
stabilize the arrangement of the ES-centered tetrahedra.
Hence, dia-C stabilizes both essential length scales close to $\epsilon_F$ in separate subspaces.
The other 17 members of the group tend to fail the separation of the subspaces below $\epsilon_F$.
Instead, gaps open down in the valence band to ensure the stable arrangement of the centered tetrahedra.

Hybridization prefers spectral ranges without perturbing planar interferences,
such as along the hat-shaped feature of the $sp$DOS of dia-C around $G^2$ = 7 (Figure \ref{fig:fig04s3}a). 

Stabilizing charge redistribution aims at space-filling networks made of equally stiff interatomic links.
Relaxation steps towards structure stability transform former $s$-weight into $p$-weight.
The ranking of the 18 examined phases due to the measured bulk moduli follows this concept.

Three stabilizing contributions can be ascribed to the empty spheres: 
(i) They absorb the electron charge which is released by the atoms for optimizing 
both above interference.
(ii) They complete the electron network by new centers for directed $p$-orbitals.
(iii) They enable polyvalent atoms to build structure elements which require lower valences.
\section{\bm Electronic interference in $\gamma$-brasses}

In the following we expand on three phases of the space group I$\overline{4}$3m (number 217),
namely $\gamma$-$\rm Ag_5Li_8$ \cite{Noritake07a}, $\gamma$-$\rm V_5Al_8$ \cite{Brandon77},
and  $\gamma$-$\rm Cu_5Zn_8$ \cite{Gourdon07}.
The references specify the origins of the applied structure data.
Powder diffraction reveals very strong reflexes (330,411) in all cases.
Only $\gamma$-$\rm Ag_5Li_8$ exhibits moreover a similarly strong reflex (211).
Mott and Jones \cite{Mott58a} reported that the stability of $\gamma$-$\rm Cu_5Zn_8$ 
must be ascribed to the dominating planar interferences $G^2$ = 18
where the active $k$-space sphere contains 1.538 electron states per atom (Table \ref{tab:tab14s6}).

Within the empirical Hume-Rothery concept one estimates the $e/a$-ratio from the $sp$-electron configurations 
of the free atoms.
For the prototype alloy, $\gamma$-$\rm Cu_5Zn_8$, one obtains 21/13 $\approx$ 1.615.
This value is generally considered characteristic of $\gamma$-brasses.
The alloys $\gamma$-$\rm Ag_5Li_8$ and $\gamma$-$\rm V_5Al_8$ do not meet this expectation. 
For $\gamma$-$\rm Ag_5Li_8$ one obtains 1,
and $\gamma$-$\rm V_5Al_8$ results 34/13 $\approx$ 2.615 with plenty 
$d$VV$^\prime$-loaded band states close to $\epsilon_F$. 
The present study has a special focus on the role of the $d$Ag-band in the lower valence band, 
on the $d$V-band around the Fermi energy, and on the interacting $d$-bands of Zn in the lower and Cu in the upper
valence band. 

\subsection{\bm $\gamma$-$\rm Ag_5Li_8$}

At given energy, the FLAPW-Fourier method \cite{Asahi05a,Mizutani11b,Mizutani17a} 
extracts a representative $k$-space sphere from the FLAPW wave functions 
between the muffin-tin spheres.
The diameter of this sphere around $\epsilon_F$ decides on the possibility
of planar interferences $G^2$ = 18.
Application to $\gamma$-$\rm Ag_5Li_8$ (Mizutani $et \; al.$ \cite{Mizutani08a})
confirms the estimated ratio of $e/a$ $\approx$ 1
which proves the alloy an exception to the empirical Hume-Rothery rule.
Instead, they announce hybridization-based stabilization in the bonding $d$AgAg$^\prime$-band. 

The present approach considers pronounced hybridization 
as due to pronounced radial interference which joins a 
strong planar interference, e.g. $G^2$ = 18.
Such joined interferences drive the content of the SC unit cell 
towards 27 nearly bcc-like subcells.
This is common to $\gamma$-brasses (cf. Table \ref{tab:tab02s2}).
However, the stabilizing joined interferences $G^2$ = 18 do not necessarily act 
in an $sp$-dominated subspace close to $\epsilon_F$
as supposed by the empirical Hume-Rothery rule. 
In the case of $\gamma$-$\rm Ag_5Li_8$ it is foremost  
the activity in the total EQS at the upper edge of the bonding $d$AgAg$^\prime$-band. \\[1ex]
\noindent
{\bm $The$ $Noritake$ $model$ $versus$ 3$\times$3$\times$3-$bcc$}
One can describe the structure of $\gamma$-$\rm Ag_5Li_8$ with reference to the SC lattice.
Two 26-atom clusters are located at the center respectively a corner of the SC unit cell \cite{Bradley26a}.
As the center of the 26-atom cluster is not occupied,
the resulting model of $\gamma$-$\rm Ag_5Li_8$ carries no atoms 
at the bcc-positions on the scale of the SC unit cell.
Furthermore, the atoms in the SC unit cell are close to the sites
of an inscribed 3$\times$3$\times$3-bcc sublattice \cite{Bradley26a}.
The decoration of this sublattice is a simple task (Appendix 6.4).

We use the structure model as proposed by Noritake $et \; al.$ \cite{Noritake07a}
with 52 atoms in the SC unit cell, side length $a$ = 9.6066 \AA. 
The 26-atom cluster is a concentric arrangement of four subclusters, namely the inner tetrahedron IT(4 Li),
the outer tetrahedron  OT(4 Ag), the octahedron OH(6 Ag$^\prime$), and the cuboctahedron CO(12 Li$^\prime$) (Appendix 6.4).
Partial occupancies as reported by Noritake $et \: al.$ \cite{Noritake07a} are not considered. 
Mizutani $et \, al.$ \cite{Mizutani08a} employ the same structure model in their FLAPW-based study.

Suppose that the transition from 3$\times$3$\times$3-bcc to the $\gamma$-phase is the final step in a relaxation sequence.
What is changed in this step and what is improved? 
Except for the IT(Li) the subclusters are less modified (Appendix 6.4).
However, the diversity of the interatomic distances grows (Table \ref{tab:tab11s6}) which indicates that mainly 
the radial interferences and hybridizations are optimized in this step.
We focus on just this aspect.

\begin{figure}
\centering
\includegraphics[width=8.5cm]{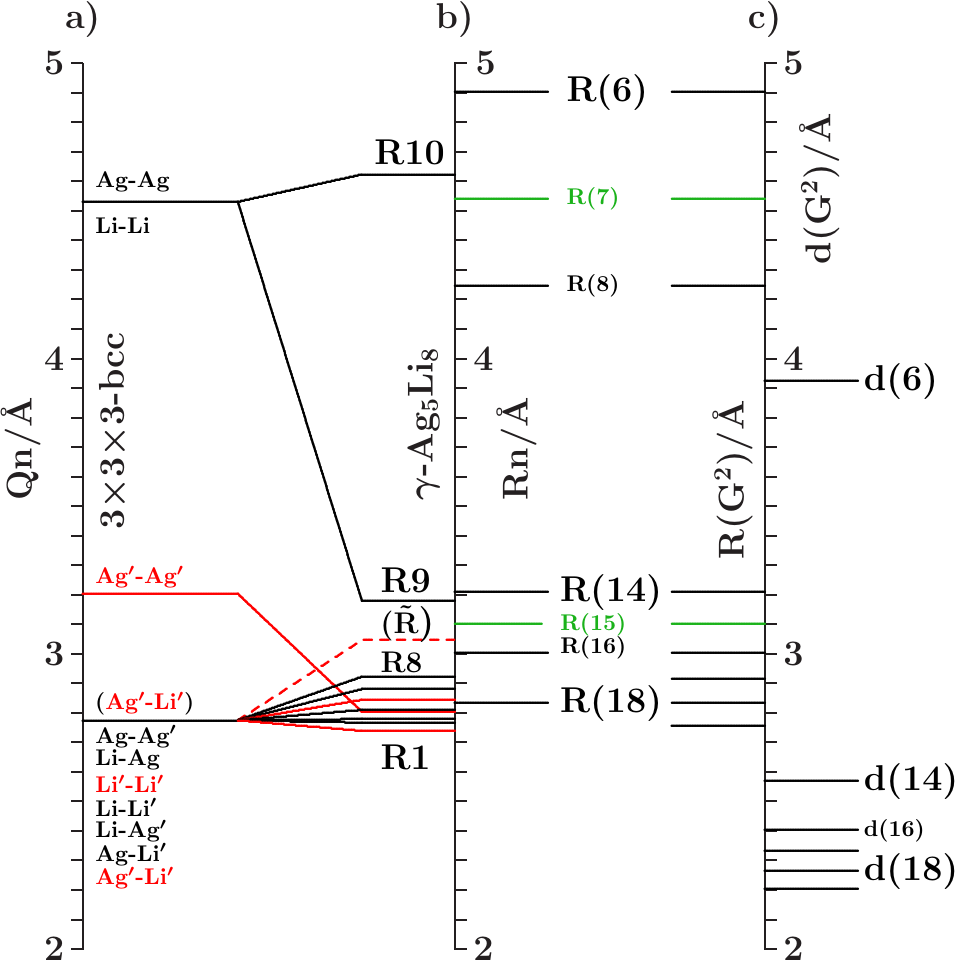}
\caption{The 10 smallest interatomic distances $R$1 - $R$10 between Li, Ag, Ag$^\prime$, Li$^\prime$ and
$\tilde{R}$(Ag$^\prime$-Li$^\prime$) (red: links between the 26-atom clusters).
a) 3$\times$3$\times$3-bcc. b) $\gamma$-$\rm Ag_5Li_8$.
c) SC reference lattice, interplanar distances $d(G^2)$ and the radial equivalents $R(G^2)$ (\ref{RHKL1}).}
\label{fig:fig16s4}
\end{figure}

Figure \ref{fig:fig16s4} compares the 10 shortest interatomic distances between Li, Ag, Ag$^\prime$, and Li$^\prime$
in the $\gamma$-phase with their former values in 3$\times$3$\times$3-bcc (Table \ref{tab:tab11s6}).
Interatomic links between the 26-atom clusters are distinguished from those inside by red color.
To the right the equivalent radii $R(G^2)$ (\ref{RHKL1}) together with the assigned interplanar distances $d(G^2)$ are shown,
including the radii $R(7)$ and $R(15)$ where no planar interferences belong.

The most obvious result of the hypothetical final relaxation step is the separation of the IT(Li) from the OT(Ag) 
which is supported by joined interferences [$d(14)$,$R$9(Li-Li)]] and [$d(6)$,$R$10(Ag-Ag)] with $G^2$=14 respectively $G^2$=6.
This follows from Figure \ref{fig:fig16s4} on applying (\ref{RIPL})
which suggests the interatomic distances $R9$(Li-Li) and $R$10(Ag-Ag) between $d(14)$ and $R(14)$ respectively $d(6)$ and $R(6)$. 
Note that in particular $R$10(Ag-Ag) is close to $R(7)$ where no additional planar reflexes acts.
This ensures freedom for radial interference and hybridization.
The joined interferences [$d(6)$,$R$10(Ag-Ag)] control the bcc-like arrangement of the OT(Ag) in the SC unit cell 
together with the size of the OT(Ag).

As the second remarkable result of the hypothetical final relaxation step we announce the formation
of the planar-radial interference block around the planar interference $G^2$ = 18.
This concerns the length scale $a$/3 of the 27 bcc-subcells in the SC unit cell (cf. Table \ref{tab:tab02s2}).
Obviously, the radial interferences and hybridizations along interatomic distances close to $R$(18) 
are tuned to the requirements of optimized hybridizations.
Hence, in particular the joined interferences [$d(18)$,$R$4(Ag$^\prime$-Ag$^\prime$)] and [$d(18)$,$R$6(Li$^\prime$-Li$^\prime$)],
both red in Figure \ref{fig:fig16s4}, exhibit a strong planar-radial coupling.
This way the short-range contacts between the 26-atom clusters via $R$4(Ag$^\prime$-Ag$^\prime$) and $R$6(Li$^\prime$-Li$^\prime$) 
can respond to the long-range arrangement of the clusters.

Invoking only structure parameters and the conditions of planar and radial interferences we have shown that 
certain joined interferences are formed in $\gamma$-$\rm Ag_5Li_8$
which improve the interference status.
Two questions arise: 
(i) How does the spectral distribution of the valence states reveal the improved interference status? 
(ii) Is the search for ''maximum interference`` equivalent to the search for ''lowest band-structure energy``?
In the following we will examine both issues.

\begin{figure}
\centering
\includegraphics[width=8.0cm]{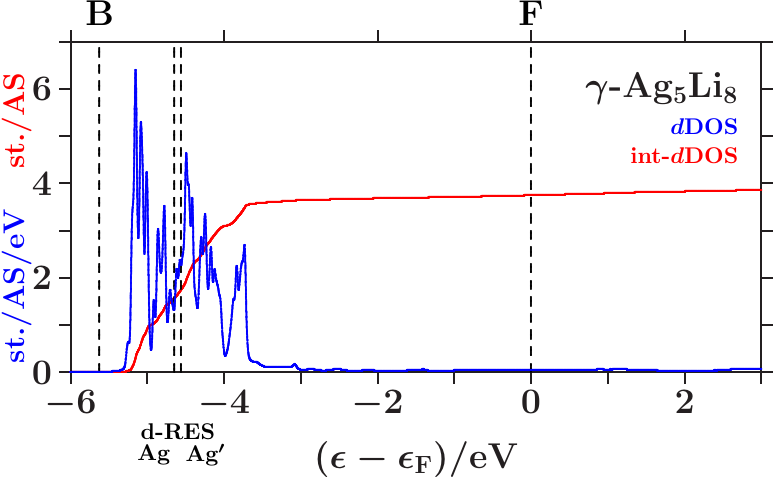}
\caption{Low flat $d$DOS confirms missing $d$-based activity around $\epsilon_F$ (B, bottom of the valence band).}
\label{fig:fig17s4}
\end{figure}

Besides the Noritake model of $\gamma$-$\rm Ag_5Li_8$ \cite{Noritake07a} we use two modified trial models, 
namely 3$\times$3$\times$3-bcc (equal decorations) and the model $\gamma$-$\rm Ag_5Li_8$-ITOT which is obtained upon
interchanging the decorations of IT and OT in the Noritake model, i.e. Ag on IT and Li on OT (equal structures).
Three systems are thus available which resemble each other regarding the planar interferences 
(cf. Appendix 6.4) but differ regarding the radial interferences and hybridizations.

The AS-models for application in the LMTO-ASMT calculations are obtained via Voronoi tessellation of the structure models.
Each Voronoi polyhedron is replaced by an AS of the same volume, and the radius of the inscribed muffin-tin sphere
is the largest to avoid overlaps with neighboring muffin-tin spheres.
This way we prepare effective four-component systems.
Valence orbitals up to l = 2 are considered in each AS.
In the $k$-space treatment we confine to the 6th special $k$-set \cite{Chadi73,Monkhorst76} 
in the irreducible wedge of the reciprocal SC lattice (5984 $k$-points).
A local exchange-correlation potential \cite{Barth72} is used. \\[1ex]
\noindent
{\bm $Are$ $there$ $active$ $subspaces?$}
We search for parts of the EQS where the interference-controlled interplay 
in the individual band states is confined.
Hence, the band states have fluctuating projections to parts of this active subspace
whereas the projections outside the active subspace are almost constant,
such as fluctuating $s$- and $p$-projections at nearly constant $d$-projections.
In that case, the MS collects only the active $sp$-weights.

For proof we proceed in two steps.
In a first step 

\begin{figure}
\centering
\includegraphics[width=5.5cm]{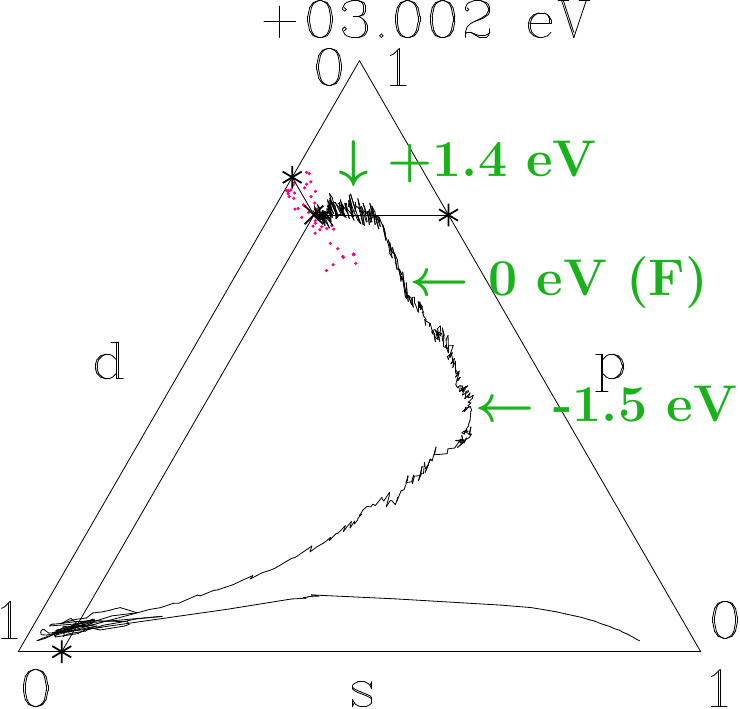}
\caption{3$\times$3$\times$3-bcc decorated according to $\gamma$-$\rm Ag_5Li_8$:
Presentation analog to Figure \ref{fig:fig19s4}.
The maximum average $p$-weight = 0.74 is reached around +1.4 eV.}
\label{fig:fig18s4}
\end{figure}

\begin{figure}
\centering
\includegraphics[width=5.5cm]{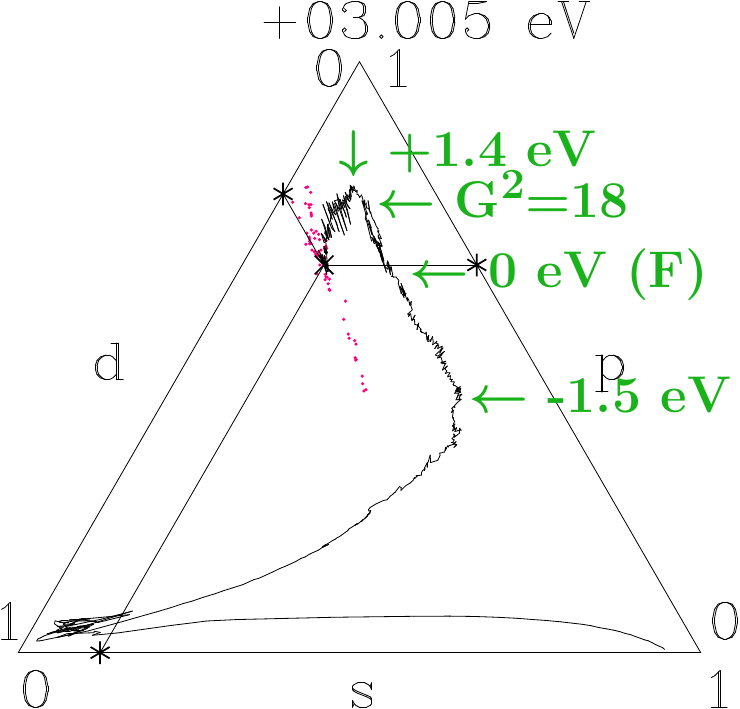}
\caption{$\gamma$-$\rm Ag_5Li_8$: The red dots show the $spd$-decompositions of the band states within 6 meV off +3.005 eV.
The tracking curve results from the interval averages throughout the valence band. 
Energies refer to $\epsilon_F$.
Between -1.5 eV and +1.5 eV the band states develop gradually $p$-dominated networks.
Around +1.4 eV the maximum average $p$-weight = 0.78 is reached.}
\label{fig:fig19s4}
\end{figure}

\begin{figure}
\centering
\includegraphics[width=5.5cm]{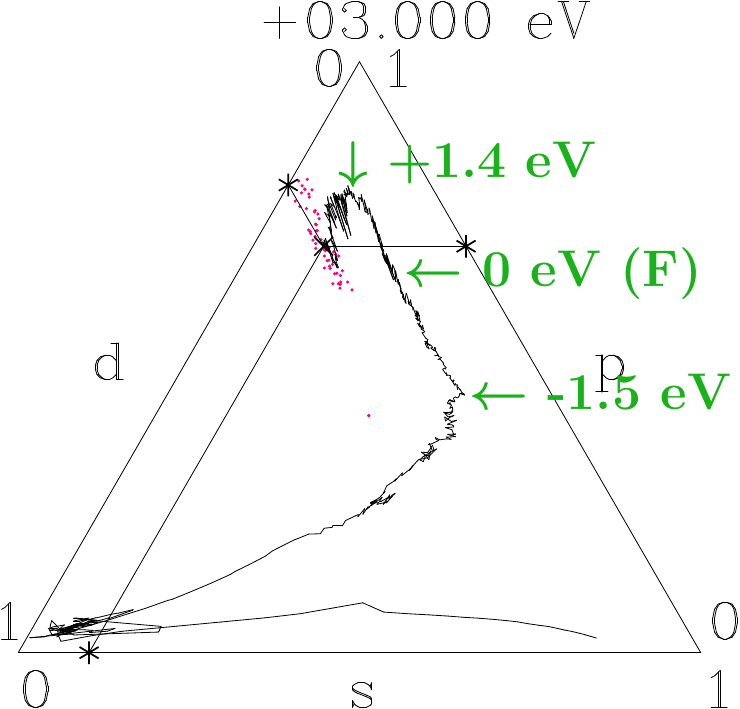}
\caption{$\gamma$-$\rm Ag_5Li_8$-ITOT with IT(Ag) and OT(Li):
Presentation analog to Figure \ref{fig:fig19s4}.
The  maximum average $p$-weight = 0.78 is reached around +1.4 eV.}
\label{fig:fig20s4}
\end{figure}

\noindent
we evaluate only the average compositions of band 
states in small energy intervals around given energies
whereas the second step examines band-state resolved information.
Projected densities of states are the most prominent quantities which 
provide the required information for the first step.
The small and flat $d$DOS around $\epsilon_F$ (Figure \ref{fig:fig17s4}) indicates absent $d$-based activities.
Vertical dashed lines in the $d$AgAg$^\prime$-band refer to the $d$-resonances.
They denote the spectral range of the non-bonding states which separates the bonding states 
below from the anti-bonding states above.

The Figures \ref{fig:fig18s4} to \ref{fig:fig20s4} use the visualization technique of Appendix 6.6.
We decompose the norm One of each band state into the $spd$-parts. 
The averages of small energy intervals provide the tracking curves shown.
At the bottom of the valence band the tracking curves start in the $s$-dominated corners, then roam in the
$d$-dominated corners ($d$AgAg$^\prime$-band),
and roughly from -1.5 eV to +1.5 eV around $\epsilon_F$ they join lines of nearly 
constant average $d$-weight where $s$-weight is continuously transformed into $p$-weight.
Similar to dia-C, $p$-dominated networks evolve within 3 eV around $\epsilon_F$ 
in the three systems due to the planar relationship. 
Differences occur in the $d$AgAg$^\prime$-bands and 
around the topmost points close to +1.4 eV 
where both $\gamma$-related systems reach maximum $p$-contents of 0.78 
whereas 3$\times$3$\times$3-bcc is lacking a small additional $p$-excess.
Without resolving individual band states we assume that 
the $d$-subspace separates from the active $sp$-subspace
between -1.5 eV and +1.5 eV.

\begin{figure}
\centering
\includegraphics[width=8.5cm]{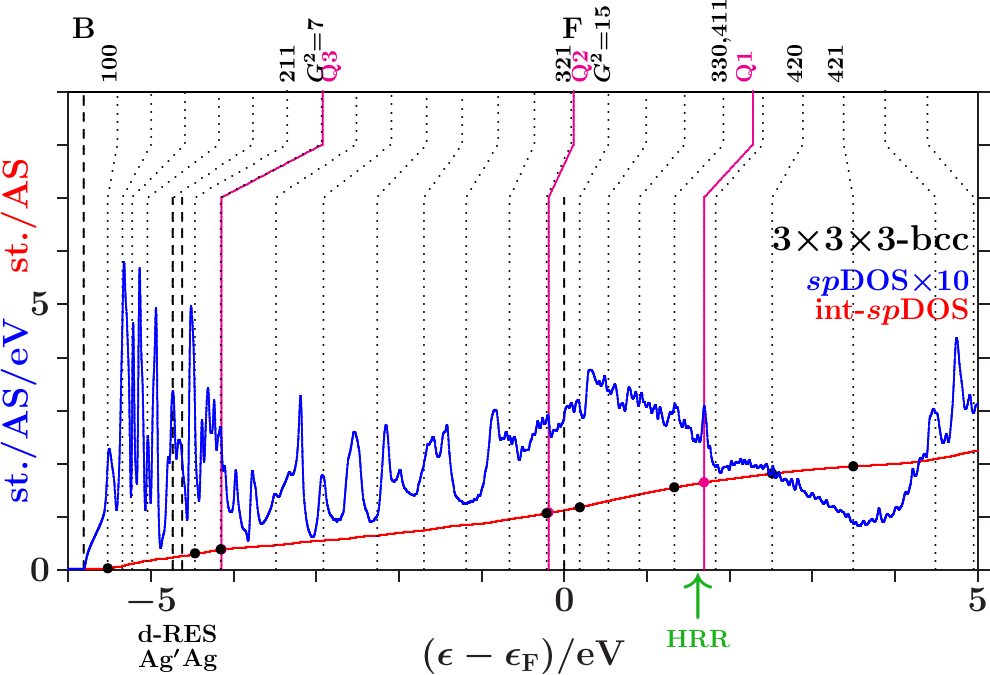}
\caption{3$\times$3$\times$3-bcc decorated according to $\gamma$-$\rm Ag_5Li_8$: 
A wide band splitting, centered around +3.5 eV,  due to planar interferences
is the background for additional radial interference and hybridization in the Figures \ref{fig:fig22s4} and \ref{fig:fig23s4}.}
\label{fig:fig21s4}
\end{figure}

\begin{figure}
\centering
\includegraphics[width=8.5cm]{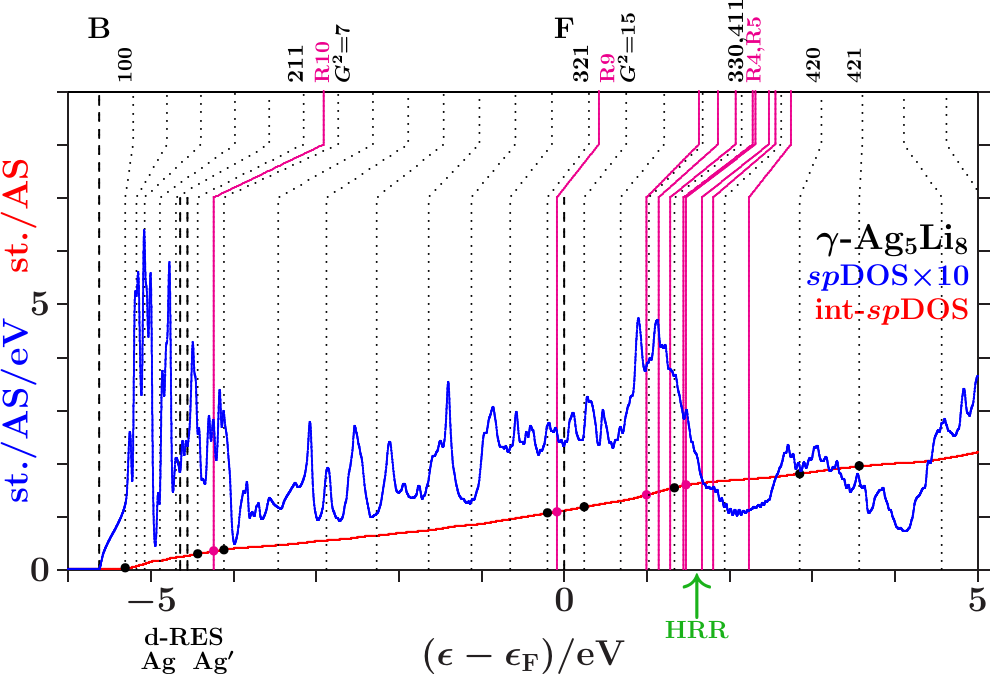}
\caption{Additional band splitting due to radial interference and hybridization
in the interference block around $G^2$ = 18.}
\label{fig:fig22s4}
\end{figure}

\begin{figure}
\centering
\includegraphics[width=8.5cm]{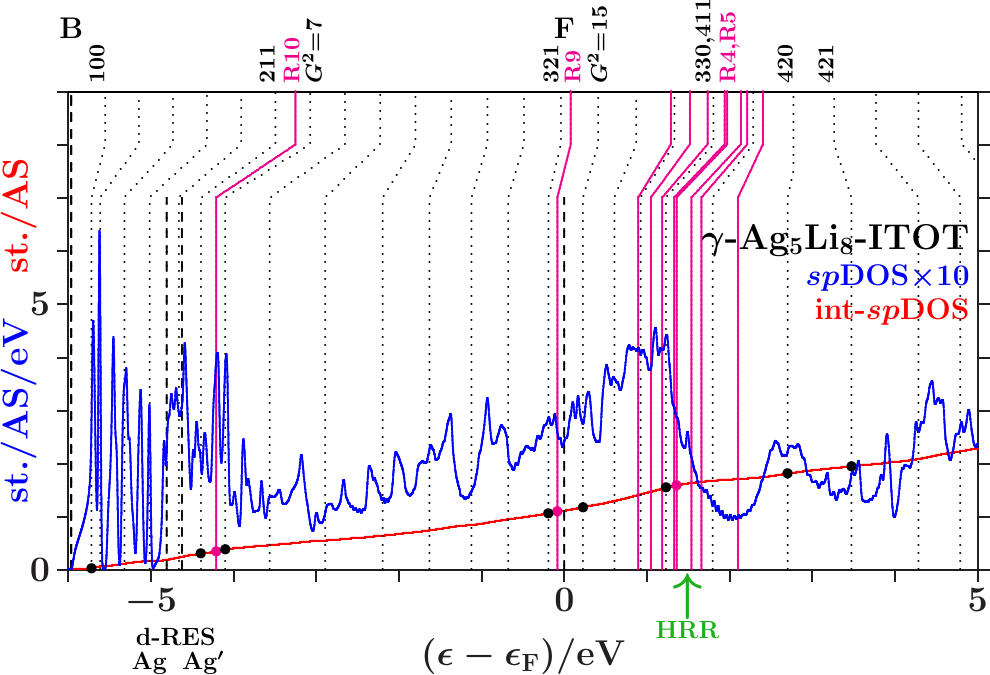}
\caption{After interchanging the decorations of IT and OT
the bonding $d$AgAg$^\prime$-band decays.}
\label{fig:fig23s4}
\end{figure}

In the second step we check this conjecture using the fluctuation patterns (red dots) 
which visualize band-state resolved information.
Preferences of the 
fluctuation patterns prefigure restrictions of the $spd$-interplay (Appendix 6.6). 
Along the tracks of nearly constant average $d$-weight the fluctuation patterns start around -1.5 eV 
without preference.  
At rising energy two preferences appear,
namely linear patterns perpendicular to the $s$-axis of the triangle ($\Delta$p = -2$\Delta$s, $\Delta$d = $\Delta$s) and 
linear patterns along the tracking curve ($\Delta$p = -$\Delta$s, $\Delta$d = 0).
The first mode allows for maximum $p$-gain on the expense of equal $s$- and $d$-losses 
whereas the second mode separates the $d$-subspace from the $sp$-interplay.
Above $\epsilon_F$ the $\gamma$-related systems tend towards the second mode whereas 3$\times$3$\times$3-bcc follows less strictly.
Finally the second mode dominates all the three systems (cf. Figures \ref{fig:fig18s4} to \ref{fig:fig20s4}).
In the first 1.5 eV above $\epsilon_F$, at least $\gamma$-$\rm Ag_5Li_8$ and $\gamma$-$\rm Ag_5Li_8$-ITOT
reveal the  $sp$-subspace active. \\[1ex]
\noindent
{\bm $Interference$ $in$ $the$ $sp$-$subspace$.}
The Figures \ref{fig:fig21s4} to \ref{fig:fig23s4} show the $sp$DOS, the integrated 
$sp$DOS, and the energies of interferences due to the MS of the $sp$-subspace (the lower parts of the vertical lines) 
respectively to the free-electron $sp$-sphere (the upper 

\newpage

\begin{figure}
\centering
\includegraphics[width=6.0cm]{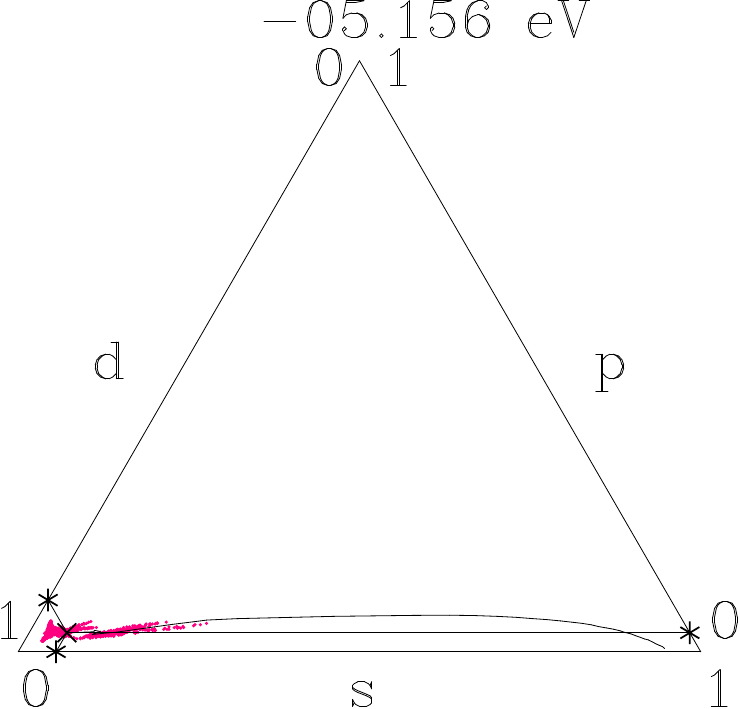}
\caption{$\gamma$-$\rm Ag_5Li_8$: Band states in the lower bonding $d$Ag band,
decomposition into the partial weights.}
\label{fig:fig24s4}
\end{figure}

\noindent
parts).  Upwards pointing green arrows designate
the energies HRR where the integrated 
$sp$DOS reaches the value of 21/13 $\approx$ 1.615.
According to the standard Hume-Rothery rule for $\gamma$-brasses this should happen at $\epsilon_F$,
just between the planar interferences $G^2$ = 18 below and the resulting pseudogap above.

The system 3$\times$3$\times$3-bcc (Figure \ref{fig:fig21s4}) follows a less restrictive version which
does not insist on the coincidence of HRR with the Fermi level.
As the spectral signature of the radial interferences $Q$1 is very weak, we conclude that
the linear descent of the $sp$DOS around HRR collects
the bonding states of a wide splitting due to planar interferences.
The center of this wide splitting is close
to the planar interference (421) ($G^2$ = 21) near +3.5 eV where the kinks of 
the dotted lines change from pointing upwards to pointing downwards.
Above this energy, the population of the $sp$-subspace acts anti-bonding in comparison with free electrons.
The other two systems resemble 3$\times$3$\times$3-bcc with respect to planar interferences (cf. Appendix 6.4).
This explains in Figures \ref{fig:fig22s4} and \ref{fig:fig23s4} the corresponding wide splittings centered close to +3.5 eV.
On this background they evolve additional redistributions due to radial interferences and hybridizations.

As discussed in connection with Figure \ref{fig:fig16s4} a block of planar-radial interferences and hybridizations forms 
around the planar interference $G^2$ = 18.
All the members interact via a common momentum pool.
This block acts bonding, i.e. high $sp$DOS is accumulated  below HRR, and the assigned pseudogap appears above.

In the anti-bonding $d$AgAg$^\prime$-band, the $sp$-subspace adopts starting information from the $d$-dominated 
processes.
This is accomplished via joined interferences [$d(6)$,$Q$3] (Figure \ref{fig:fig21s4}) respectively  [$d(6)$,$R$10] 
(Figures \ref{fig:fig22s4} and \ref{fig:fig23s4}).
The strong planar interference (211) ($G^2$ = 6) controls the bcc-like arrangement of the OT on the scale of the SC unit cell,
and $R$10(Ag-Ag) describes the size of the OT.
Well-tuned systems should concentrate high spectral $sp$-weight to the interference energies.
In particular $\gamma$-$\rm Ag_5Li_8$ (Figure \ref{fig:fig22s4}) meets this demand with respect to both components of [$d(6)$,$R$10]. 
Note that $G^2$ = 7 (no SC reflex) above this range ensures the necessary freedom for local processes.

The Fermi energy just between the joined interferences [$d$(14),$R$9(Li-Li)] below and $G^2$ = 15 (no SC reflex) above
suggests that the radial interference and hybridization $R$9(Li-Li) attempts joining exclusively the planar interference $G^2$ = 14.
Inspecting the course of the $sp$DOS around the Fermi energy in the Figures \ref{fig:fig21s4} to \ref{fig:fig23s4}, we find that 
$\gamma$-$\rm Ag_5Li_8$ (Figure \ref{fig:fig22s4}) replaces the upwards pointing course of the competing systems 
(Figures \ref{fig:fig21s4} and \ref{fig:fig23s4}) by a plateau,
i.e. spectral weight has been moved downwards, a stabilizing contribution.
This plateau disappears in Figure \ref{fig:fig23s4} after interchanging the decorations of IT and OT.
The compressed weakly scattering IT(Li) are thus required to leave the 
strongly scattering OT(Ag) centered at the bcc-positions on the scale of the SC unit cell. \\[1ex]
\noindent
{\bm $Interference$ $in$ $the$ $low$ $valence$ $band$}.
We focus on processes in the lower left corner of Figure \ref{fig:fig24s4} which represents
the bonding $d$AgAg$^\prime$-band.
All band states are clearly $d$-type,
and the fluctuating red dots indicate $sd$-interplay at small $p$-content.

\begin{figure}
\centering
\includegraphics[width=9.0cm]{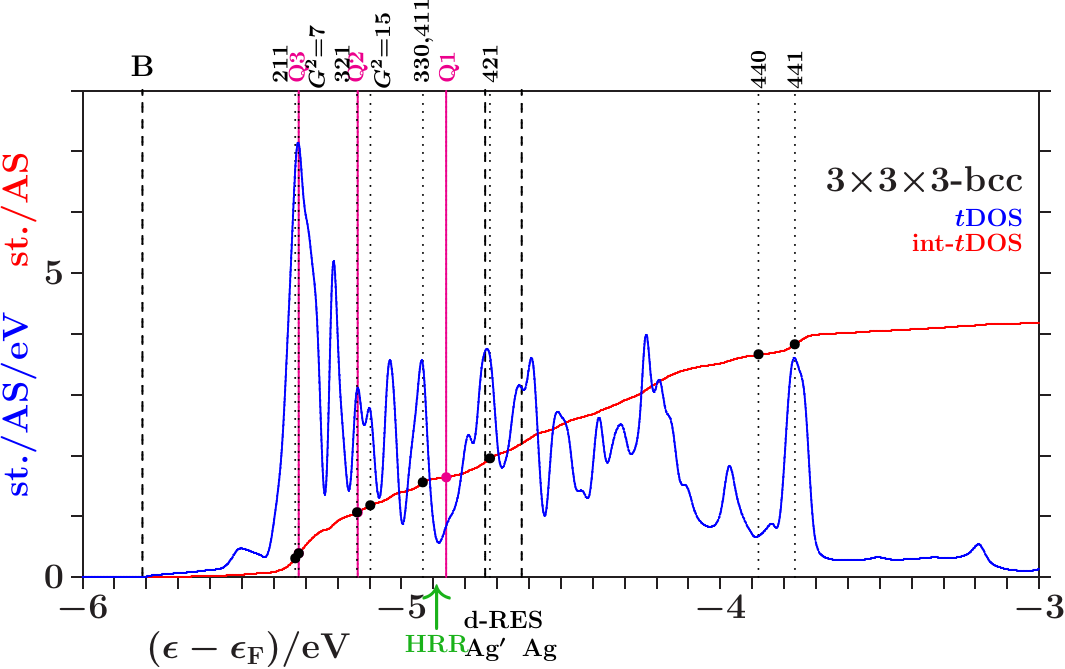}
\caption{3$\times$3$\times$3-bcc of $\gamma$-$\rm Ag_5Li_8$: Interference in the total EQS.
Note the extinction of the radial processes $Q$1 due to low spectral density
whereas the planar processes such as (211) appear at high spectral density.}
\label{fig:fig25s4}
\end{figure}

\begin{figure}
\centering
\includegraphics[width=9.0cm]{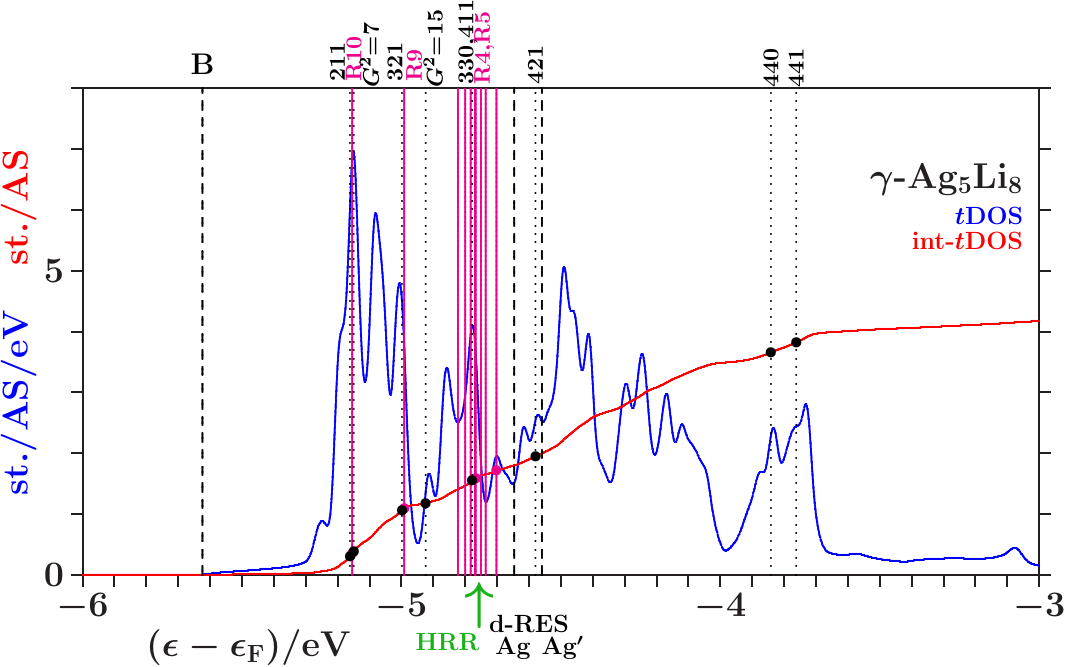}
\caption{Interference in the total EQS.
The interference block around $G^2$ = 18 acts at the upper bound of the bonding range,
optimized radial processes at high spectral densities.}
\label{fig:fig26s4}
\end{figure}

\begin{figure}
\centering
\includegraphics[width=9.0cm]{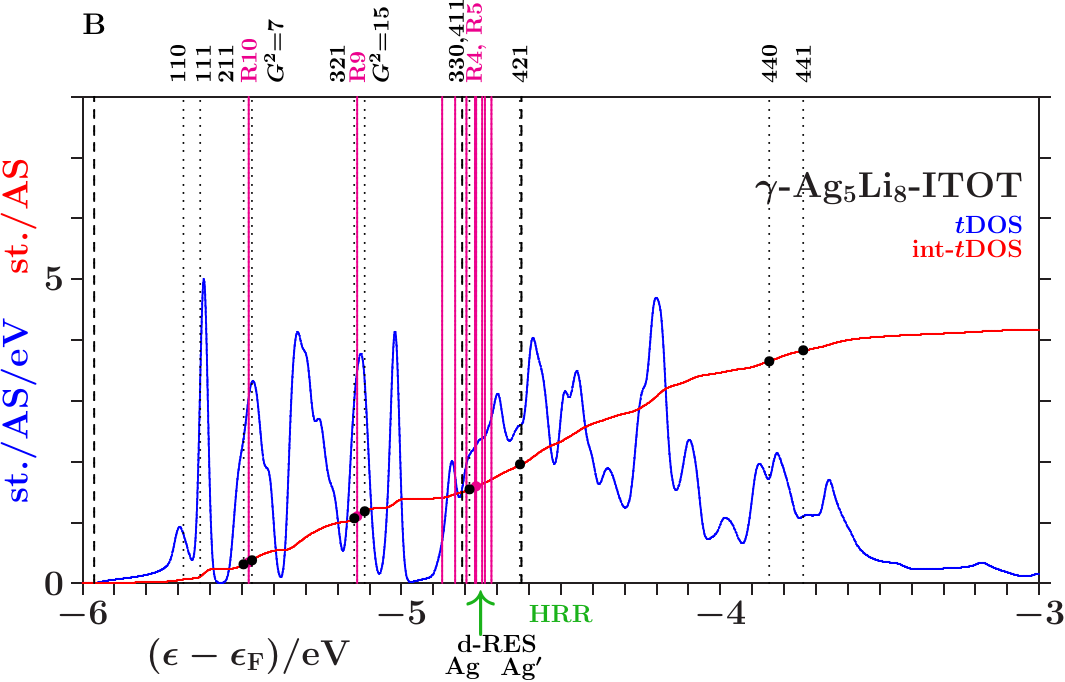}
\caption{Interference in the total EQS,
the bonding $d$Ag band after decay due to the interchanged decoration of IT and OT.
The interference block around $G^2$ = 18 acts among the non-bonding states.}
\label{fig:fig27s4}
\end{figure}

The Figures \ref{fig:fig25s4} to \ref{fig:fig27s4} show the $t$DOS of the systems together with the integrated $t$DOS.
The interference energies (vertical lines) result from (\ref{PLZ},\ref{RAZ}) 
using the integrated $t$DOS.
Green upwards pointing arrows (labels HRR) show where the integrated tDOS reaches the level of 21/13 $\approx$ 1.615.

Suppose, $\gamma$-$\rm Ag_5Li_8$ is tuned to stabilizing $d$-dominated interference
and hybridization in the bonding $d$AgAg$^\prime$-band.
In that case, the interference block around $G^2$ = 18 should act in the
bonding $d$AgAg$^\prime$-band at high $t$DOS, followed by a distinct pseudogap.
Inspecting the Figures \ref{fig:fig25s4} to \ref{fig:fig27s4} we find 
that the $d$-band splitting of $\gamma$-$\rm Ag_5Li_8$ meets just the above expectation.
Different 3$\times$3$\times$3-bcc (Figure \ref{fig:fig25s4}) where the radial interference $Q$1 occurs 
at very low $t$DOS in contrast to the planar interferences $G^2$ = 18. 
Consequently, the radial effects on the scale $Q$1 are suppressed,
just as observed in the $sp$-subspace (Figure \ref{fig:fig21s4}), too.
Also different $\gamma$-$\rm Ag_5Li_8$-ITOT (Figure \ref{fig:fig27s4}),
the interference block around $G^2$ = 18 acts in the non-bonding range 
where spectral weight should be removed rather than piled up.

Down in the bonding $d$AgAg$^\prime$-band, the joined interferences [$d$(6),$Q$3(Ag-Ag)] 
(Figure \ref{fig:fig25s4}) and [$d$(6),$R$10(Ag-Ag)] (Figure \ref{fig:fig26s4}) 
are well centered at the respective lowest pronounced peaks of the $t$DOS.
This equivalence manifests the memory of the bcc-like planar order in $\gamma$-$\rm Ag_5Li_8$.
With OT(Li) and IT(Ag), on the contrary,
$\gamma$-$\rm Ag_5Li_8$-ITOT (Figure \ref{fig:fig27s4}) 

\noindent
suffers from the decay of the bonding $d$AgAg$^\prime$-band with lower spectral weight
at the joined interferences around $G^2$ = 6.

Otherwise than in the $sp$-subspace of $\gamma$-$\rm Ag_5Li_8$ (Figure \ref{fig:fig22s4}) 
where the joined interferences [$d$(14),$R$9(Li-Li)] cause 
no striking spectral features they accumulate
notable spectral weight in the total EQS (Figure \ref{fig:fig26s4}).

As to $\gamma$-$\rm Ag_5Li_8$, we can answer the first question posed 
above in this Subsection 4.1:
Structure relaxation aims at increasing electronic influence 
due to enhanced electronic interference,
i.e. spectral weight in the active space is allocated 
to the essential planar-radial 
interferences and hybridizations. \\[1ex]
\noindent
{\bm $Low$ $band$-$structure$ $energy$}.
We calculate the band-structure energy per atom (AS),

\begin{equation}
          E_b(\epsilon) = \int\limits_{\epsilon_b}\limits^{\epsilon} d{\epsilon}^\prime \; ({\epsilon}^\prime - \epsilon_b) \, n_{tot}(\epsilon^\prime),
\Label[EB1]
\end{equation}

\noindent
as a function of the energy $\epsilon$ throughout the valence band
where $n_{tot}(\epsilon)$ is the total density of states per atom (tDOS),
and $\epsilon_b$ denotes the bottom of the valence band.

\begin{figure}
\centering
\includegraphics[width=8.5cm]{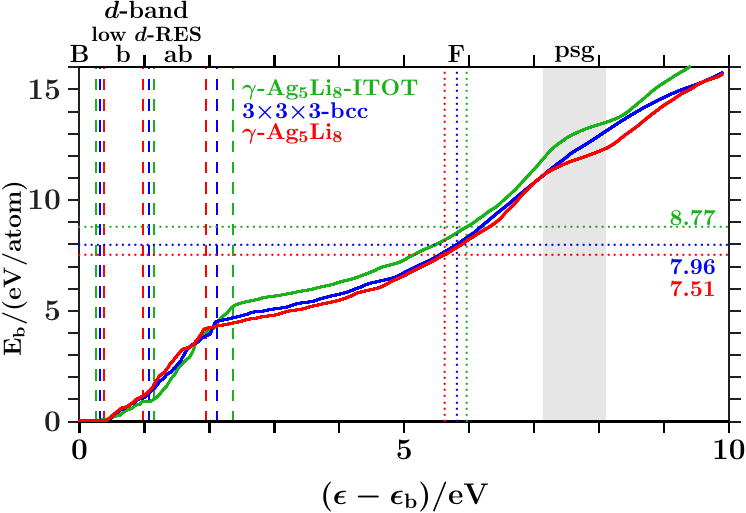}
\caption{Among the three competing models, the band energy (\ref{EB1}) of $\gamma$-$\rm Ag_5Li_8$ is the lowest
from the upper edge of the $d$AgAg$^\prime$-band up to at $\epsilon_F$.
This is due to the dense stacking of band states in the narrow $d$AgAg$^\prime$-band (Figure \ref{fig:fig30s4}).}
\label{fig:fig28s4}
\end{figure}

The results are shown in Figure \ref{fig:fig28s4}.
Colored dotted lines indicate the values of $\epsilon_F$ and $E_b(\epsilon_F)$.
The edges of the $d$AgAg$^\prime$-bands in the $t$DOS as well as the lower $d$-resonances 
are marked by colored vertical dashed
lines, and the shaded area above $\epsilon_F$ remembers the pseudogap of $\gamma$-$\rm Ag_5Li_8$ (Figure \ref{fig:fig22s4}).

\begin{figure}
\centering
\includegraphics[width=8.5cm]{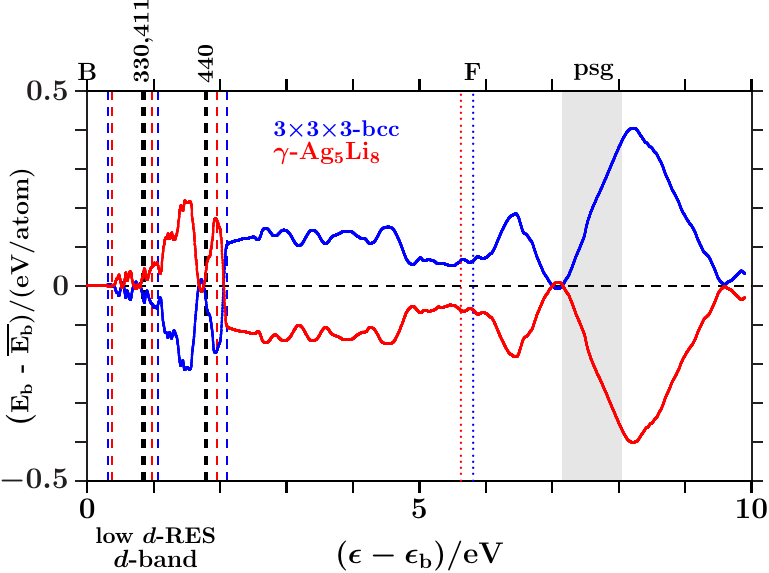}
\caption{The distances of both band energies from the average, $\overline{E_b} = (E_b(\gamma) + E_b(bcc))/2$, are
plotted versus the energy which retains the true band-energy differences.
Note the inversion of the ranking at the upper edge of the $d$AgAg$^\prime$-band.}
\label{fig:fig29s4}
\end{figure}

Despite of different interference tunings of $\gamma$-$\rm Ag_5Li_8$ and
3$\times$3$\times$3-bcc in the bonding $d$AgAg$^\prime$-bands (Figures \ref{fig:fig25s4} and \ref{fig:fig26s4}) the
band energies respond notably not until the anti-bonding range (Figures \ref{fig:fig28s4} and \ref{fig:fig29s4}).
The trial system $\gamma$-$\rm Ag_5Li_8$-ITOT (Figure \ref{fig:fig27s4}) with OT(Li)
has the bonding range heavily distorted.
The system leaves the $d$AgAg$^\prime$-band with the highest band energy,
and this stays valid up to the Fermi energy.
Hence, the ranking of the systems results mainly from different stacking
modes of band states in the $d$AgAg$^\prime$-band.

\begin{figure}
\centering
\includegraphics[width=8.5cm]{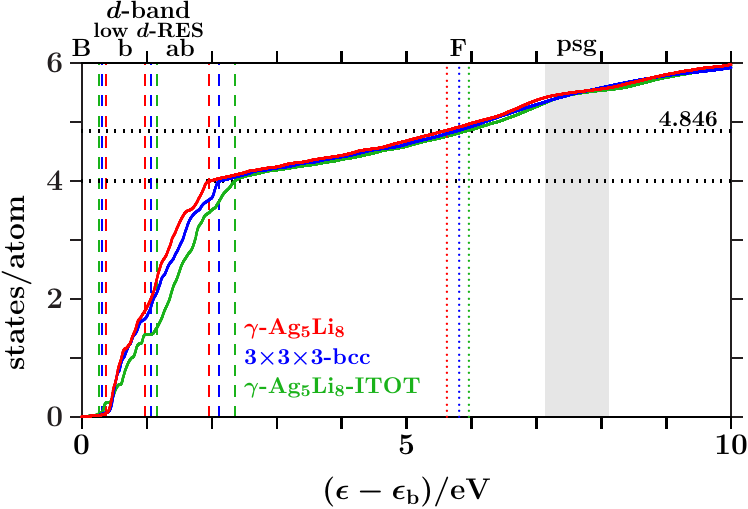}
\caption{At each energy, the $\gamma$-phase achieves the largest int-$t$DOS throughout
the occupied valence band, except for a tiny range close to the bottom.}
\label{fig:fig30s4}
\end{figure}

For evidence we show the integrated $t$DOS of the systems in Figure \ref{fig:fig30s4}.
The systems agree in that $N(\epsilon_F)$ = (5$\times$11+8$\times$1)/13 $\approx$ 4.846 states/AS are reached up to the
respective Fermi energy and nearly 4 states/AS up to the upper edge of the
$d$AgAg$^\prime$-band.
Apart from a tiny exception at the lower $d$-band edge,
the $\gamma$-phase has the largest integrated $t$DOS and thus the
most efficient stacking mode of band states throughout the occupied valence band.

Figure \ref{fig:fig28s4} allows only for a limited insight into the competition between
3$\times$3$\times$3-bcc and $\gamma$-$\rm Ag_5Li_8$ at given energies.
For improved insight we show in Figure \ref{fig:fig29s4} the distances of the band energies from the average,
$\overline{E_b} = (E_b^\gamma + E_b^{\rm bcc})/2$.
This retains the true differences.
%$E_b(\gamma) - E_b(bcc) = (E_b(\gamma) - \overline{E_b}) - (E_b(bcc) - \overline{E_b})$.
Except for a small environment of 1.7 eV, the band energy of the $\gamma$-phase is the larger

\newpage

\begin{figure}
\vspace{0.3cm}
\centering
\includegraphics[width=8.5cm]{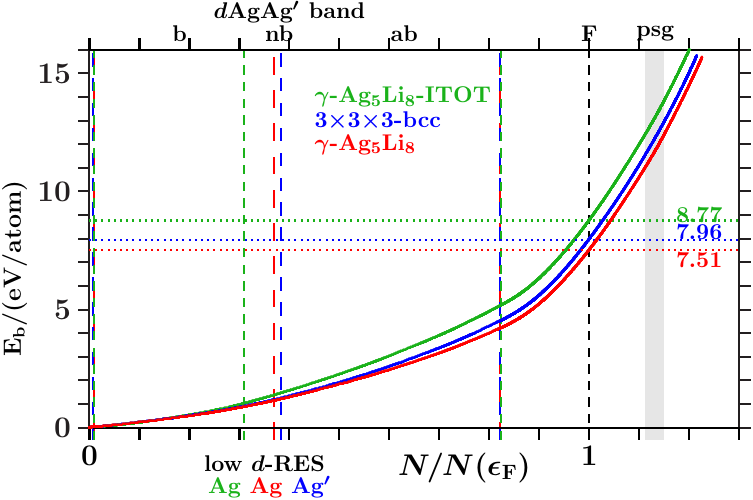}
\caption{The band energies versus the relative band filling $N(\epsilon)/N(\epsilon_F)$.}
\label{fig:fig31s4}
\end{figure}

\begin{figure}
\vspace{1.2cm}
\centering
\includegraphics[width=8.5cm]{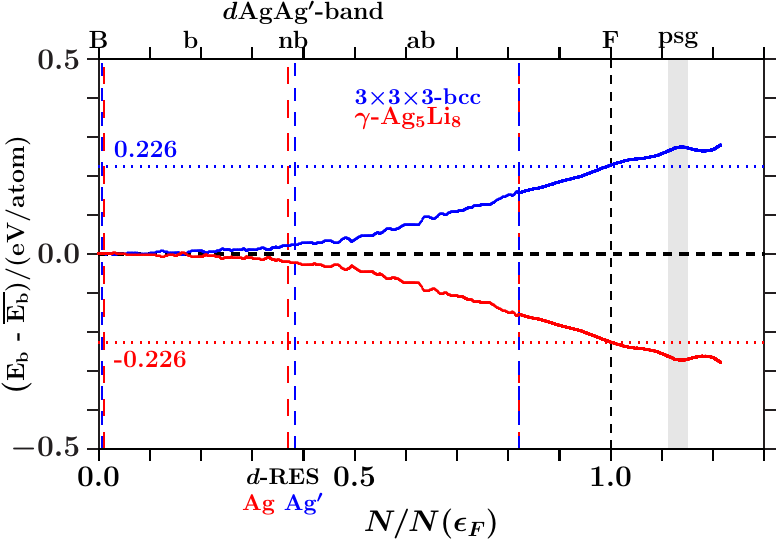}
\caption{The distances of the band energies from the average versus the relative valence-band filling,
"b`` bonding, "nb`` non-bonding,"ab`` anti-bonding.}
\label{fig:fig32s4}
\end{figure}

\noindent
one throughout the $d$AgAg$^\prime$-band.  This changes to
the opposite at the upper edge of the $d$AgAg$^\prime$-band,
because high spectral weight around the planar interferences (440) and (441) (Figures \ref{fig:fig25s4} and \ref{fig:fig26s4}) 
acts in the $\gamma$-phase at lower energy above the bottom of the valence band.

The final ranking of the systems at the respective Fermi energies 
is derived from equivalent situations,
both systems occupy equal numbers of band states. 
To compare the systems in equivalent situations throughout the valence band % we change the representation.
we replace $E_b(\epsilon)$ by $E_b(N)$,
i.e. by the band-structure energy up to $N$ included valence band states per atom.
The energies $\epsilon(N)$ are implicitly determined by

\begin{equation}
          N = \int\limits_{\epsilon_b}\limits^{\epsilon(N)} d{\epsilon} \, n_{tot}({\epsilon}).
\Label[EB2]
\end{equation}

\noindent
Figure \ref{fig:fig31s4} shows the band energies versus the relative valence band filling,
$N$/$N(\epsilon_F)$.
At the resolution of this plot the upper $d$AgAg$^\prime$-band edges of the systems coincide.
Starting around the $d$AgAg$^\prime$-resonances the band energies grow without change of the ranking.

\begin{figure}
\onecolumn
\centering
\includegraphics[width=13cm]{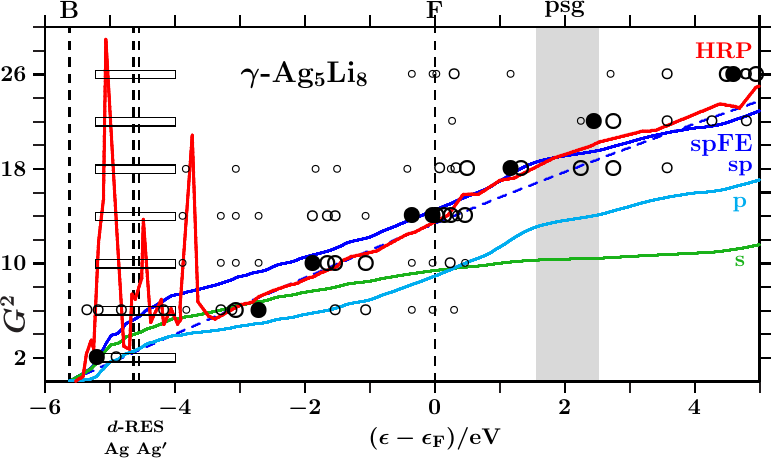}
\caption{The present VEC-based approach versus the FLAPW-Fourier approach, see text.
Full curves, Hume-Rothery plot (red) and the interferences $G^2(\epsilon-\epsilon_F)$ in the indicated subspaces.
The blue dashed curve is due to the free-electron $sp$DOS.}
\label{fig:fig33s4}
\twocolumn
\end{figure}

Figure \ref{fig:fig32s4} applies the presentation mode of Figure \ref{fig:fig29s4}.
The advantage of $\gamma$-$\rm Ag_5Li_8$ over 3$\times$3$\times$3-bcc 
at equal band fillings grows monotonously up to -0.45 eV/atom at $\epsilon_F$.

To answer the second question at the top of Subsection 4.1 we argue as follows: 
$\gamma$-$\rm Ag_5Li_8$ achieves the lower band energy due to the concentration of spectral weight
to joined interferences in the narrow $d$AgAg$^\prime$-band.
This configures the stacking of band states up to and beyond $\epsilon_F$.
Enhanced electronic interference causes reduced band energy. \\[1ex]
\noindent
{\bm $Comparison$ $with$ $the$ $Hume$-$Rothery$ $plot$}. 
The FLAPW-based analysis of interference in $\gamma$-$\rm Ag_5Li_8$ \cite{Mizutani08a}
starts with the inspection of the Fourier representation of the FLAPW wave functions in the interstitial between the muffin-tin spheres,

\begin{equation}
         \Psi^{\nu}(\vec{r},\vec{k}) = \sum_{\vec{g}} \; c^{\nu}(\vec{k}+\vec{g}) \; e^{i(\vec{k}+\vec{g})\vec{r}}.
\Label[FLAPW]
\end{equation}

\noindent
The key argument is that large $|c^{\nu}(\vec{k}+\vec{g})|^2$ are due to the coupling with 
certain fulfilled Bragg conditions at symmetry points, such as a point N where $\vec{k}_N = (2\pi/a)(1/2,1/2,0)$. 

The favored Bragg conditions specify constructive interference in the true backscattering of plane waves, $(\vec{k}_N+\vec{g})$,
by appropriate (hkl) lattice planes, i.e. $(\vec{k}_N+\vec{g})$ + $\vec{g}_{hkl}$ = -$(\vec{k}_N+\vec{g})$.
As depicted in 

\newpage

\hspace*{1cm}
\vspace{9.35cm}

\noindent
Figure \ref{fig:fig01s2}, this can be seen as a transition on the surface
of a sphere in the extended $k$-space, diameter 
\noindent
$2|\vec{k}_N+\vec{g}| = |\vec{g}_{hkl}| = (2\pi/a)\sqrt{G^2}$.
Following this concept the authors extract a continuous function $G^2(\epsilon)$ (the Hume-Rothery plot, HRP) 
from the dominating spectral components throughout the Brillouin zone
which defines the diameter, $(2\pi/a)\sqrt{G^2(\epsilon)}$, of a representative $k$-space sphere.
From $G^2(\epsilon_F) \approx$ 13.4 they deduce 1 $e/a$ inside the sphere
and conclude that $\gamma$-$\rm Ag_5Li_8$ is an exception to the 
empirical Hume-Rothery rule because it fails 21/13 $e/a$ $\approx$ 1.615 $e/a$.

Figure \ref{fig:fig33s4} shows the lower part of the Fourier spectrum at symmetry points N which has been extracted from
results published by Mizutani and coworkers \cite{Mizutani14b,MizutaniV1}.
Three groups are formed depending on the cumulated weights (short weights) of the $|c^{\nu}(\vec{k}_N+\vec{g})|^2$ 
with $(2(\vec{k}_N+\vec{g}))^2 = (2\pi/a)^2 \, G^2$.
Big circles indicate weights $\ge$ 0.5,
the medium-sized circles 0.5 $>$ weights $\ge$ 0.1, and
the small circles including the elongated boxes in the $d$AgAg$^\prime$-band refer to weights $<$ 0.1.  
Big bullets highlight the largest weights along the respective lines $G^2$.
The full curves show the Hume-Rothery plot (red, HRP) and the 
continuous functions $G^2(\epsilon)$ of the present VEC-based approach
applying (\ref{PLZ},\ref{INTL1}) combined with 
the respective integrated partial DOS of the subspaces $sp$ (blue), $s$ (green), and $d$ (cyan).
For comparison we add the blue dashed curve labelled as $sp$FE which is obtained
upon using in each AS the integrated free-electron $sp$DOS above 
the bottom of the valence band
(symbol B, $\epsilon_b - \epsilon_F$ = -5.62 eV, LMTO-ASMT).
Note that the energies indicated by the upper parts of the vertical dotted lines in Figure \ref{fig:fig22s4}
are just the energies where $sp$FE passes integer values of $G^2$.

\begin{figure}
\centering
\includegraphics[width=8.5cm]{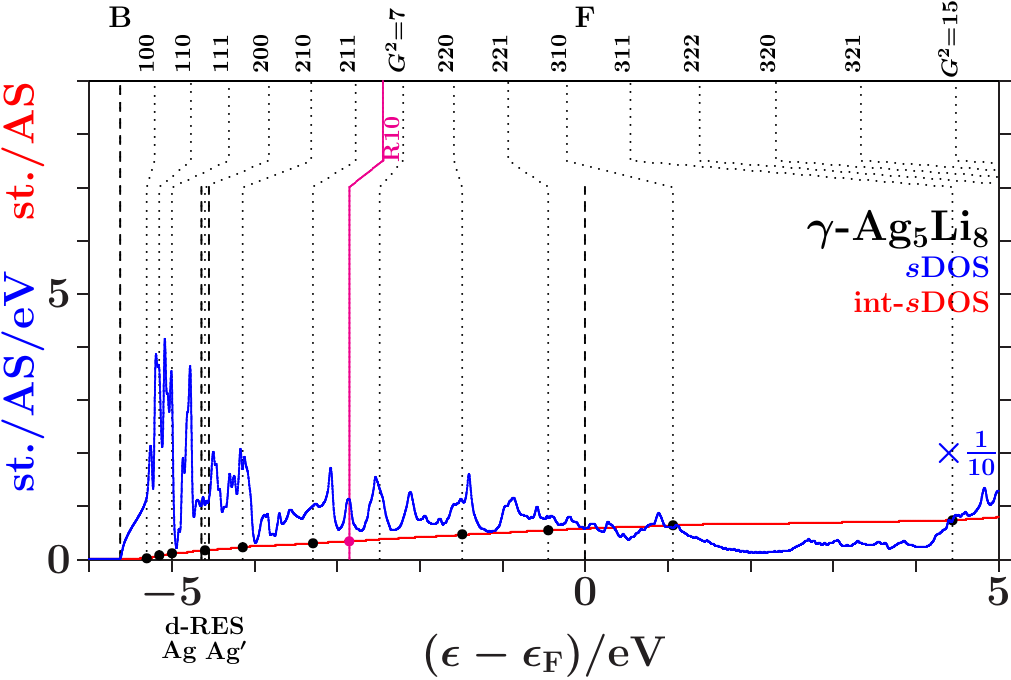}
\caption{The b/ab-splitting in the $s$-subspace around -2 eV accumulates bonding weight around -3 eV, supported by the joined
interferences [$d$(6),$R$10(Ag-Ag)}
\label{fig:fig34s4}
\end{figure}

Several conclusions can be drawn from Figure \ref{fig:fig33s4}: \\
(i) Around the dominating planar interferences $G^2$ = 18 
the HRP matches the blue curve derived from the $sp$DOS,
i.e. the HRP extracts the total accumulated $sp$-weight
including the contributions of radial interference and hybridization. 
Both approaches are thus equivalent with respect to the internal length
scale of the 27 bcc-like subcells. \\
(ii) The bullets at $G^2$ = 2, 10, 14, and 18 are close
 to the blue full curve (label $sp$).
Even $G^2$ = 6 may be added because the $d$DOS slightly
 below -4 eV is small (Figure \ref{fig:fig17s4}). 
Note that the bullets indicate vicinity to planar interferences 
in the true backscattering at symmetry points N.
The interference energies are thus determined by the MS of the $sp$-subspace.
The HRP, on the contrary, matches the blue dashed curve $sp$FE where
no spectral charge redistribution is considered.
At least regarding $G^2$ = 6 one may doubt because of 
the strong x-ray peak (211). \\
(iii) The planar interferences (211) (no common measure) 
control the bcc-like arrangement of the subclusters OT(Ag) 
on the length scale $a$ of the SC unit cell.
Bullet and big circle at $G^2$ = 6 around -3 eV close to the intersection with the 
green $s$-derived curve suggest that the HRP is really $s$-based.
For verification we refer to Figure \ref{fig:fig34s4} where
a wide b/ab-splitting of the $s$DOS (center close to -2 eV) accumulates bonding weight
around -3 eV, supported by the joined interferences [$d$(6),$R$10]. \\
(iv) The intersection between the $s$- and the $p$-related full curves slightly 
above the Fermi level in Figure \ref{fig:fig33s4} 
allows for enhanced $sp$-interplay which promotes the hybridization 
around $G^2$ = 15 as mentioned above in connection with Figure \ref{fig:fig22s4}. \\[1ex]
\noindent
{\bm $Charge$ $transfer$ $and$ $the$ $ratio$ $e/a$}.
We start from a hypothetical initial state where the AS contain free atoms (configurations Ag(1 0 10), Li(1 0 0)).
The 20 AgAg$^\prime$ contribute 220 valence electrons and further 32 are added by the 32 LiLi$^\prime$.
In the final state, the charges inside the AS result from the LMTO-ASMT calculation.
We find that $\sim$5.3 valence electrons move from AgAg$^\prime$ to LiLi$^\prime$,
i.e. 2.1 \% of the total 252 valence electrons per SC unit cell. 

As shown in Section 3, the $s$-to-$p$ transfer is sensitive to structure stability.
Table \ref{tab:tab03s4} gives a compilation of the charge transfers between the involved AS-orbitals. 
The last two lines refer to the average changes in the first step from free atoms to the perfect planar order
and in the second step to the adjusted radial order. 

Starting from free atoms, on the average 0.527 + 0.005 = 0.532 $p$-states per AS are occupied in two steps,
 mainly on the expense of decreasing $s$-occupation.
The chief contribution is due to bonding into planar order (step 1), the adjustment of the 

\begin{table}
\small
\centering
\caption{$\gamma$-$\rm Ag_5Li_8$, the changes of the AS.charges in the transition from free atoms
to the $\gamma$-phase. 
The last two lines show the average changes in two steps, see text.} 
\vspace{0.2cm}
        \begin{tabular}{ l | l | l | l }
        \hline
        AS-types \hspace*{1.0cm}        & $s$ \hspace*{0.8cm}   & $p$ \hspace*{0.8cm}   & $d$ \hspace*{0.8cm} \\
        \hline
	Li(IT)				& -0.555		& +0.616		& +0.132                 \\	
	Ag(OT)				& -0.227		& +0.415		& -0.492                 \\	
	Ag$^\prime$(OH)			& -0.229		& +0.445		& -0.460                 \\	
	Li$^\prime$(CO)			& -0.546		& +0.586		& +0.118                 \\		
        \hline
%	averages(8/8/12/24)		& 			& 			&		         \\
	to 3$\times$3$\times$3-$bcc$	& -0.414		& +0.527		& -0.113	         \\
	to $\gamma$-$\rm Ag_5Li_8$	& -0.011		& +0.005		& +0.006	         \\
        \hline
        \end{tabular}
\label{tab:tab03s4}
\end{table}

\noindent
radial order (step 2) adds only a small enhancement of the network-forming $p$-weight
(cf. Figures \ref{fig:fig18s4} and \ref{fig:fig19s4}).
Note that this charge partition at $\epsilon_F$ represents an intermediate state of the growing $p$-network
along the line in Figure \ref{fig:fig19s4}.

The ratio $e/a$ is intended to describe the size of the active $k$-space sphere at $\epsilon_F$.
Mizutani $et \; al.$ \cite{Mizutani08a} derive $G^2(\epsilon_F)= 13.4$ from the HRP
(cf. Figure \ref{fig:fig33s4}, red curve) and insert it into
($\pi$/156)($G^2(\epsilon_F))^{3/2}$ = $e/a$ ((\ref{PLZ}) with $N_{act}$ = 52).
The result is $e/a$ = 0.99 (Table \ref{tab:tab04s4}).

Using the integrated free-electron $sp$DOS (Figure \ref{fig:fig33s4}, the blue dashed curve, label $sp$FE), we obtain $e/a$ = 0.98.
Except for the $d$AgAg$^\prime$-band the Hume-Rothery plot proves $sp$FE-like up to $\epsilon_F$.
This disregards the wide splitting of the $sp$-band which is centered close to +3.5 eV.

\begin{table}
\small
\centering
\caption{The ratios $e/a$ of $\gamma$-$\rm Ag_5Li_8$ derived from the HRP (Figure \ref{fig:fig33s4}) and
from the integrated free-electron $sp$DOS ($sp$FE) respectively the integrated $sp$DOS ($sp$).}
\vspace{0.2cm}
        \begin{tabular}{ l | l | l }
        \hline
        		\hspace*{0.3cm} & $G^2(\epsilon_F)$  \hspace*{0.6cm} 	&  $e/a$ \hspace*{0.6cm} \\
        \hline
        HRP \cite{Mizutani08a}         	& 13.4                                  & 0.99 		       \\
        HRP \cite{Mizutani17a}, not discussed & 13.51                           & 1.00 		       \\
        $sp$FE                         	& 13.33                                 & 0.98 		       \\
        $sp$                           	& 14.46                                 & 1.11 		       \\
        \hline
        \end{tabular}
\label{tab:tab04s4}
\end{table}

With the integrated $sp$DOS of the present approach (Figure \ref{fig:fig33s4}, the blue full curve, label $sp$) 
one obtains $G^2(\epsilon_F)$ = 14.46 and hence $e/a$ = 1.11, just between 1.0549 ($G^2$ = 14) and 1.1699 ($G^2$ = 15) 
(Table \ref{tab:tab14s6}).
Obviously, the joined interferences [$d$(14),$R$9(Li-Li)] in the $sp$-subspace are addressed.
They act in Figure \ref{fig:fig19s4} around $\epsilon_F$ along the line of nearly constant average $d$-weight.
Hence, the ratio $e/a$ = 1.11 is significant as it refers to the enhancement of the $p$-network around $\epsilon_F$.
Rather than a pseudogap the spectral signature is a plateau of the $sp$DOS (Figure \ref{fig:fig22s4}). \\[1ex]
\noindent
{\bm\bf $Summarizing$ $\gamma$-$\rm Ag_5Li_8$}.
The alloy is electronically stabilized in the extended $k$-space by
joined planar-radial interferences around $G^2$ = 18, 14, and 6 which is common with other $\gamma$-brasses.

Joined interferences $G^2$ = 18 around the planar interference (330) include the radial interferences and hybridizations
on the shortest interatomic distances.
This interference block keeps the content of the SC unit cell close to 27 bcc-related subcells.

The joined interferences $G^2$ = 6 and $G^2$ = 14 with the guiding planar interferences (211) respectively (321)
keep the OT(Ag) nearly conform to the bcc-positions on the scale $a$/3 whereas the IT(Li) are notably compressed.

As the planar interferences $G^2$ = 7 and 15 are not excited in SC lattices the
energy range for radial adjustments in the joined interferences $G^2$ = 6 respectively 14 is enlarged.

In the atomic-site angular-momentum representation
the electronic stabilization of $\gamma$-$\rm Ag_5Li_8$ arises  
from interferences in various parts of the electron quantum space at various energies.
The empirical Hume-Rothery rule, on the contrary, supposes that the main contribution, 
the joined interferences $G^2$ = 18, arise in the $sp$-subspace close to $\epsilon_F$.
This way, the empirical Hume-Rothery rule selects a subclass of $\gamma$-brasses which achieve structure stability
mainly in the $sp$-subspace below $\epsilon_F$,
$\gamma$-$\rm Ag_5Li_8$ does not belong to this subgroup.
However, the chief information of the rule stays valid, 
namely that the atoms in the SC unit cell are driven towards
the sites of a 3$\times$3$\times$3-bcc sublattice if e/a = 21/13 is achieved in an active electron space.

The systems $\gamma$-$\rm Ag_5Li_8$ and 3$\times$3$\times$3-bcc are nearly equivalent regarding planar interferences,
differences occur in the radial interferences and hybridizations.
Hence, a spectral DOS-feature of $\gamma$-$\rm Ag_5Li_8$ which does not occur in 3$\times$3$\times$3-bcc
must arise from the readjustment of the radial order, i.e. from radial interference and hybridization.
This applies to the peak/dip-feature (width 2 eV) in the $sp$DOS of $\gamma$-$\rm Ag_5Li_8$ 
around the joined interferences $G^2$ = 18 above $\epsilon_F$.
The planar interference $G^2$ = 18 links efficiently the local radial processes.

Similar to dia-C (Figure \ref{fig:fig07s3}) the band states de-

\begin{figure}
\vspace{0.5cm}
\centering
\includegraphics[width=9.0cm]{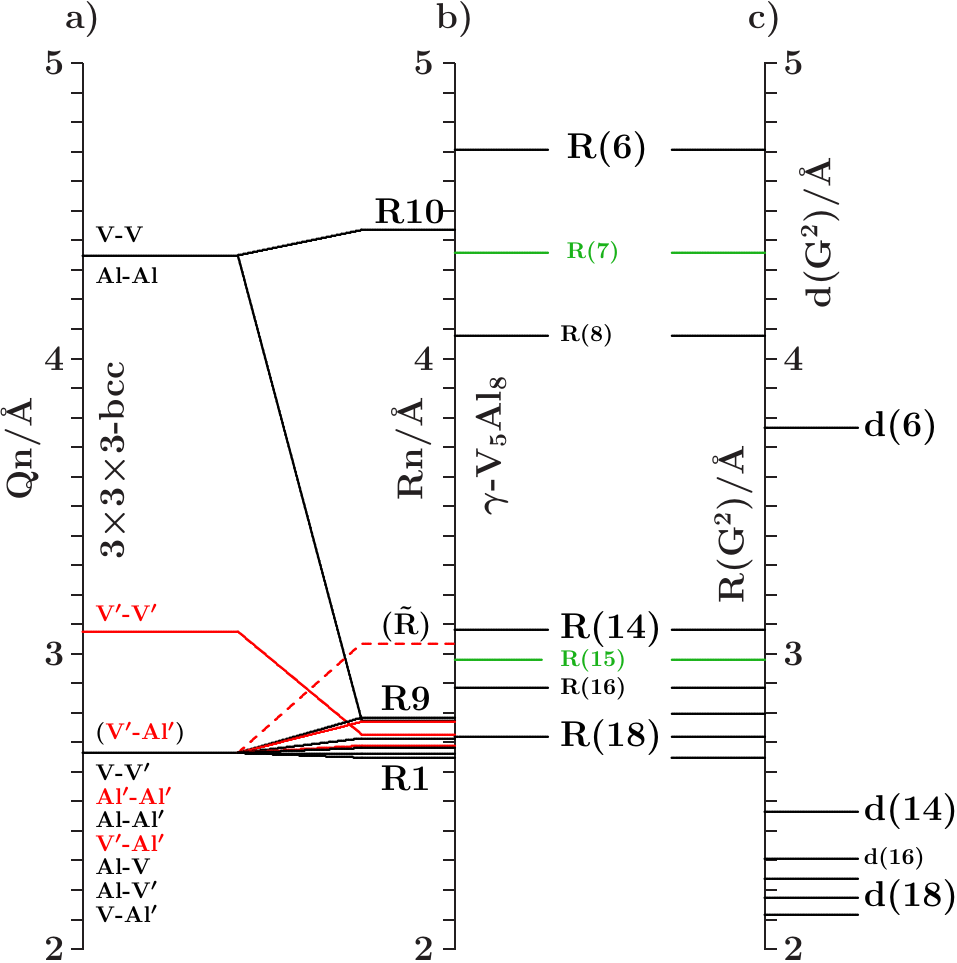}
\caption{The 10 smallest interatomic distances between Al, V, V$^\prime$, Al$^\prime$ and the larger $\tilde{R}$(V$^\prime$-Al$^\prime$) (red: links between the 26-atom clusters of $\gamma$-$\rm V_5Al_8$).
a) 3$\times$3$\times$3-bcc. b) $\gamma$-$\rm V_5Al_8$.
 c) SC reference lattice, interplanar distances $d(G^2)$ and the radial equivalents $R(G^2)$ (\ref{RHKL1}).}
\label{fig:fig35s4}
\end{figure}

\noindent
velop $p$-dominated networks.
However, at $\epsilon_F$, the $p$-dominated networks of $\gamma$-$\rm Ag_5Li_8$ are neither strictly $p$-type
nor fully developed.

Enhanced spectral density at fulfilled interference conditions indicates improved interference 
and the band energy is reduced.

\subsection{\bm $\gamma$-$\rm V_5Al_8$}

The early transition metal vanadium in $\gamma$-$\rm V_5Al_8$ contributes occupied $d$-orbitals 
around the Fermi energy which must affect the structure stabilization.
For this reason, Mizutani $et \; al.$ \cite{Mizutani06a} have examined $\gamma$-$\rm V_5Al_8$ by means of the FLAPW-Fourier method
employing the structure model of Brandon $et \; al.$ \cite{Brandon77} with the lattice constant $a$ = 9.223 \AA.

They extrapolate the Hume-Rothery plot outside the $d$VV$^\prime$-band to the Fermi energy and conclude from $G^2(\epsilon_F)$ = 21
that $\gamma$-$\rm V_5Al_8$ must be stabilized at $e/a$ = 1.94 (cf. Table \ref{tab:tab14s6}).
This is clearly above $e/a$ = 21/13 $\approx$ 1.615 and therefore in contrast to the empirical Hume-Rothery rule.

However, the structure of $\gamma$-brasses suggests 
that above all the interplanar length scale $d(18)$ and the assigned radial length scales must be stabilized.
This requires joined interferences around $G^2$ = 18 at $Z(d(18))$ = 1.538 (Table \ref{tab:tab14s6}).
Two questions arise: In which active space? At which energy? 

\newpage

\begin{figure}
\onecolumn
\centering
\includegraphics[width=13.0cm]{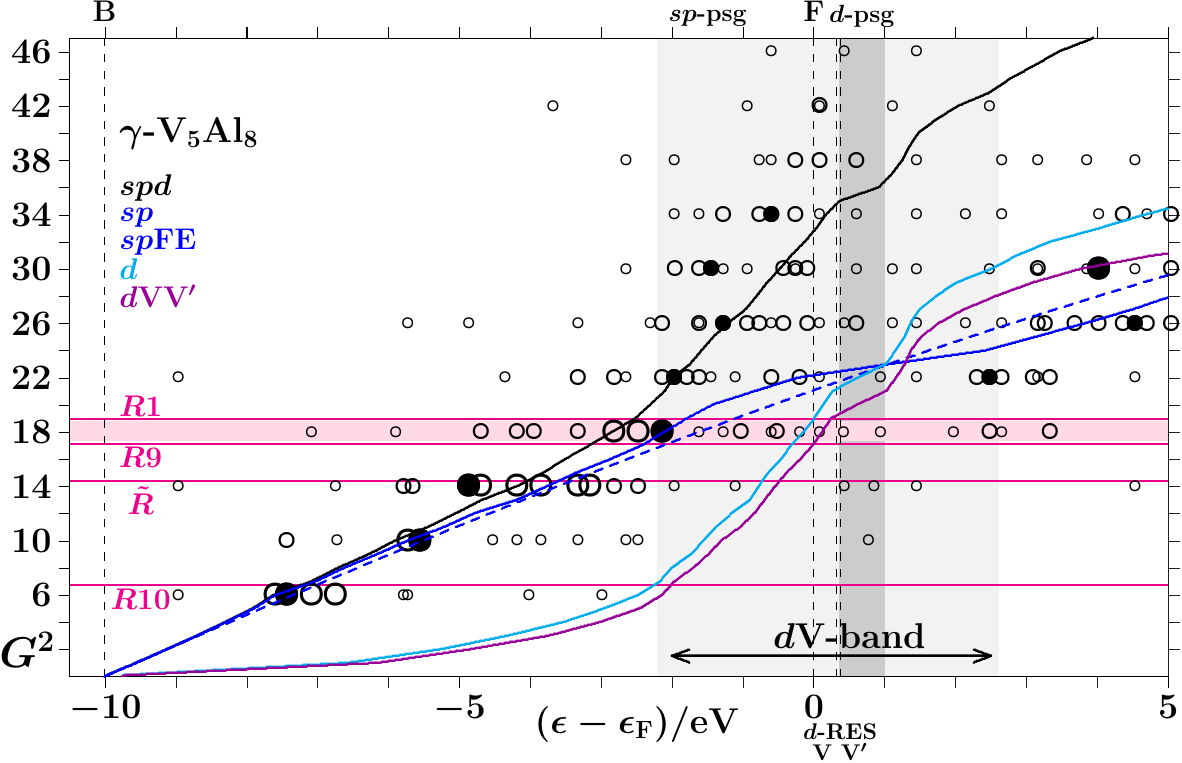}
\caption{ The present VEC-based approach versus the FLAPW-Fourier-approach \cite{MizutaniV1,Mizutani11b}, see text.
Full curves, the interferences $G^2(\epsilon-\epsilon_F)$ in the indicated subspaces.
The blue dashed curve is due to the integrated free-electron $sp$DOS.
Radial interferences (magenta) are denoted by the equivalent values due to (\ref{GQR}).
Note the different widths of the plane-wave spectra outside and
inside the $d$VV$^\prime$-band.}
\label{fig:fig36s4}
\twocolumn
\end{figure}

\noindent
{\bm\bf $The$ $Brandon$ $model$ $versus$ 3$\times$3$\times$3-bcc}.
We use the Brandon model \cite{Brandon77} just as done by Mizutani $et \; al.$ \cite{Mizutani06a},
i.e. $a$ = 9.223 \AA.
No mixed occupations of the subclusters IT(Al), OT(V), OH(V$^\prime$), and CO(Al$^\prime$) are considered.
The shortest interatomic distances among the four components
are listed in the Table \ref{tab:tab12s6}.
For comparison we add the reference system 3$\times$3$\times$3-bcc. 

Joined planar-radial interferences of $\gamma$-$\rm V_5Al_8$ (Figure \ref{fig:fig35s4}) arise
by analogy with $\gamma$-$\rm Ag_5Li_8$ (Figure \ref{fig:fig16s4}).
Main processes are again the compression of IT(Al) 
and the formation of the interference block around $G^2$ = 18.
Different from $\gamma$-$\rm Ag_5Li_8$, the distance $R$8(Al-Al) (size of the
IT(Al), cf. Table \ref{tab:tab12s6}) now joins the interference block.
Only $\tilde{R}$(V$^\prime$-Al$^\prime$) remains assigned to the planar interference $G^2$ = 14.
The radial interference and hybridization along $R$6(V$^\prime$-V$^\prime$) 
(OH-OH, link between 26-atom clusters) joins again closely the planar interference $G^2$ = 18.

We conclude that $\gamma$-$\rm V_5Al_8$ and $\gamma$-$\rm Ag_5Li_8$ are almost equivalent regarding the 
essential length scales which must be stabilized by joined planar-radial interferences. \\[1ex]
\noindent
{\bm\bf $Stabilization$ $in$ $subspaces$}.
Figure \ref{fig:fig36s4} compares the interference status due to the present VEC-based approach with results of the FLAPW-Fourier analysis.
We use the same three classes of circles and bullets as in Figure \ref{fig:fig33s4} to reproduce
the FLAPW-based weighting of planar interferences at symmetry points N \cite{MizutaniV1,Mizutani11b}.

\newpage
\hspace*{1cm}
\vspace{11.0cm}

The continuous curves are derived from (\ref{PLZ},\ref{INTL1}) with 
different choices of the integrated projected DOS, in detail the int-$t$DOS (black, label $spd$), the int-$sp$DOS
(blue, label $sp$), the int-$d$DOS 
(cyan, label $d$), and the int-$d$VV$^\prime$DOS of the 
vanadium subsystem (dark violet, label $d$VV$^\prime$).
The blue dashed curve (label $sp$FE) is derived from the integrated free-electron $sp$DOS 
(symbol B, bottom of the valence band, $\epsilon_b = \epsilon_F$ - 10.01 eV).  

Magenta horizontal lines in Figure \ref{fig:fig36s4} result from the inversion of (\ref{RHKL1}),

\begin{equation}
	G^2(R{\rm n}) = \frac{25}{16}\left(\frac{a}{R{\rm n}}\right)^2,
\label{GQR}
\end{equation}

\noindent
which defines a hypothetical planar interference $G^2(R{\rm n})$ (non-integer) with the same momentum transfer 
as the radial interference along $R$n.
The interference block around $G^2$ = 18 between $R$1 and $R$9 in Figure \ref{fig:fig36s4} is replaced by a magenta shaded stripe.

Suppose $G^2(R{\rm n})$ slightly above an integer $\tilde{G}^2$.
A joined interference [$d(\tilde{G}^2)$,$R$n] arises.
This enables the interplay between the radial order in the SC unit cell and the global planar order.
On the one hand, at given size of the SC unit cell (side length $a$) 
certain radial interferences $R{\rm n}$ are selected for global support.
The global order acts on the local order.
On the other hand, a given strong radial interference $R$n, e.g. for chemical reasons,
may cause the size of the SC unit cell to adjust for $G^2(R{\rm n})$ slightly above an integer $\tilde{G}^2$.
The local order acts on the global order.
Hence, along the magenta horizontal lines in Figure \ref{fig:fig36s4}, appropriate conditions can emerge for the interplay between
the local radial order and the global planar order.

All subspace-related curves have intersections with the magenta lines at certain 
energies where the above mentioned joined interferences may be active in the chosen subspace.
The question arises whether the FLAPW-derived weighting of interferences in the extended $k$-space
follows the expectation, i.e. whether the large circles and the bullets are close to the intersections.

Up to the lower bound of the $d$VV$^\prime$-band in Figure \ref{fig:fig36s4}, the FLAPW-derived weighting 
reveals indeed large circles and bullets close to the intersections between 
magenta lines and the blue curve with the label $sp$.
This means that each strong planar interference $G^2 \le 19$ in the $sp$-subspace indicates strong
radial interference and hybridization on an assigned radial scale, and this 
is relevant to the charge redistribution.

Inside the $d$VV$^\prime$-band, however, the $d$-type radial interferences and hybridizations on the scales $R$1 - $R$9 
are represented by wide flat spectra of planar interferences in the extended $k$-space, starting from $G^2$ = 18
up to large $G^2$ in the total EQS (Figure \ref{fig:fig36s4}, label $spd$).
Different from below the $d$VV$^\prime$-band, no
radial interferences are assigned to individual planar interferences $G^2 > 19$ via (\ref{GQR})
because there are no interatomic distances which are short enough.
Individual planar interferences $G^2 > 19$ are thus less reliable indicators regarding hybridization 
on the atomic scale.
Hence, the information content of $e/a$ seems problematic if derived from a single interference $G^2 > 19$.

To analyze the processes on the shortest radial scales $R$1 - $R$9
we follow in Figure \ref{fig:fig36s4} the shaded stripe around $G^2$ = 18 towards increasing energies,
knowing that the interference block gets active in successively reduced subspaces.
Starting at the lower bound of the $d$VV$^\prime$-band with the total EQS (label $spd$) the maximum
weight of the Fourier component $G^2$ = 18 (bullet) occurs at the intersection 
between the blue curve ($sp$-subspace) and the line $G^2$ = 18.
We conclude that an important stabilizing contribution must arise from the interference block
around $G^2$ = 18 in the $sp$-subspace at the lower bound of the $d$VV$^\prime$-band. 
The question is how such essential interferences clearly below the Fermi energy can support the stabilization of the structure.

\begin{figure}
\centering
\includegraphics[width=8.4cm]{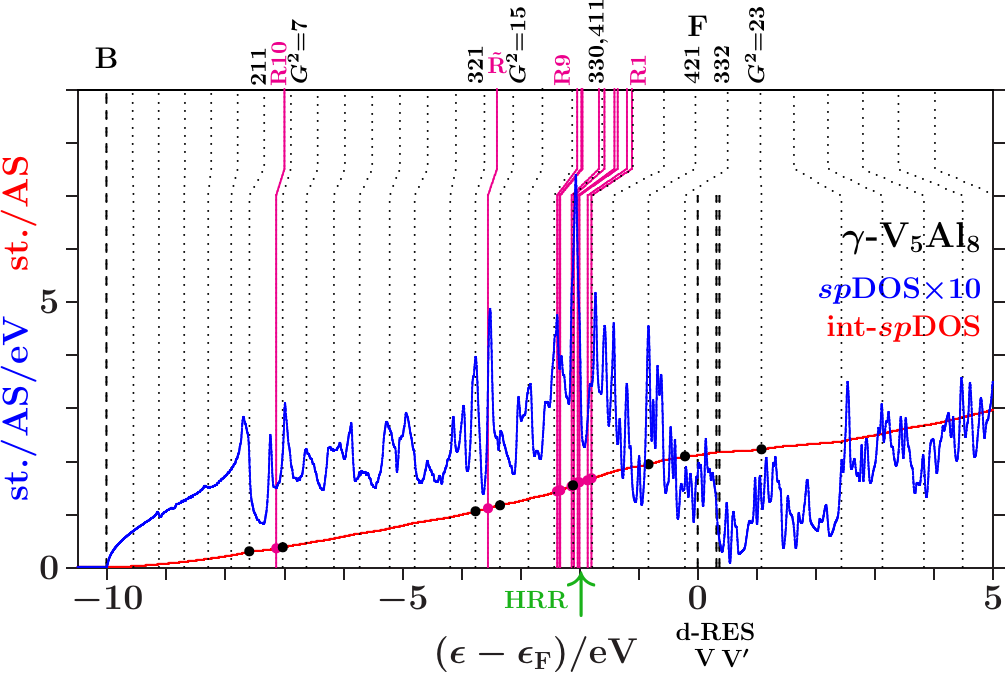}
\caption{Joined interferences in the $sp$-subspace around $G^2$ = 17,18,19 at the lower bound of the $d$VV$^\prime$-band
 support the 4 eV wide pseudogap about the $d$VV$^\prime$-resonances.}
\label{fig:fig37s4}
\end{figure}

The $sp$DOS of $\gamma$-$\rm V_5Al_8$ (Figure \ref{fig:fig37s4}) offers an answer to this question.
Joined interferences $G^2$ = 18 around -2.1 eV, the lower bound of the $d$VV$^\prime$-band, support the opening of
a wide $sp$-pseudogap throughout the $d$VV$^\prime$-band.
This kind of pseudogaps is commonly ascribed to the displacement of spectral $sp$-weight by $d$-states.
The complete spectral feature is centered at $G^2$ = 23, again a gap in  

\begin{figure}
\centering
\includegraphics[width=8.4cm]{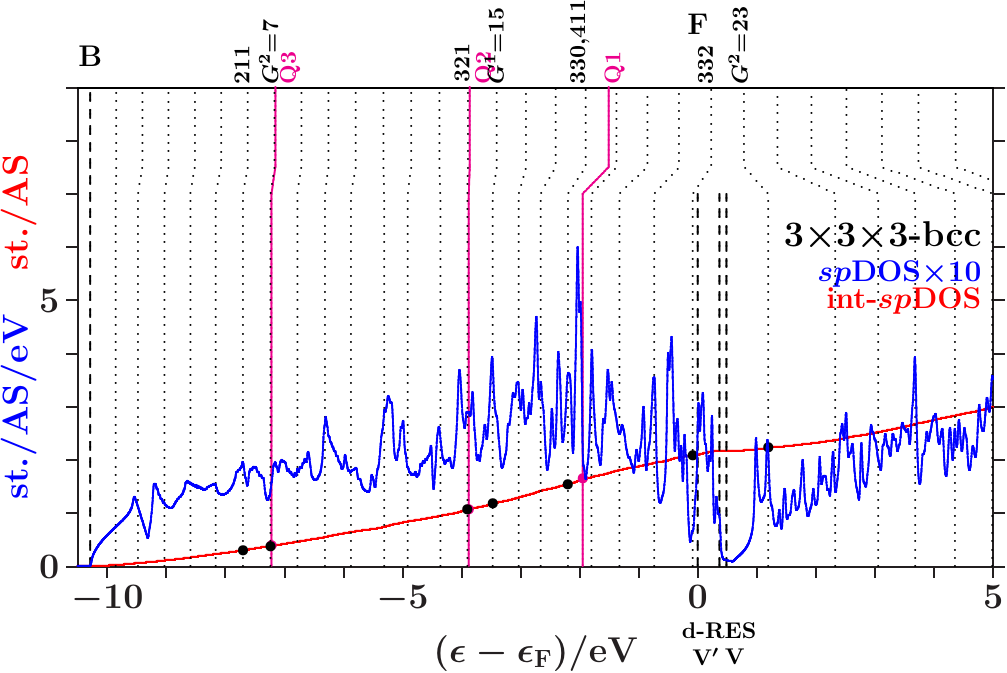}
\caption{$\times$3$\times$3-bcc of $\gamma$-$\rm V_5Al_8$: Note the reduced tuning of spectral weight
to fulfilled interference conditions due to the still mismatched interatomic distances.}
\label{fig:fig38s4}
\end{figure}

\begin{figure}
\centering
\includegraphics[width=6.0cm]{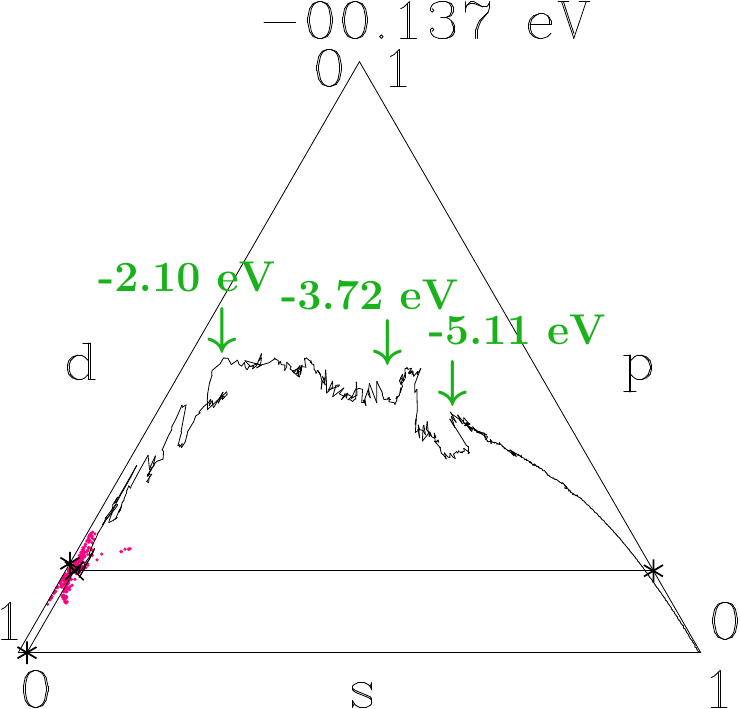}
\caption{$\gamma$-$\rm V_5Al_8$: Note the $sp$-type features between -5,11 eV and -3.72 eV.
Joined interferences $G^2$ = 18 act in the $sp$-subspace 
at the lower bound of the $d$VV$^\prime$-band (-2.1 eV).}
\label{fig:fig39s4}
\end{figure}
 
\begin{figure}
\centering
\includegraphics[width=8.4cm]{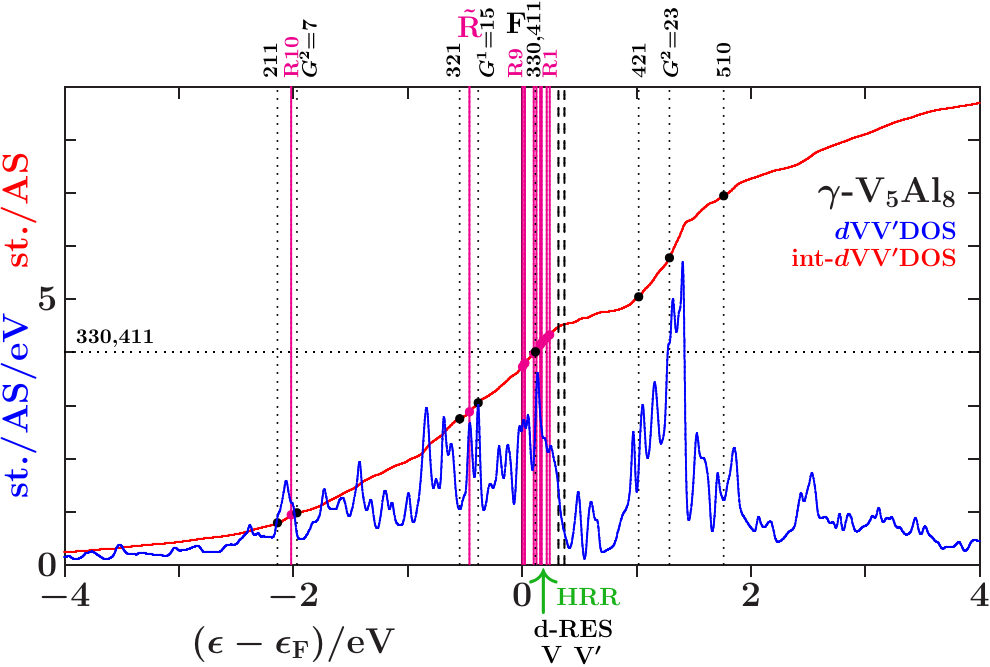}
\caption{The joined interferences $G^2$ = 18 in the $d$VV$^\prime$-subspace
support the $d$-band splitting.
Due to the wide $sp$-pseudogap (\ref{fig:fig39s4}) the active space shrinks to the $d$VV$^\prime$-subspace.}
\label{fig:fig40s4}
\end{figure}

\noindent
the sequence of the SC planar interferences.
Comparison with 3$\times$3$\times$3-bcc (Figure \ref{fig:fig38s4}) confirms the enhanced concentration of $sp$-weight
at the interference energies in $\gamma$-$\rm V_5Al_8$.
Even the joined interferences $G^2$ = 14 and $G^2$ = 6 on larger length scales are included.
Hence, the radial adjustment in the hypothetical final relaxation step improves the interference status.

Figure \ref{fig:fig39s4} reveals the view of the band states.
We show the $spd$-decompositions.
Starting at the lower bound of the $d$VV$^\prime$-band around -2.1 eV,
the band states get increasingly $d$-dominated.
Towards $\epsilon_F$, the $s$-subspace is nearly excluded, only a small interplay of the network-forming $pd$-components remains.
This way, joined interferences $G^2$ = 18 in the $sp$-subspace fairly below the Fermi energy enable 
nearly undisturbed $dd$-networking towards the $d$VV$^\prime$-resonances.

If we reduce to the subspace of $d$VV$^\prime$ (Figure \ref{fig:fig40s4}, in Figure \ref{fig:fig36s4} the dark violet full curve) 
the planar interference $G^2$ = 18 must be active at the energy where the 20 VV$^\prime$
contribute each (52/20) 1.538 $\approx$ 4 $d$VV$^\prime$-states on the average.
This happens at +0.118 eV, just below +0.187 eV where the (52/20)(21/13) = 4.2 $d$VV$^\prime$-states
are are reached due to the Hume-Rothery rule.
In the $d$VV$^\prime$DOS (Figure \ref{fig:fig40s4}) a deep pseudogap opens above the lowest
$d$VV$^\prime$-resonance at +0.318 eV.
Such pseudogaps are known to result from bonding/anti-bonding splittings of transition metal $d$-bands.
Again, the joined interferences around $G^2$ = 18 support the formation of this hybridization phenomenon.

\begin{figure}
\centering
\includegraphics[width=6.0cm]{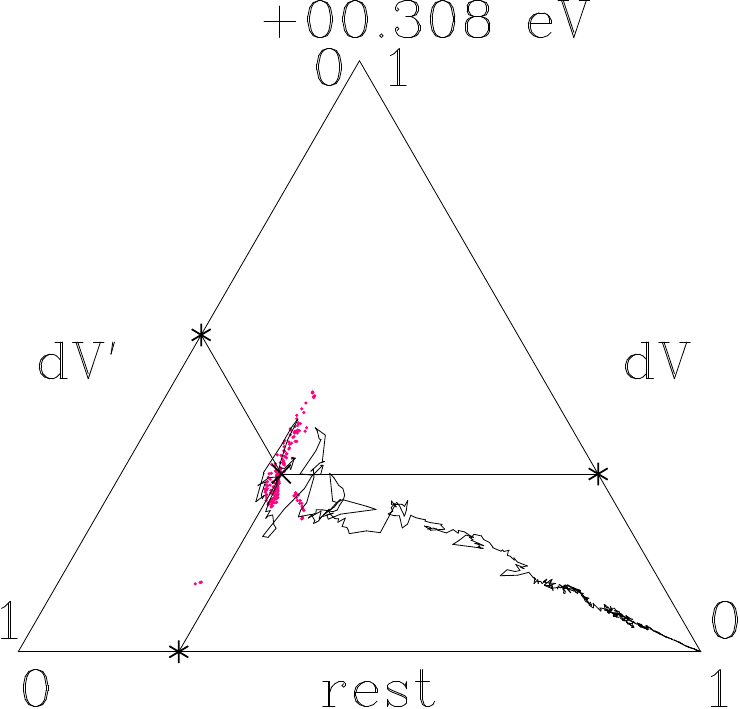}
\caption{$\gamma$-$\rm V_5Al_8$: The $d$V-dV$^\prime$ interplay in the band states close to the $d$VV$^\prime$-resonances.
\vspace{0.45cm}}
\label{fig:fig41s4}
\end{figure}

The band states close to the lowest $d$VV$^\prime$-resonance at +0.318 eV
prove $d$-type with moderate $d$V-$d$V$^\prime$ interplay and minor participation of the residual EQS (Figure \ref{fig:fig41s4}).
This confirms that the wide $sp$-pseudogap (Figure \ref{fig:fig37s4}) 
facilitates just this kind of nearly undisturbed $dd$-networking.

\begin{figure}
\vspace{-0.1cm}
\centering
\includegraphics[width=8.4cm]{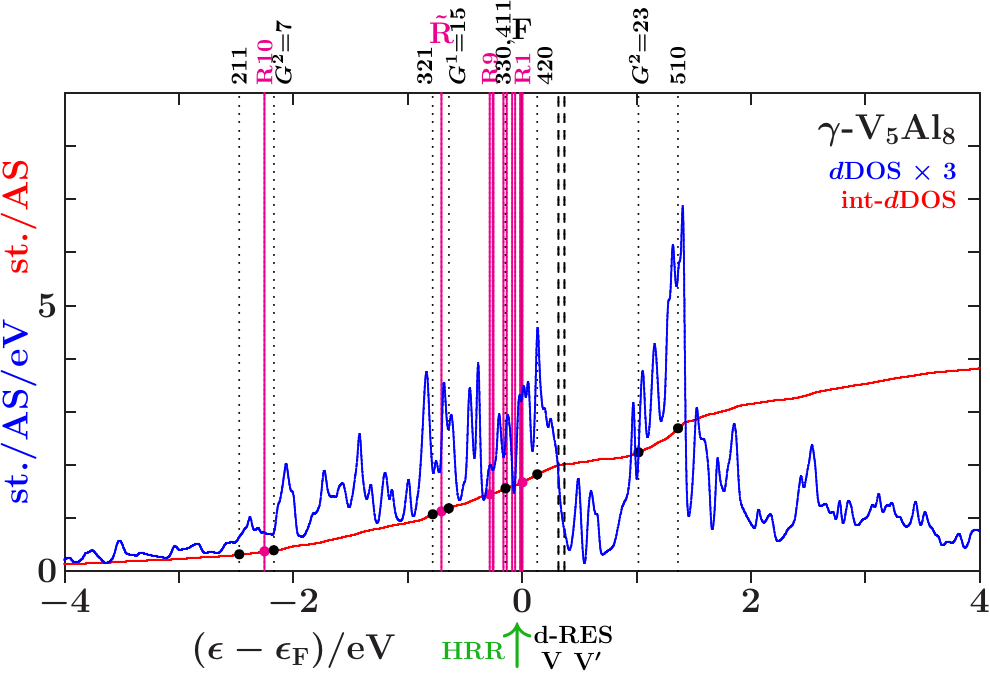}
\caption{Joined interferences $G^2$ = 18 act in the $d$-subspace of $\gamma$-$\rm V_5Al_8$
somewhat below $\epsilon_F$.
The energy HRR close to $\epsilon_F$ indicates the validity of the Hume-Rothery rule in the $d$-subspace.}
\label{fig:fig42s4}
\end{figure}

\begin{figure}
\centering
\includegraphics[width=8.6cm]{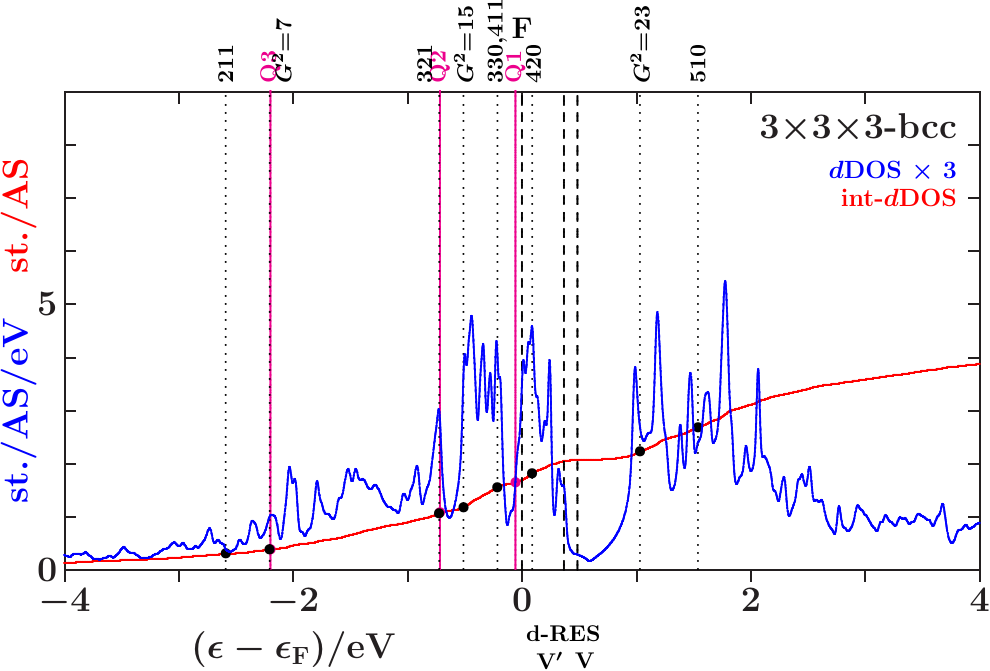}
\caption{3$\times$3$\times$3-bcc of $\gamma$-$\rm V_5Al_8$: The joined interferences [$d$(18),$Q$(1)]
acquire only small spectral density which indicates insufficient electron-structure tuning.}
\label{fig:fig43s4}
\end{figure}

The situation around $\epsilon_F$ proves intermediate on
the way to the interference-supported $d$-band splitting (Figure \ref{fig:fig42s4}).
The cyan full curve (label $d$) in Figure \ref{fig:fig36s4} passes the line $G^2$ = 18 (1.538 per site) at -0.144 eV
whereas 1.659 per site are accumulated up to $\epsilon_F$.
This is close to 21/13 $\approx$ 1.615 as supposed by the empirical Hume-Rothery rule at the occupation edge.
Hence, along with the growing $d$VV$^\prime$-network, joined interferences $G^2$ = 18
arise in the total $d$-subspace close to $\epsilon_F$.
The Hume-Rothery rule applies to the total $d$-subspace (cf. the comments on Table \ref{tab:tab02s2}).
Figure \ref{fig:fig42s4} shows that the interference block around $G^2$ = 18 of $\gamma$-$\rm V_5Al_8$
acquires more spectral weight than [$d$(18),$Q$1] in 3$\times$3$\times$3-bcc (Figure \ref{fig:fig43s4}).

\begin{figure}
\vspace{-0.3cm}
\centering
\includegraphics[width=8.4cm]{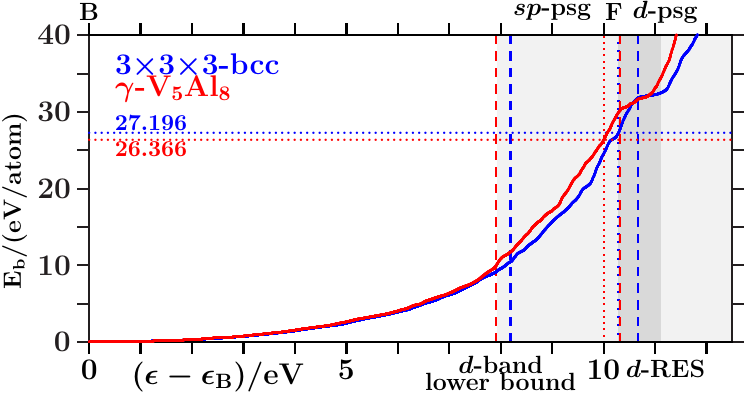}
\caption{The electronic band energies $E_b$ (\ref{EB1})
versus the energy above the bottom of the valence band.}
\label{fig:fig44s4}
\end{figure}

\begin{figure}
\centering
\includegraphics[width=8.4cm]{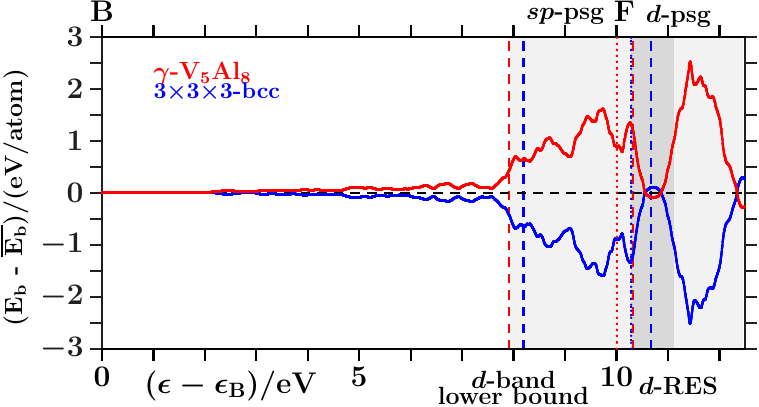}
\caption{The distances of the electronic band energies $E_b$ (\ref{EB1})
from the common average versus the energy above the bottom of the valence band.}
\label{fig:fig45s4}
\end{figure}

\begin{figure}
\centering
\includegraphics[width=8.4cm]{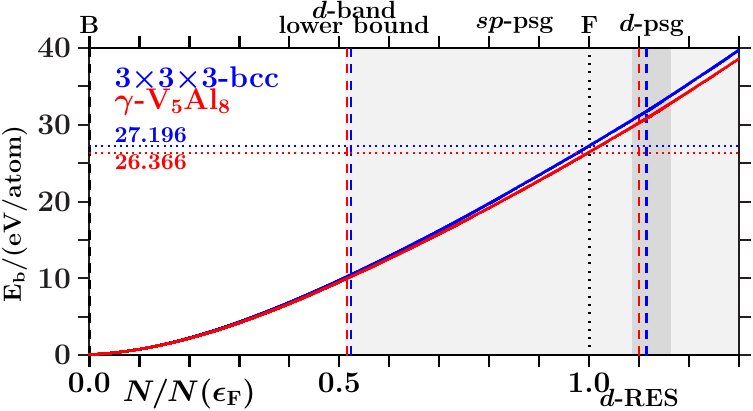}
\caption{The electronic band energies $E_b$ (\ref{EB1})
versus the relative valence-band filling, $N/N(\epsilon_F)$.}
\label{fig:fig46s4}
\end{figure}

\begin{figure}
\centering
\includegraphics[width=8.4cm]{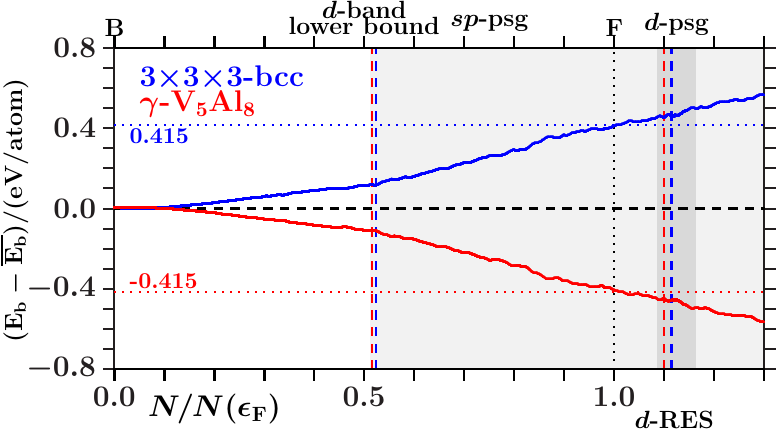}
\caption{The distances of the electronic band energies $E_b$ (\ref{EB1})
from the common average versus the relative valence-band filling, $N/N(\epsilon_F)$.}
\label{fig:fig47s4}
\end{figure}

Summarizing joined interferences $G^2$ = 18 we note
activities at the bottom of the $d$VV$^\prime$-band in $sp$-dominated subspaces
and towards the $d$VV$^\prime$-resonances in $d$-dominated subspaces.
The former open a wide $sp$-pseudogap for low $sp$-presence in the band states towards the $d$VV$^\prime$-resonances,
the latter act in the $d$VV$^\prime$-subspace for $d$-band splitting
and in the total $d$-subspace below the Fermi energy for bcc-related subcells on the scale $a$/3. \\[1ex]
\noindent
{\bm\bf $Band$ $energy$}.
The Figures \ref{fig:fig44s4} to \ref{fig:fig47s4} use the same presentation modes of the band energies $E_b(\epsilon)$ (\ref{EB1})
of $\gamma$-$\rm V_5Al_8$ and 3$\times$3$\times$3-bcc as demonstrated with $\gamma$-$\rm Ag_5Li_8$.
Vertical dashed lines denote the lower bounds of the $d$VV$^\prime$-bands
and the lower $d$VV$^\prime$-resonances whereas
dotted lines highlight the Fermi energies and the resulting band energies.
The shaded areas remember both pseudogaps of the $\gamma$-phase.

The plots versus the energy above the bottom of the valence band (Figures \ref{fig:fig44s4} and \ref{fig:fig45s4})
compare the systems in non-equivalent situations due to different band fillings $N(\epsilon)$.
Large differences arise in particular in the $d$VV$^\prime$-band due to
specific b/ab-splittings on small energy scales.
To get rid of such details we plot the band energies in the Figures \ref{fig:fig46s4} and \ref{fig:fig47s4}
versus the relative band filling, $N(\epsilon)/N(\epsilon_F)$.
Similar to $\gamma$-$\rm Ag_5Li_8$ (Figure \ref{fig:fig32s4}) a monotonous increase of the band-energy difference remains.

With the improved interference status, the $\gamma$-phase ends up
with the lower band energy at the Fermi energy (26.366 eV/atom versus 27.196 eV/atom, Figure \ref{fig:fig44s4}).
The difference of -0.830 eV/atom results mainly from the $d$-subspace (-0.683 eV/atom)
whereas the $sp$-subspace contributes only -0.147 eV/atom.
Nevertheless, the stable radial order of $\gamma$-$\rm V_5Al_8$ results from the cooperation
of joined interferences $G^2$ = 18 in both the $d$- and the $sp$-subspaces. \\[1ex]
\noindent
{\bm\bf $Charge$ $transfer$ $and$ $e/a$-$ratio$}.
We proceed as with $\gamma$-$\rm Ag_5Li_8$.
In the initial state free atoms are supposed in the AS (configurations V(2 0 3),Al(2 1 0)).
The 20 VV$^\prime$ contribute 100 valence electrons,
and the 32 AlAl$^\prime$ add further 96.
Inspecting the 

\noindent
final state shows that $\sim$ 0.18 valence electrons per SC unit cell have moved from 
VV$^\prime$ to AlAl$^\prime$, only 0.1 \% of the total 196 valence electrons per SC unit cell. \\

\begin{table}
\small
\centering
\caption{$\gamma$-$\rm V_5Al_8$, the changes of the AS-charges in the transition from free atoms
to the $\gamma$-phase.
The last two lines show the average changes in two steps.}
\vspace{0.2cm}
        \begin{tabular}{ l | l | l | l }
        \hline
        AS-types \hspace*{1.0cm}        & $s$ \hspace*{0.8cm}   & $p$ \hspace*{0.8cm}   & $d$ \hspace*{0.8cm}    \\
        \hline
        Al(IT)                          & -0.941                & +0.491                & +0.349                 \\
        V(OT)                           & -1.494                & +0.685                & +0.707                 \\
        V$^\prime$(OH)                  & -1.441                & +0.777                & +0.717                 \\
        Al$^\prime$(CO)                 & -0.914                & +0.570                & +0.385                 \\
        \hline
%        averages(8/8/12/24)             & 	                & 	                & 	                 \\
        to 3$\times$3$\times$3-bcc      & -1.136                & +0.610                & +0.526                 \\
        to $\gamma$-$\rm V_5Al_8$       & +0.007                & +0.014                & -0.021                 \\
        \hline
        \end{tabular}
\label{tab:tab05s4}
\end{table}

Table \ref{tab:tab05s4} shows the changes of the partial AS-charges in the transition from free atoms to the $\gamma$-phase.
Enhanced $p$- and $d$-occupations support the 
opening of the stabilization-relevant pseudogaps in the respective subspaces
(cf. Figures \ref{fig:fig37s4} and \ref{fig:fig40s4}).  

The last two lines present the average changes due to bonding into the perfect planar order (line 5) and 
due to readjusted radial order (line 6).
Hence, 1.129 = 1.136 - 0.007 $s$-electrons per AS change into 0.624 = 0.610 + 0.014 $p$-electrons 
and 0.505 = 0.526 - 0.021 $d$-electrons with the partitions according to the two steps mentioned. 
The chief contributions arise from bonding into the perfect planar order.

Table \ref{tab:tab06s4}  displays $e/a$-ratios of $\gamma$-$\rm V_5Al_8$ 
which result from the FLAPW-Fourier method \cite{Mizutani06a,Mizutani17a}
respectively from the present approach. 
The top two lines (label HRP) refer to extrapolations of high-energy Hume-Rothery plots 
to the center of the $d$VV$^\prime$-band.
The former one \cite{Mizutani06a} provides $G^2(\epsilon_F)$ = 21 in perfect agreement with the
free-electron value (label $sp$FE) of the present approach.
A subsequent attempt \cite{Mizutani17a} results $G^2(\epsilon_F)$ = 22.7
but this extrapolation suffers from wide fluctuations. 
Without arbitrariness the present VEC-based approach (label $sp$) derives $G^2(\epsilon_F)$ = 22.17 
from the integrated $sp$DOS (Figure \ref{fig:fig36s4}, blue full curve).

\begin{table}
\small
\centering
\caption{$\gamma$-$\rm V_5Al_8$, the ratios $e/a$           
due to the Hume-Rothery plot (HRP) and to the integrated free-electron $sp$DOS ($sp$FE) 
respectively the actual integrated $sp$DOS ($sp$).}
\vspace{0.2cm}
        \begin{tabular}{ l | l | l }
        \hline
         					& $G^2(\epsilon_F)$  \hspace*{0.6cm}	&  $e/a$ \hspace*{0.6cm} \\
        \hline
        HRP \cite{Mizutani06a}         		& 21                                  	& 1.94 		       \\
        HRP \cite{Mizutani17a}, not shown    	& 22.7                                	& 2.18 		       \\
        $sp$FE                         		& 21.04                               	& 1.94		       \\
        $sp$                           		& 22.17                               	& 2.10 		       \\
        \hline
        \end{tabular}
\label{tab:tab06s4}
\end{table}

However, none of the offered $e/a$-ratios in Table \ref{tab:tab06s4} is really significant
regarding the electronic stabilization of $\gamma$-$\rm V_5Al_8$. 
As pointed out in connection with Figure \ref{fig:fig36s4},
one cannot deduce the status of charge redistribution on existing interatomic length scales 
from an individual planar interference $G^2$ $\ge$ 20. 
Individual planar interferences $G^2$ $\le$ 19, on the contrary, are indicators of stabilizing charge redistributions
on the shortest radial scales via (\ref{GQR}).

The alloy $\gamma$-$\rm V_5Al_8$ prefers stabilization in the total $d$-subspace close to $\epsilon_F$.
By analogy to the empirical Hume-Rothery rule 
one could try to estimate from free atoms the capacity to accumulate $d$-weight.
This results e/a = 60/52 = 1.154 $<$ 21/13 $\approx$ 1.615, not sufficient for planar interferences $G^2$ = 18
in the total $d$-subspace. 
Thanks to the considerable $s$-to-$d$ transfer of +0.505 (Table \ref{tab:tab05s4}),
the calculated occupation of the total $d$-subspace up to $\epsilon_F$ is 1.659 
close to 21/13 $\approx$ 1.615.
The corresponding demands the empirical Hume-Rothery rule regarding planar interferences $G^2$ = 18 in the 
$sp$-subspace slightly below the Fermi energy.
However, one cannot estimate the $s$-to-$d$ transfer by simple means.
Hence, a practicable prognostic tool is not obtained this way. \\[1ex]
\noindent
{\bm\bf $Summarizing$ $\gamma$-$\rm V_5Al_8$}.
The alloy is electronically stabilized by joined planar-radial interferences and hybridizations $G^2$ = 18, 14, and 6.

The interference block around $G^2$ = 18 contributes threefold:
(i) $G^2$ = 18 acts in the $sp$-subspace at the lower bound of the $d$VV$^\prime$-band 
which ensures low $sp$-presence around the $d$VV$^\prime$-resonances.
(ii) $G^2$ = 18 acts in the $d$VV$^\prime$-subspace below the $d$VV$^\prime$-resonance for $d$-band splitting.
(iii) $G^2$ = 18 acts in the total $d$-subspace below $\epsilon_F$ in an intermediate state
of the growing $d$-network.

The interference block $G^2$ = 18 in the $sp$-subspace configures favorable conditions 
for the interference block $G^2$ = 18 in $d$-dominated subspaces towards the $d$VV$^\prime$-resonances.

Two findings regarding the electron-per-atom ratio are announced in view of $\gamma$-$\rm V_5Al_8$:
(i)  Estimating e/a due to the $sp$-configurations of the free atoms is not significant
because the stabilization is rather $d$-like.
(ii) Estimating e/a due to an individual planar interference $G^2 \ge 20$ is not significant
because the shortest interatomic distances couple to $G^2$ = 19 at the most.
 
Joined interferences $G^2$ = 6 stabilize the OT(V) which is common with $\gamma$-$\rm Ag_5Li_8$.
Different from $\gamma$-$\rm Ag_5Li_8$,
the compression of the IT(Al) is assigned with joined interferences $G^2$ = 18 instead of $G^2$ = 14.

The band-energy differences at given band fillings grow monotonously throughout the valence band.
This is common with $\gamma$-$\rm Ag_5Li_8$.

\subsection{\bm $\gamma$-$\rm Cu_5Zn_8$}

The alloy is the prototype of the group-one $\gamma$-brasses in the classification 
scheme after Mizutani $et \; al.$ \cite{Mizutani17a}.
Generic properties are the estimated ratios e/a = 21/13 $\approx$ 1.615
and distinct pseudogaps around the Fermi energy.
This is confirmed by the FLAPW-Fourier method, in particular by the Hume-Rothery plot \cite{Mizutani17a}.
Powder diffraction experiments show dominant peaks (330,411) and much smaller peaks nearby.
Paxton $et \; al.$ \cite{Paxton97} have performed the first LMTO-ASA calculations.
They ascribe structure stability mainly due to the planar interferences (330,411) whereas
deviations from bcc enter via the weak planar interferences (420), (332), and (422).
Reinvestigation by Mizutani $et \; al.$ \cite{Mizutani04a}, again using the LMTO-ASA,
reveals a moderate contribution only by (420) in addition to the dominating contributions by (330,411).
More important, they demonstrate that the Hume-Rothery stabilization 
is efficiently assisted by $sp$-$d$ hybridization.

In the following we focus on both above topics: (i) We analyze the relaxation of the properly decorated 
3$\times$3$\times$3-bcc substructure to the experimentally verified structure.
(ii) We merge planar interference and hybridization to result in
joined planar-radial interferences. 

The alloy explains how two interacting $d$-bands control the environment of the Fermi energy. \\[1ex]
\noindent
{\bm $The$ $Gourdon$ $model$ $versus$ 3$\times$3$\times$3-$bcc$}.
Gourdon $et \; al.$ \cite{Gourdon07} compare different stoichiometries 
around the ideal system $\gamma$-$\rm Cu_5Zn_8$.
We employ the ideal structure due to neutron diffraction with the side length $a$ = 8.866 \AA $\;$ of the SC unit cell. 

\begin{figure}
\centering
\includegraphics[width=8.5cm]{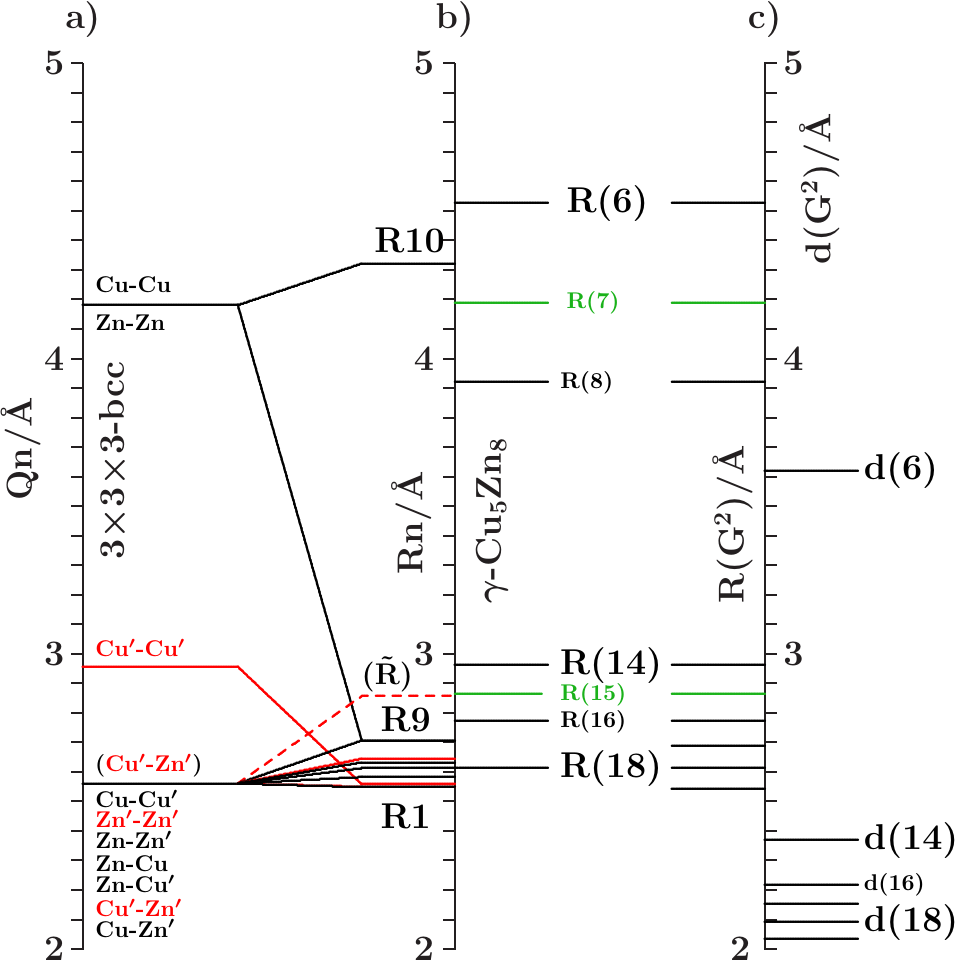}
\caption{The 10 smallest interatomic distances between Zn, Cu, Cu$^\prime$, Zn$^\prime$ and the larger
$\tilde{R}$(Cu$^\prime$-Zn$^\prime$) (red: links between the 26-atom clusters).
a) 3$\times$3$\times$3-bcc. b) $\gamma$-$\rm Cu_5Zn_8$.
 c) SC reference lattice, interplanar distances $d(G^2)$ and the radial equivalents $R(G^2)$ (\ref{RHKL1}).}
\label{fig:fig48s4}
\end{figure}

By analogy with Figure \ref{fig:fig16s4} ($\gamma$-$\rm Ag_5Li_8$) and Figure \ref{fig:fig35s4} ($\gamma$-$\rm V_5Al_8$) 
we characterize in Figure \ref{fig:fig48s4} the relaxation of the shortest interatomic distances on going
from 3$\times$3$\times$3-bcc to $\gamma$-$\rm Cu_5Zn_8$ (cf. Table \ref{tab:tab13s6}).
Note again the compression of the IT(Zn) and the formation of 
a block of joined planar-radial interferences around $G^2$ = 18, 
all in close analogy with $\gamma$-$\rm V_5Al_8$ (Figure \ref{fig:fig35s4}). \\[1ex]
\noindent
{\bm $Electronic$ $stabilization$ $in$ $subspaces$}.
We search for joined interferences $G^2$ = 18 in the momentum space,
i.e. for the active subspaces in the atomic-site angular momentum representation and for the interference energies. 

\begin{figure}
\centering
\includegraphics[width=8.5cm]{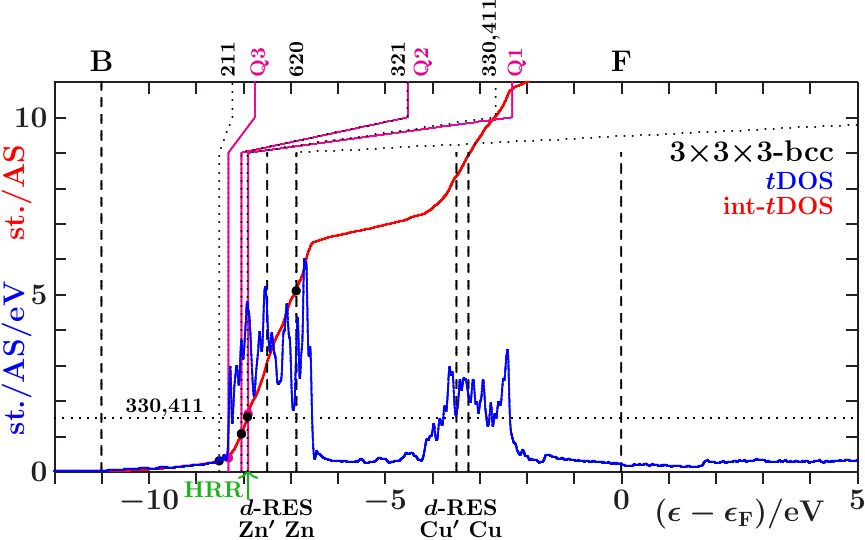}
\caption{3$\times$3$\times$3-bcc: Interferences in the $d$ZnZn$^\prime$-band.}
\label{fig:fig49s4}
\end{figure}

\begin{figure}
\centering
\includegraphics[width=8.5cm]{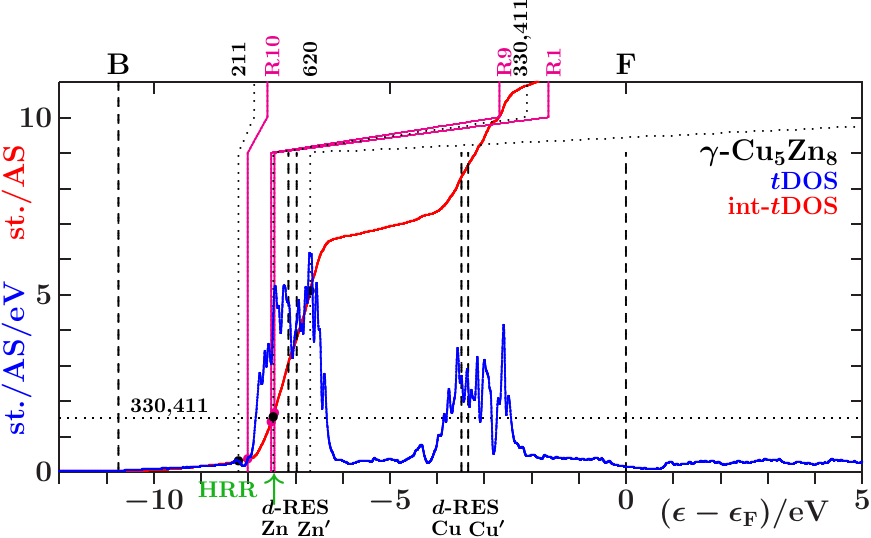}
\caption{$\gamma$-$\rm Cu_5Zn_8$: Interferences in the $d$ZnZn$^\prime$-band.
Note the reduced width of the $d$ZnZn$^\prime$-band.}
\label{fig:fig50s4}
\end{figure}

\begin{figure}
\centering
\includegraphics[width=5.5cm]{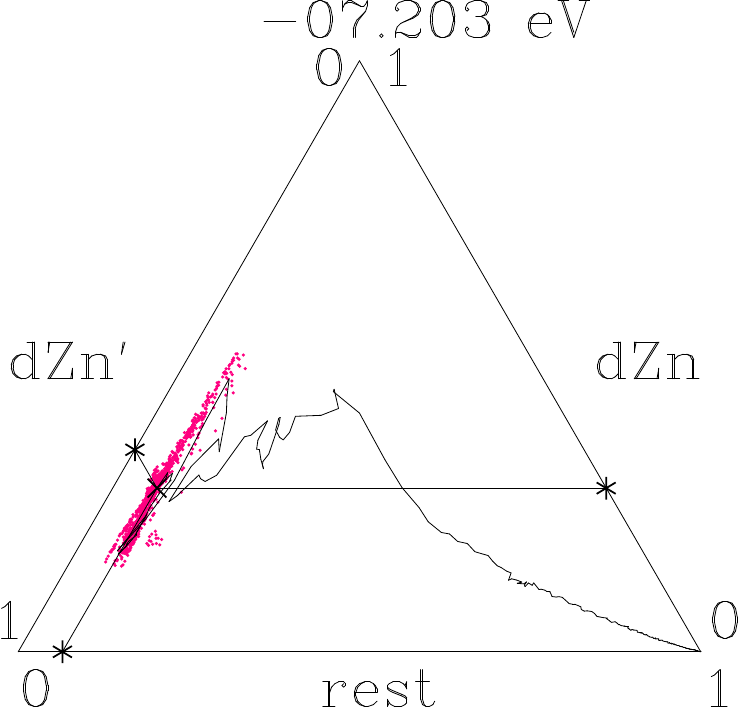}
\caption{$\gamma$-$\rm Cu_5Zn_8$: The decomposition of band states below the $d$ZnZn$^\prime$-resonances.
The interplay is confined to the active $d$ZnZn$^\prime$-subspace.}
\label{fig:fig51s4}
\end{figure}

The Figures \ref{fig:fig49s4} and \ref{fig:fig50s4} show the total DOS of 3$\times$3$\times$3-bcc and $\gamma$-$\rm Cu_5Zn_8$
with some indicated interferences.
In both cases, the joined interferences $G^2$ = 18 act in the bonding $d$ZnZn$^\prime$-band.
Different from 3$\times$3$\times$3-bcc, the $\gamma$-phase has 
the $d$-resonances of Zn (-7.14 eV) and Zn$^\prime$ (-6.97 eV) close together.
The width of the $d$ZnZn$^\prime$-band shrinks and distinct bonding and anti-bonding peaks appear, 
separated by a narrow pseudogap around the $d$-resonances.
Note that the joined interferences $G^2$ = 18 act rather far below the $d$-resonances.
Similar to $\gamma$-$\rm V_5Al_8$ (Figure \ref{fig:fig40s4}), one may suppose active subspaces where 
they get closer to the $d$ZnZn$^\prime$-resonances such as the $d$ZnZn$^\prime$-subspace (Figure \ref{fig:fig51s4}). 

\begin{figure}
\centering
\includegraphics[width=8.1cm]{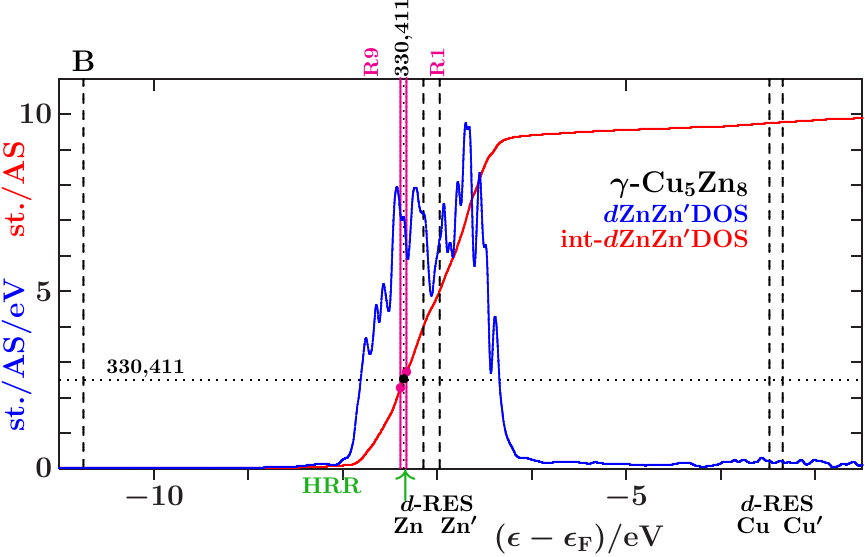}
\caption{$d$ZnZn$^\prime$-subspace: Joined interference $G^2$ = 18.}
\label{fig:fig52s4}
\end{figure}

\begin{figure}
\centering
\includegraphics[width=8.1cm]{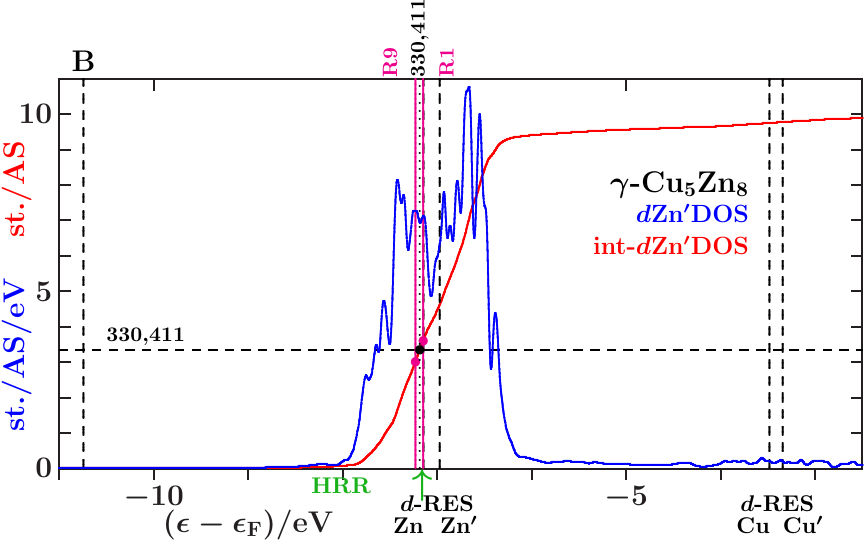}
\caption{$d$Zn$^\prime$-subspace: Joined interference $G^2$ = 18.}
\label{fig:fig53s4}
\end{figure}

\begin{figure}
\centering
\includegraphics[width=8.1cm]{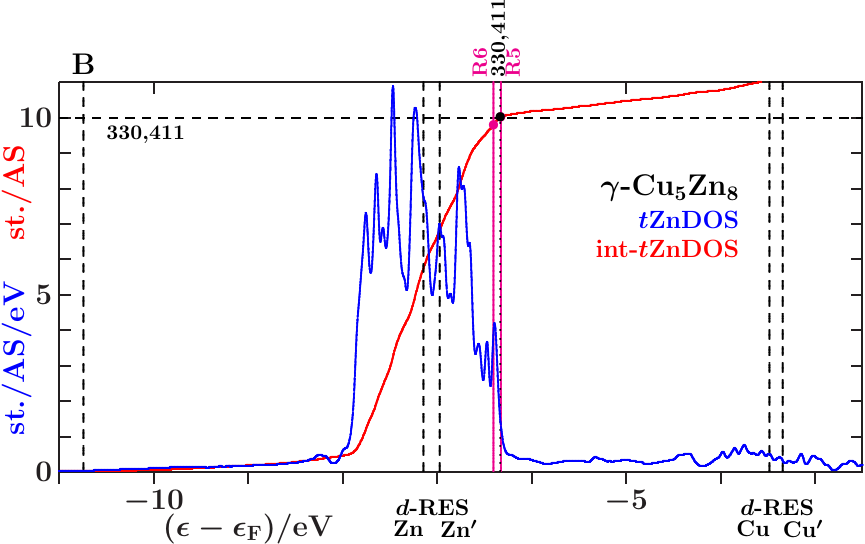}
\caption{$t$Zn-subspace: Joined interference $G^2$ = 18.}
\label{fig:fig54s4}
\end{figure}

Going into detail we reduce the observed space 
in the Figures \ref{fig:fig50s4} and \ref{fig:fig52s4} to \ref{fig:fig54s4}
from the total EQS to

\newpage

\noindent
the subspaces $d$ZnZn$^\prime$, $d$Zn$^\prime$, and finally to $t$Zn.
The latter three subspaces contain only 32, 24, respectively 8 of the 52 AS in the SC unit cell.
At interference the MS requires contributions per active AS which are enlarged 
by factors of 52/32, 52/24, respectively 52/8 in comparison with Figure \ref{fig:fig50s4}.
Hence, joined interferences $G^2$ = 18 in the subspaces of $d$ZnZn$^\prime$ (Figure \ref{fig:fig52s4}) and
$d$Zn$^\prime$ (Figure \ref{fig:fig53s4}) dominate the lower respectively the upper subpeak 
in the bonding $d$ZnZn$^\prime$-band.
The small spectral feature at the upper edge of the anti-bonding $d$ZnZn$^\prime$-band 
results from joined interferences $G^2$ = 18 in the $t$Zn-subspace (Figure \ref{fig:fig54s4}).
We conclude that joined interferences $G^2$ = 18 in particular in the $d$Zn$^\prime$-subspaces
support the $d$-band splitting.

\begin{figure}
\centering
\includegraphics[width=8.1cm]{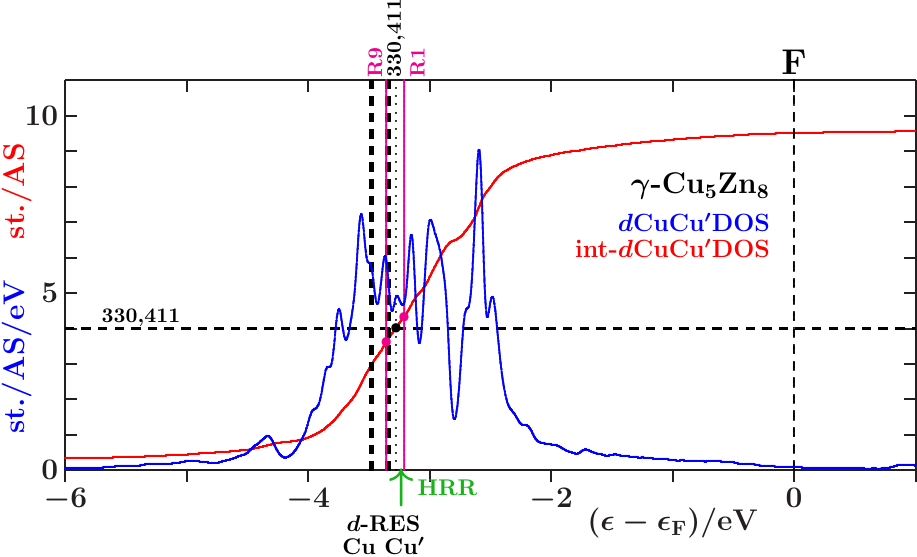}
\caption{$d$CuCu$^\prime$-subspace: Joined interference $G^2$ = 18.}
\label{fig:fig55s4}
\end{figure}

\begin{figure}
\centering
\includegraphics[width=8.5cm]{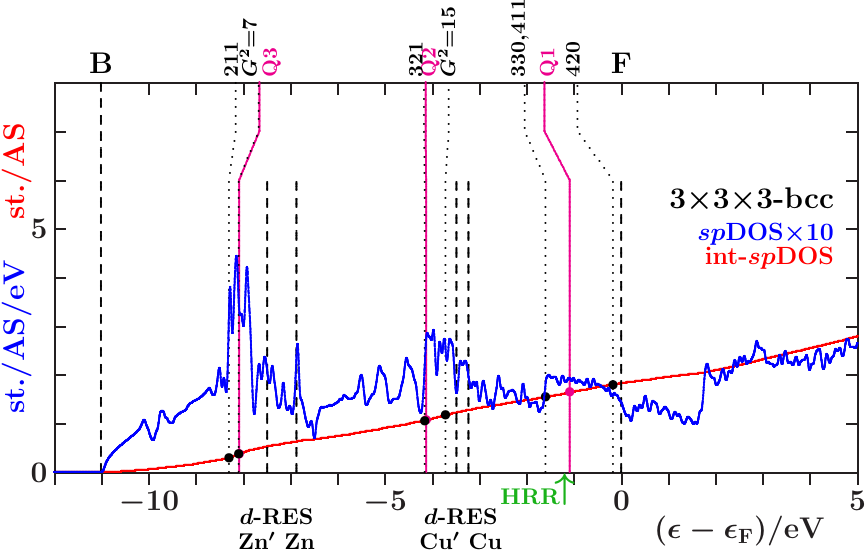}
\caption{Interferences prior to the radial adjustment.}
\label{fig:fig56s4}
\end{figure}

\begin{figure}
\centering
\includegraphics[width=8.4cm]{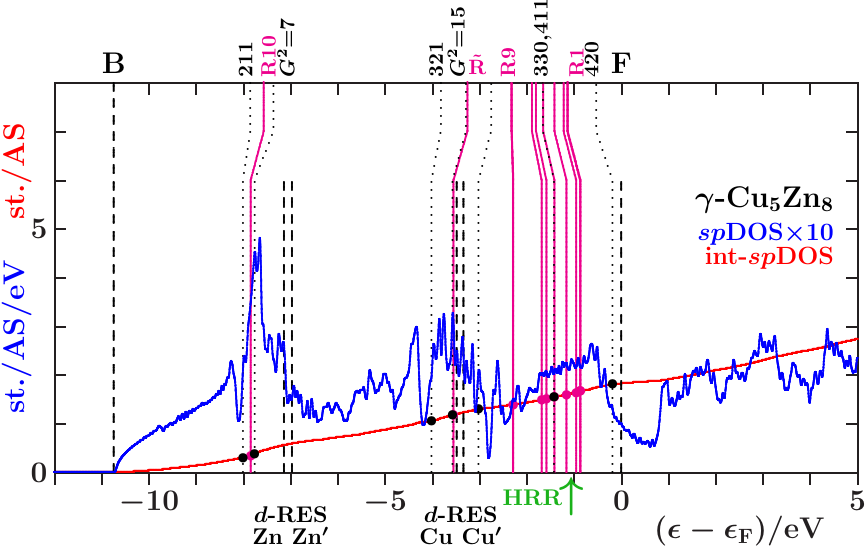}
\caption{Interferences after the radial adjustment.}
\label{fig:fig57s4}
\end{figure}

Contrary to the $d$ZnZn$^\prime$-band the $d$CuCu$^\prime$-band (Figure \ref{fig:fig55s4}) appears quite strange.
Joined interferences $G^2$ = 18 at low spectral density and high spectral
density at the $d$CuCu$^\prime$-resonances are just the opposite of the expected allocations.
Obviously, the zinc $d$-band preconfigures the copper $d$-band for low efficiency of
joined interferences $G^2$ = 18 in the $d$CuCu$^\prime$-subspace.
Instead, apart from the $d$CuCu$^\prime$-resonances,
a deep pseudogap opens around -2.8 eV.
Moreover, centered just around -2.8 eV, a wide spectral depression appears
in the $sp$DOS of $\gamma$-$\rm Cu_5Zn_8$ (Figure \ref{fig:fig57s4})
after radial adjustment starting from 3$\times$3$\times$3-bcc (Figure \ref{fig:fig56s4}).
Hence, $sp$-$d$ coupling must be important,
and the stabilization of $\gamma$-$\rm Cu_5Zn_8$ is neither simply NFE-like
nor is it confined to the pseudogap around the Fermi level.

\begin{figure}
\centering
\includegraphics[width=6.0cm]{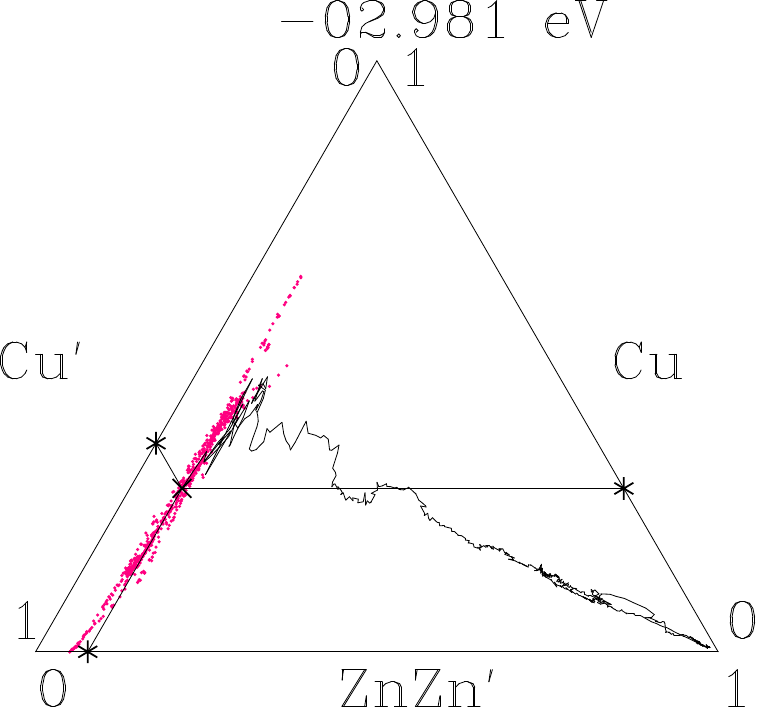}
\caption{$\gamma$-$\rm Cu_5Zn_8$: The decomposition of band states below the pseudogap at -2.8 eV (Figure \ref{fig:fig54s4}).}
\label{fig:fig58s4}
\end{figure}

Searching for the origin of the deep pseudogap around -2.8 eV (Figure \ref{fig:fig55s4}),
we inspect the partial weights of the band states in the subspaces ZnZn$^\prime$, Cu, and Cu$^\prime$.
Figure \ref{fig:fig58s4} applies to the peak below the pseudogap and reveals
extensive Cu-Cu$^\prime$ interplay, well separated from the ZnZn$^\prime$-subspace.
However, the MS of the total CuCu$^\prime$-subspace is too large for joined interferences $G^2$ = 18
which occur even in the
$d$CuCu$^\prime$-subspace clearly below -2.8 eV (Figure \ref{fig:fig55s4}).

\newpage

\begin{figure}
\centering
\includegraphics[width=8.4cm]{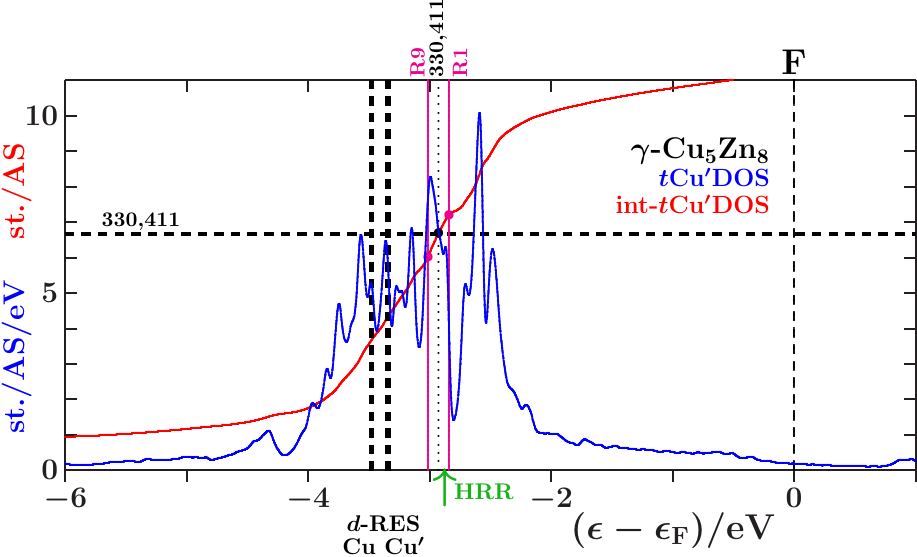}
\caption{Joined interferences $G^2$ = 18 in the subspace of Cu$^\prime$(OH).}
\label{fig:fig59s4}
\end{figure}

To find the really active subspace we follow Figure \ref{fig:fig69s6}c.
Like the subclusters OH(Ag$^\prime$) of $\gamma$-$\rm Ag_5Li_8$,
the OH(Cu$^\prime$) occupy sites in the SC unit cell close to the (330)-planes.
Joined interferences $G^2$ = 18 must thus be possible provided the MS of the Cu$^\prime$-subspace is large enough.
Figure \ref{fig:fig59s4} confirms this conjecture. 
Hence, the deep pseudogap around -2.8 eV arises from the total Cu$^\prime$-subspace 
due to joined interferences $G^2$ = 18 in the spectral peak around -2.9 eV.

\begin{figure}
\centering
\includegraphics[width=6.0cm]{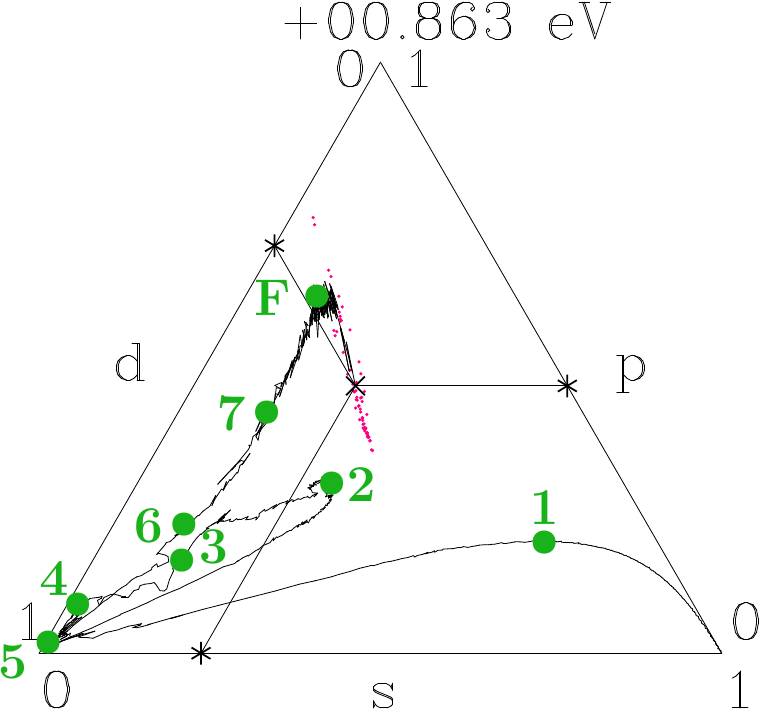}
\caption{$\gamma$-$\rm Cu_5Zn_8$: $spd$-decomposition of band states,
the energies (eV) (1 - F): -8.5, -5, -4.1, -3.4, -2.8, -2, -1.4, 0.0 Fermi energy.}
\label{fig:fig60s4}
\end{figure}

What is the most essential process above the copper $d$-band?
The trajectory in Figure \ref{fig:fig60s4} shows the average $spd$-decompositions of the band states
up to the upper bound of the pseudogap above $\epsilon_F$.
Band states above the upper bound of the copper $d$-band (-2 eV, bullet 6)
form $p$-dominated networks supported by joined interferences $G^2$ = 18
in the $sp$-subspace around -1.42 eV (Figure \ref{fig:fig57s4}).
The fluctuation patterns get $sp$-type (red dots).
Approaching the Fermi energy, the main process is the formation of $p$-dominated networks,
just as observed in the cases of dia-C (Figure \ref{fig:fig07s3}) and $\gamma$-$\rm Ag_5Li_8$ (Figure \ref{fig:fig19s4}).
The $sp$DOS of 3$\times$3$\times$3-bcc (Figure \ref{fig:fig56s4}) and $\gamma$-$\rm Cu_5Zn_8$ (Figure \ref{fig:fig57s4})
indicate joined interferences $G^2$ = 18, 14, and 6 as stabilization-relevant.
Around $G^2$ = 18 the situation resembles $\gamma$-$\rm Ag_5Li_8$ (Figures \ref{fig:fig21s4} and \ref{fig:fig22s4}).
However, planar interferences $G^2$ = 18

\noindent
in $\gamma$-$\rm Cu_5Zn_8$ occur in the occupied valence band at -1.42 eV.
Hence, structure stabilization can benefit from the fully pronounced $p$-dominated network 
(Figures \ref{fig:fig60s4} and \ref{fig:fig19s4}).
Note that the spectral appearance resembles dia-C (Figure \ref{fig:fig04s3}), too, 
but without a point of inflection towards a gap because there is no 
critical multiple scattering like C-C.

Planar interferences $G^2$ = 14 join the radial interferences $\tilde{R}$(Cu$^\prime$-Zn$^\prime$) along links
to neighboring 26-atom clusters in adjacent SC unit cells. 
The radial adjustment benefits from the absence of planar interferences $G^2$ = 15.

\begin{figure}
\centering
\includegraphics[width=8.0cm]{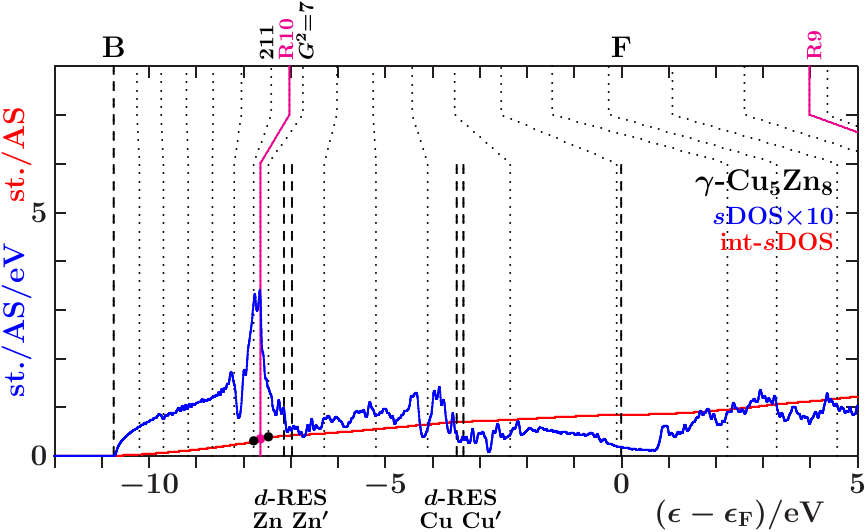}
\caption{Interference in the $s$-subspace.}
\label{fig:fig61s4}
\end{figure}

Joined interferences [d(6),R10(Cu-Cu)] with $G^2$ = 6 act in the $s$-subspace around -7.8 eV 
at enhanced spectral density (Figure \ref{fig:fig61s4})
just in the bonding range of a $s$-band splitting centered somewhat below -5 eV.
A related spectral feature with the joined interferences $G^2$ = 6 appears in the $sp$-subspace (Figure \ref{fig:fig57s4})
at slightly lower energy which confirms the $s$-character.
All that remembers $\gamma$-$\rm Ag_5Li_8$ (Figure \ref{fig:fig34s4}) where the size and the arrangement 
of the subclusters OT are stabilized by similar means. 

\begin{figure}
\centering
\includegraphics[width=8.0cm]{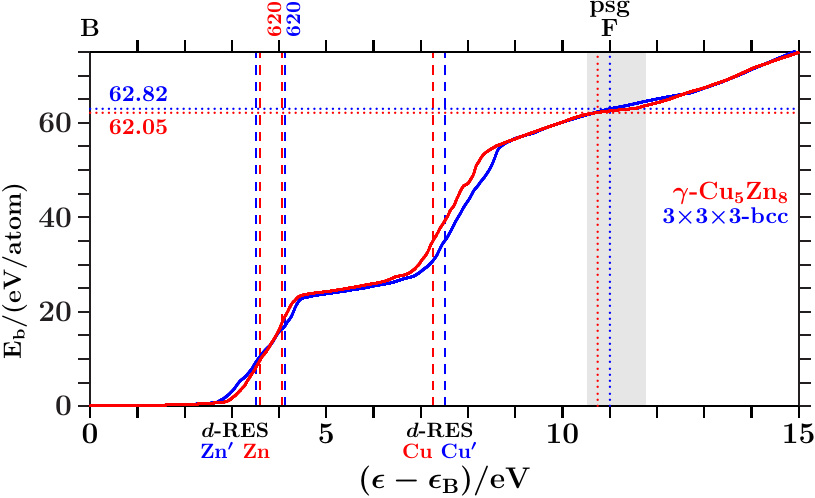}
\caption{Band energies above the bottom of the valence band.}
\label{fig:fig62s4}
\end{figure}

\begin{figure}
\centering
\includegraphics[width=8.1cm]{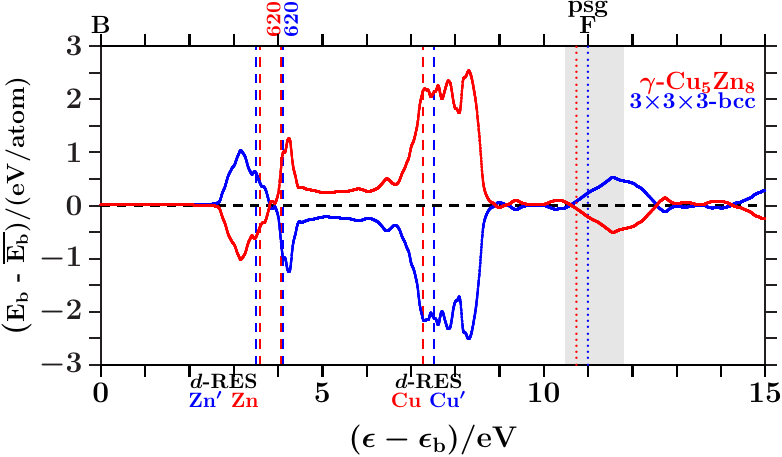}
\caption{The distances of the band energies from the average.}
\label{fig:fig63s4}
\end{figure}

\begin{figure}
\centering
\includegraphics[width=8.1cm]{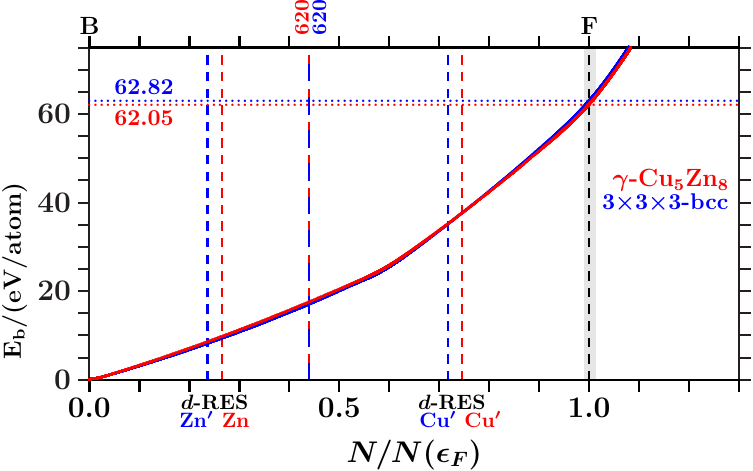}
\caption{Band energies versus the relative valence band filling.}
\label{fig:fig64s4}
\end{figure}

\begin{figure}
\onecolumn
\centering
\includegraphics[width=13.0cm]{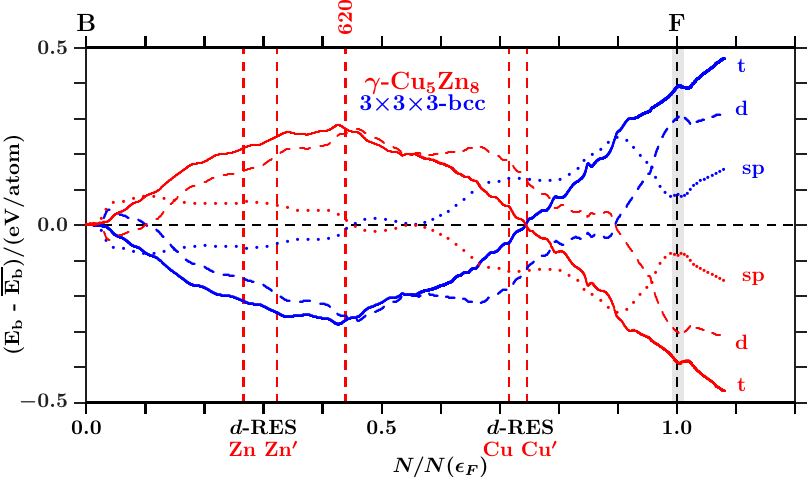}
\caption{The total and the partial band energies
with reference to the respective averages, plotted versus
the the relative band filling, $N/N(\epsilon_F)$.
Vertical dashed lines indicate the $d$-resonances of the $\gamma$-phase.
In the zinc $d$-band, the $\gamma$-phase first acquires excessive band energy which is removed
above the planar interference (620).}
\label{fig:fig65s4}
\twocolumn
\end{figure}

The planar interference (420) with $G^2$ = 20 is the first one above the interference block around $G^2$ = 18
which includes the planar interferences $G^2$ = 17, 18, and 19 together with the radial interferences
on the shortest interatomic distances.
Hence, no radial interference and hybridization joins $G^2$ = 20 via (\ref{GQR}).
$G^2$ = 20 is the first lonely planar interference with limited impact on structure stabilization.
This explains the identical appearance in 3$\times$3$\times$3-bcc (Figure \ref{fig:fig56s4})
and $\gamma$-$\rm Cu_5Zn_8$ (Figure \ref{fig:fig57s4}) which differ mainly by radial adjustments. \\[1ex]
\noindent
{\bm $Band$ $energy$}.
The Figures \ref{fig:fig62s4} to \ref{fig:fig65s4} ($\gamma$-$\rm Cu_5Zn_8$) use the same
presentation modes of the band energies $E_b(\epsilon)$ (\ref{EB1})
as the Figures \ref{fig:fig29s4} to \ref{fig:fig32s4} ($\gamma$-$\rm Ag_5Li_8$) and
the Figures \ref{fig:fig44s4} to \ref{fig:fig47s4} ($\gamma$-$\rm V_5Al_8$).
Vertical dashed lines denote the lower $d$-resonances whereas
dotted lines indicate the Fermi energies and the resulting band energies.
Shaded areas around $\epsilon_F$ refer to the pseudogap of the $\gamma$-phase.
With the specified partitions into $d$- and $sp$-contributions we obtain
$E_b^{\rm bcc}(\epsilon_F)$ = (52.05 + 10.77) eV/atom  for 3$\times$3$\times$3-bcc and
$E_b^\gamma(\epsilon_F)$ = (51.45 + 10.60) eV/atom for $\gamma$-$\rm Cu_5Zn_8$ (Figure \ref{fig:fig62s4}).
The step from 3$\times$3$\times$3-bcc to the $\gamma$-phase lowers the band energy
by -(0,60 + 0.17) eV/atom, mainly $d$-type due to the short-range character of the charge redistribution.

Band-energy differences at given energies above the bottom of the valence band are much better evident from
Figure \ref{fig:fig63s4} where the band energies are
plotted with reference to the common averages.
The excessive band-energy of the $\gamma$-phase which emerges in the zinc $d$-band is removed again
in the copper $d$-band.
For this purpose both $d$-bands couple across the interstitial $sp$-type spectral range.
The Figures \ref{fig:fig64s4} and \ref{fig:fig65s4} compare the band energies of the systems in equivalent situations
at equal valence-band fillings, $N/N(\epsilon_F)$.
Figure \ref{fig:fig62s4} changes into Figure \ref{fig:fig64s4}

\newpage

\noindent
which reveals a fundamental difference to $\gamma$-$\rm Ag_5Li_8$ (Figures \ref{fig:fig31s4} and \ref{fig:fig32s4})
and $\gamma$-$\rm V_5Al_8$ (Figures \ref{fig:fig46s4} and \ref{fig:fig47s4}).
In both cases the band-energy differences grow monotonously towards the Fermi energy in favor of the $\gamma$-phase
whereas $\gamma$-$\rm Cu_5Zn_8$ starts at low valence-band filling with the larger band energy.
This turns into the opposite around the $d$Cu$^\prime$-resonance.
For further analysis we show in Figure \ref{fig:fig65s4} besides the total band energies (full
curves) the $d$-parts (dashed curves) and the $sp$-parts 
(dotted curves), too, each with reference to the respective average, 
$\overline{E_b^x} = (E_b^x(\gamma) + E_b^x({\rm bcc}))/2$.
The superscript "$x$`` specifies optionally to $t$, $d$, or $sp$.
Both partial band-energy differences add up to the total difference,
$E_b^t - \overline{E_b^t} = (E_b^d - \overline{E_b^d}) + (E_b^{sp} - \overline{E_b^{sp}})$.
Dashed lines denote the $d$-resonances of the $\gamma$-phase.
The planar interference (620) at $N/N(\epsilon_F) \approx 0.44$ 
occurs in the total DOS of the $\gamma$-phase just at the main peak of the anti-bonding zinc $d$-band
close to the upper bound (Figure \ref{fig:fig50s4}).

With emphasis on the band energy, Figure \ref{fig:fig65s4} gives a detailed description of the transition from 
3$\times$3$\times$3-bcc to $\gamma$-$\rm Cu_5Zn_8$.
Two items are evident:
(i) Due to appropriate stacking of band states in the zinc $d$-band, 
the $\gamma$-phase accumulates band energy over 3$\times$3$\times$3-bcc.
(ii) After removal of the excess between the upper bound of the zinc $d$-band and the $d$Cu$^\prime$-resonance
the band-energy splitting acts in favor of the $\gamma$-phase.

The removal of the band-energy excess of the $\gamma$-phase occurs on large and short length scales.
For separate access to large respectively short length scales 
we dissect the band-energy splitting into $sp$- 

\newpage
\hspace{1cm}
\vspace{9.8cm}

\noindent
and $d$-parts. 
Hence, band-energy differences on large length scales start acting in favor of the $\gamma$-phase just
beyond the upper bound of the zinc $d$-band and continue acting in the upper $>$50 $\%$ of the valence band.
Short length scales, on the contrary, contribute this way only in the upper 10 $\%$ of the valence band.
The electronic stabilization of $\gamma$-$\rm Cu_5Zn_8$ thus depends critically on 
the $d$-$sp$ interplay throughout the valence band.
Only the short-range support is confined to the environment of the Fermi energy. \\[1ex]
\noindent
{\bm $Charge$ $transfer$ $and$ $the$ $ratio$ $e/a$}.
In the initial state we suppose free atoms inside the AS (configurations, Cu(1 0 10), Zn(2 0 10)).  
The 20 CuCu$^\prime$ contribute 220 valence electrons, and the 32 ZnZn$^\prime$ add further 384.
In the final state, due to the LMTO-ASMT calculation, $\sim$ 2.2 valence electrons per SC
unit cell have moved from ZnZn$^\prime$ to CuCu$^\prime$.
This makes $\sim$ 0.36 \% of the 604 valence electrons per SC unit cell.

Table \ref{tab:tab07s4} shows the changes of the partial charges in the transition from free atoms to the $\gamma$-phase.
The charge redistribution aims at a pronounced continuous $p$-network around the Fermi energy
which is evident from Figure \ref{fig:fig60s4}.
The last two lines present the average partial changes in two steps:
The first step (to 3$\times$3$\times$3-bcc) concerns bonding into perfect planar order
whereas the second step (to $\gamma$-$\rm Cu_5Zn_8$) adjusts the radial order.
Hence, 0.772 + 0.003 = 0.775 $s$-electrons per AS and 0.213 - 0.005 = 0.208 $d$-electrons per AS
change into 0.985 - 0.002 = 0.983 $p$-electrons per AS.
This forms the stabilizing $p$-network.
In the final state, due to the LMTO-ASMT

\begin{table}
\small
\centering
\caption{$\gamma$-$\rm Cu_5Zn_8$, the changes of the partial AS-charges in the transition from free atoms 
to the alloy. The last two lines show the average changes in two steps.}
\vspace{0.2cm}
        \begin{tabular}{ l | l | l | l }
        \hline
        AS-types \hspace*{1.0cm}        & $s$ \hspace*{0.8cm}   & $p$ \hspace*{0.8cm}   & $d$ \hspace*{0.8cm} \\
        \hline
	Zn(IT)				& -1.107		& +1.010		& -0.034                 \\	
	Cu(OT)				& -0.303		& +0.881		& -0.507                 \\	
	Cu$^\prime$(OH)			& -0.270		& +0.901		& -0.498                 \\	
	Zn$^\prime$(CO)			& -1.077		& +1.048		& -0.021                 \\
        \hline
	to 3$\times$3$\times$3-bcc	& -0.772		& +0.985		& -0.213	         \\
	to $\gamma$-$\rm Cu_5Zn_8$	& -0.003		& -0.002		& +0.005	         \\
        \hline
        \end{tabular}
\label{tab:tab07s4}
\end{table}

\noindent
calculation, $\sim$ 2.2 valence electrons per SC 
unit cell have moved from ZnZn$^\prime$ to CuCu$^\prime$.
This makes $\sim$ 0.36 \% of the 604 valence electrons per SC unit cell.

By means of $local$ $reading$, Mizutani $et \; al.$ \cite{Mizutani17a} extract
$G^2(\epsilon_F)$ = 18.55 $\pm$ 0.55 from the Hume-Rothery plot and
obtain via (\ref{PLZ}) $e/a$ = 1.61 $\pm$ 0.05 in perfect agreement with the estimated 21/13 $\approx$ 1.615.
One may expect the planar interference $G^2$ = 18 immediately below $\epsilon_F$.

In the present approach, the equivalent of the $e/a$-ratio is the average $sp$-occupation up to $\epsilon_F$.
We obtain  (1.615 - 0.775) + 0.983 = 1.823 and via (\ref{PLZ}) $G^2(\epsilon_F)$ = 20.16
which explains the planar interference (420) slightly below $\epsilon_F$ 
(Figure \ref{fig:fig57s4}, $G^2$ = 20, $Z(d(20))$ = 1.8012, Table \ref{tab:tab14s6}). \\[1ex]
\noindent
{\bm\bf $Summarizing$ $\gamma$-$\rm Cu_5Zn_8$}.
Joined interferences $G^2$ = 18 play a major part in the electronic stabilization of $\gamma$-$\rm Cu_5Zn_8$.
They support the resonance splitting of the zinc $d$-band and the formation of the deep pseudogap in the
upper copper $d$-band as well as the growing $p$-dominated network below the Fermi energy.

The ratio e/a = 21/13 = 1.615 of $\gamma$-$\rm Cu_5Zn_8$, estimated from the free atoms,
suggests the joined interferences $G^2$ = 18 to occur in the $sp$-subspace closely below the Fermi energy.
However, due to $d$-to-$p$ transfer (Table \ref{tab:tab07s4}) the total content of the $sp$-subspace amounts to 
1.823 which allows for the planar interference $G^2$ = 20 at -0.187 eV whereas the joined interferences
$G^2$ = 18 occur at -1.42 eV.
Planar interference $G^2$ = 20 have no direct access to the radial interference and hybridization on the atomic scale
because no existing interatomic distance is short enough to form joined interferences.

The spectral signature of the joined interferences $G^2$ = 18 in the $sp$DOS resembles $\gamma$-$\rm Ag_5Li_8$ and 
dia-C, except for a pseudogap above $\epsilon_F$ instead of a gap.
The three systems support the formation of stabilizing $p$-dominated networks 
by joined interferences $G^2$ = 18 respectively $G^2$ = 8. 

The transition from 3$\times$3$\times$3-bcc to $\gamma$-$\rm Cu_5Zn_8$ adjusts the radial order
(short-range) whereas the planar order (long-range) is less affected. 
The interference status is improved upon concentrating spectral weight at fulfilled interference 
conditions, and the total band energy decreases.

In the zinc $d$-band, the $\gamma$-phase accumulates excessive total band energy over 3$\times$3$\times$3-bcc
which is removed again up to the $d$-resonance of Cu$^\prime$(OH).
For specific access to the short-range respectively the long-range stabilization,
we subdivide the total band energy at given valence-band filling into the $d$-part and the $sp$-part. 
It turns out that the $sp$-part of the band-energy difference acts 
in roughly the upper half of the valence band for stability of the $\gamma$-phase 
and the $d$-part only in the top 10 $\%$.
Nevertheless, the $d$-part dominates the resulting band-energy difference of -0.77 eV/atom at the Fermi energy.
\section{Conclusions}

This study examines cubic crystals where the content of the simple cubic (SC) unit cell (side length $a$)
is close to a $n$$\times$$n$$\times$$n$-bcc substructure.
We confine to $n$ = 2 (diamond-structures: number 227 and zinc-blende structures: number 216) 
and to $n$ = 3, ($\gamma$-brasses: number 217). 
Using published structure models in LMTO-ASA calculations with inscribed muffin-tin potentials
we obtain various projected densities of states 
together with the compositions of the band states in the atomic-site angular-momentum representation.

The main purpose of the study is the detailed analysis of the first-principle results within the framework
of a VEC-based concept after Hume-Rothery, Mott, and Jones (HMJ-concept). 
At each energy, the size of the active momentum sphere (MS) in the extended $k$-space
is derived from the calculated integrated density of states in atomic-site angular-momentum representation, 
projected to the active electron space.
The stabilization-relevant interferences occur in transitions on the surface of the MS 
which involve the largest momentum transfers (along the diameter of the MS, "true backscattering``).
We derive the active electron space 
from restricted compositions of the electron band states in atomic-site angular-momentum representation.

Interference and hybridization determine the charge distribution 
on the atomic scale and control this way the formation of low-temperature equilibrium structures.
To surmount the usual contrasting treatment of interference and hybridization
we use "joined planar-radial interferences`` where hybridization results from 
well-tuned radial interference due to the sequence of neighbor shells around reference atoms.
Separate interference conditions for planar and radial interferences select momentum transfers 
as close together as to allow for interplay via the common momentum pool in the extended $k$-space.
This ensures the balance between the local radial order inside the SC unit cell 
and the global planar order outside. 
With reference to the SC unit cell certain inside-outside relations arise 
such as the empirical Hume-Rothery rules which predict crystal structures 
on the basis of the electron-per-atom ratios (e/a-ratio) in the SC unit cell,
estimated from the $sp$-configurations of free atoms.

Drawing on the example of bcc-related crystals, the present study 
demonstrates a general principle of electronic structure stabilization:
The accumulated charge in an active electron quantum space determines the length scale of stabilizing
interferences in this space and consequently the formation of the quantum states.
This can be seen as a consequence of the Hohenberg-Kohn theorem \cite{Hohenberg64}.

The examined systems reveal several structure-relevant planar and radial length scales.
Most prominent, despite of non-uniform decoration, the $n^3$ subcells in the SC unit cell
must be kept close to bcc on the scale $a/n$.
This is achieved by joined interferences $G^2$ = 2$n^2$ which describe the interplay
between the planar interferences $(nn0)$ and the radial interferences (hybridizations)
on the shortest interatomic links.
Hence, the leading interferences in diamond- and zinc-blende phases are the joined interferences $G^2$ = 8
and $G^2$ = 18 in $\gamma$-brasses.

A different kind of essential joined interferences connects planar interferences $(hkl)$ 
(no common measure) with radial interferences along interatomic links which characterize the size
of certain subclusters in the SC unit cell.
This way both the size and the arrangement of such subclusters in the SC unit cell are controlled
by electronic interference.
In diamond- and zinc-blende phases this applies to the planar interferences (111) which
act on the length scale of the elementary tetrahedron (interatomic link $a\sqrt{2}/2$).
In $\gamma$.brasses the planar interference (211) and (321) act on the subclusters OT respectively IT.
Starting from the sites in the 3$\times$3$\times$3-bcc sublattice, the IT(weak scatterers) are notably compressed
whereas the OT(strong scatterers) almost keep their bcc-positions on the scale $a$/3.

The predictive power of the empirical Hume-Rothery rules rests on the reasonable assumption that the 
allocation of partial charges to the effective atoms of the crystal is not too much different from the free atoms.
The rules estimate the size of the MS in the $sp$-subspace at $\epsilon_F$ and compares 
with the required sizes at the main planar interferences of crystals.
Stabilizing interference in the extended $k$-space is thus supposed to occur around $\epsilon_F$ and,
after projection to the atomic-site angular-momentum representation, in the $sp$-subspace.

The present study demonstrates that the essential stabilizing interferences are neither confined to the environment
of $\epsilon_F$ nor are they confined to the $sp$-subspace.
They occur in various parts of the total electron quantum space including $d$-related subspaces
In view of the long-lasting discussions concerning the electron quantum weight which should be considered
in the HMJ-concept \cite{Massalski10a,Mizutani11b}
the present study has demonstrates that the stabilization of the real-space structure
is due to interferences in the extended $k$-space regardless of the assignment to partial charges after projection to the
atomic-site angular-momentum representation.

On this background, in systems with nearly $n$$\times$$n$$\times$$n$-bcc sublattices in the SC unit cell,
electron-per-atom ratios close to e/a = 1.6 turn out to indicate 
active joined interferences $[d(2n^2),a\sqrt{3}/2n]$ in the concerned subspace just below the respective energy
which drives towards 2$n^3$ bcc-subcells.
The planar and the radial components of these joined interference act at e/a = 1.481 respectively 1.574 (cf. Table \ref{tab:tab02s2}),
in the case of $\gamma$-brasses at e/a = 1.538 respectively 1.635 due 52 instead of 54 occupied states in the SC unit cell.
 
Regarding $\gamma$-brasses, the ratio e/a, estimated from the $sp$-configurations of free atoms,
predicts significant joined interferences $G^2$ = 18 in the $sp$-subspace in any case:
(i) They must act clearly above $\epsilon_F$ if the estimated e/a $<$ 21/13, such as $\gamma$-$\rm Ag_5Li_8$.
Note that the assigned $p$-dominated network acts already around $\epsilon_F$.
(ii) They must act clearly below $\epsilon_F$ if the estimated e/a $>$ 21/13, such as $\gamma$-$\rm V_5Al_8$.
Note that favorable conditions are adjusted for joined interferences $G^2$ = 18 in $d$-type subspaces around $\epsilon_F$.
(iii) They may act rather around $\epsilon_F$ if the estimated e/a $\approx$ 21/13, such as $\gamma$-$\rm Cu_5Zn_8$.   
Note that the true interference energies depend on the actual $d$-to-$sp$ transfer.

Two different concepts are usually employed as guides along paths to low-temperature equilibrium structures, 
namely decreasing electronic band energy and increasing electronic interference. 
We analyze hypothetical relaxation steps from properly decorated $n$$\times$$n$$\times$$n$-bcc systems 
to the respective $\gamma$-phases which improve in particular the radial interferences and hybridizations.
In each case, improved interference manifests by enhanced spectral weight at fulfilled interference conditions,
and the electronic band energy decreases.

Vacancies (in the atomic-sphere approximation: empty spheres ES) 
are created in order to enable polyvalent atoms to
build structure elements in the SC unit cell which are reserved to lower valences.
Only half the lattice sites in the SC unit cell are occupied with atoms in dia-C.
We show that the ES acquire charge from the atoms for separate joined interferences close to the Fermi energy
in the sublattice of the atoms ($G^2$ = 8) and in the sublattice of the ES ($G^2$ = 3). 

Band states with internal $p$-dominated networks dominate towards the Fermi energy,
The transfer of $s$-weight into $p$-weight turns out a universal process in structure formation.
We show for the examined $n$ = 2 systems that the measured bulk moduli follow this trend.

The existing medium-range radial order of amorphous phases is included as the limiting case
where global planar order has not yet developed, e.g. for time-scale reasons. 
As yet, sufficiently detailed structure information is missing.

As an outlook, we note that the concept of joined planar-radial interferences allows for treating 
several issues of electronically supported structure formation.
The size and the shape of the unit cell of a stable crystal which both determine the translation symmetry
must act on the content of the unit cell and vice versa depending on strengths of the 
radial respectively the planar processes.
Even quasicrystalline order in two and three dimensions is involved where
other global planar arrangements develop for partnership of local radial order.  

\vspace{0.4cm}

\noindent
{\normalsize \bf Acknowledgements} 

The author gratefully acknowledges the exchange of ideas with Prof. P. H\"aussler, Dr. R. Arnold, and Dr. C. V. Landauro 
during many years at the Chemnitz University of Technology.
Also gratefully acknowledged, the financial support by the Deutsche Forschungsgemeinschaft during this period.
\section{Appendices}   

\subsection{Diamond, zinc blende, and 2$\times$2$\times$2-bcc}   

\begin{figure}
\centering
\includegraphics[width=5.5cm]{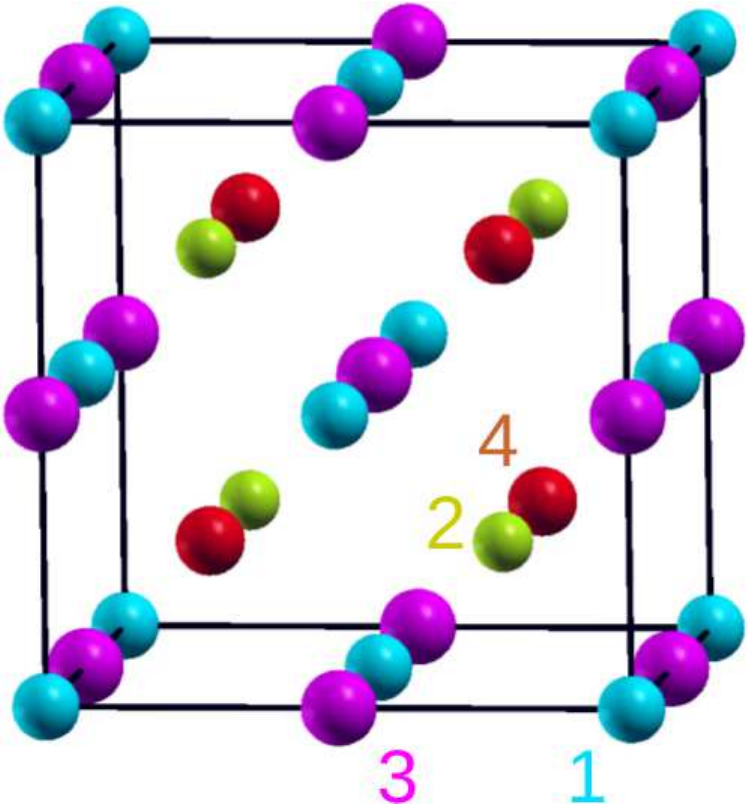}
\caption{The fcc-sublattices F1 to F4.}
\label{fig:fig66s6}
\end{figure}

We assign the 16 lattice sites in the SC unit cell (side length $a$) of 2$\times$2$\times$2-bcc
to four fcc-sublattices F1 - F4 (Figure \ref{fig:fig66s6}).
F1 belongs to the depicted SC unit cell. 
Shifting F1 by $a\sqrt{3}/4$ along a space diagonal guides to F2.
F3 and F4 are obtained on shifting F1 respectively F2 by $a/2$ along cubic axes.
F1 with F2 and F3 with F4 form two interpenetrating diamond lattices.
The fcc-sublattices are linked by $R$1 = $a\sqrt{3}$/4, $R$0 = $a$/2, and $R$3 = $a\sqrt{3}$/2
as shown by Table \ref{tab:tab08s6}.
$R$2 = $a\sqrt{2}$/2 are interatomic links inside the fcc-sublattices. 
Table \ref{tab:tab09s6} shows decorations of the fcc-sublattices in essential cases.

\subsection{Single-scattering approximation to the atomic-sphere density of states}

Within the muffin-tin scattered-wave (MT-SW) concept, 
we calculate an approximate partial DOS, $n_{sl}(\epsilon)$, of the AS number $s$ (ASs)
which includes only scattering paths with one scattering in the environment of the ASs.
The energy $\epsilon$ and $k = \sqrt{\epsilon}$ refer to the muffin-tin zero (MT0).
In this approximation to the back-scattered field, one can estimate the consequences 
of fulfilled interference conditions.

The MT orbitals are used in the straightforward representation
$\phi_L(\vec{r}) = i^l \phi_l(kr) Y_L(\vec{e})$
with complex spherical harmonics, $Y_L(\vec{e})$, where $\vec{e}$ is the unit vector 

\begin{table}
\caption{Interatomic distances which link the sublattices.}
\vspace{0.2cm}
        \begin{tabular}{ c | c  c  c  c }
        \hline
             & F1   & F2   & F3        & F4        \\
        \hline
        F1   & $R$2 & $R$1 & $R$0,$R$3 & $R$1      \\
        F2   &      & $R$2 & $R$1      & $R$0,$R$3 \\
        F3   &      &      & $R$2      & $R$1      \\
        F4   &      &      &           & $R$2
        \end{tabular}
\label{tab:tab08s6}
\end{table}

\begin{table}
\small
\centering
\caption{Typical occupations of the sublattices.}
\vspace{0.2cm}
        \begin{tabular}{ c | c  c  c  c }
        \hline
        phase  & F1 & F2 & F3 & F4 \\
        \hline
        diamond (\#227) & C  & C  & empty  & empty \\
        cubic BN (\#227) & B  & N  & empty  & empty \\
        {Heusler $\rm AlCo_2Cr$}(\#225)  & Al & Co & Cr & Co \\
        \end{tabular}
\label{tab:tab09s6}
\end{table}
\begin{figure}
\centering
\includegraphics[width=4.5cm]{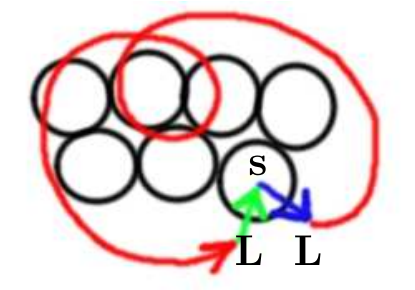}
\caption{The scattering paths of the on-site Green function. The figure shows the non-overlapping muffin-tin spheres.
The corresponding applies to the overlapping atomic spheres.}
\label{fig:fig67s6}
\end{figure}

\begin{figure}
\onecolumn
\centering
\includegraphics[width=12.0cm]{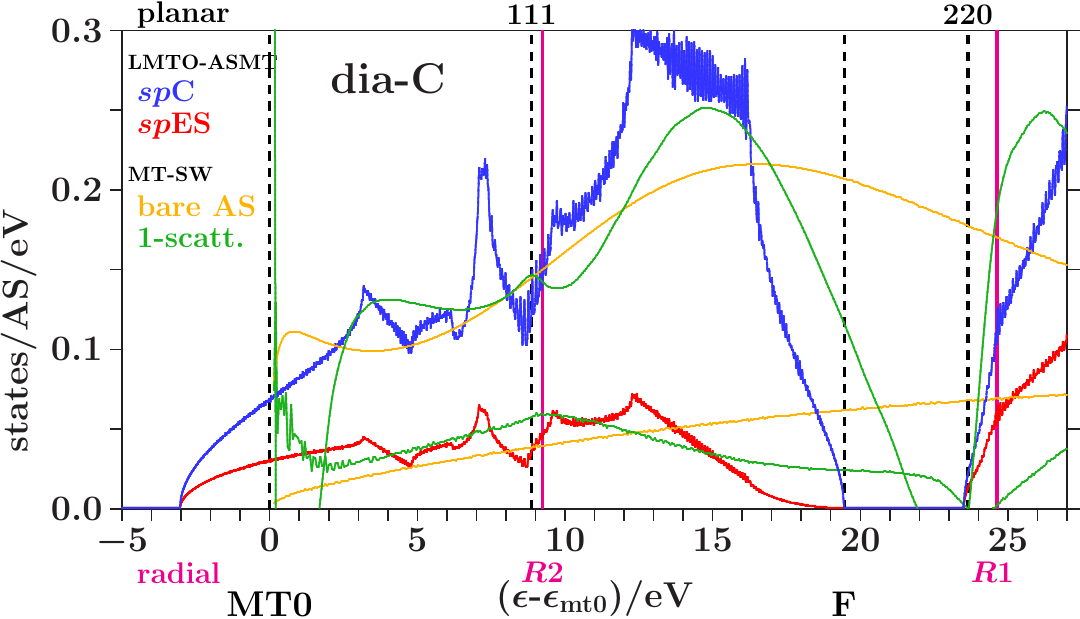}
\caption{dia-C: The calculated $sp$DOS (LMTO-ASMT) of the AS with atoms (blue) and ES (red),
the corresponding results of the bare-AS (yellow), and the approximations with only one scattering in the environment (green).}
\label{fig:fig68s6}
\twocolumn
\end{figure}

\begin{figure}
\onecolumn
\centering
\SetFigLayout{2}{2}
        \subfigure[IT(Li)]{\includegraphics[width=6.2cm]{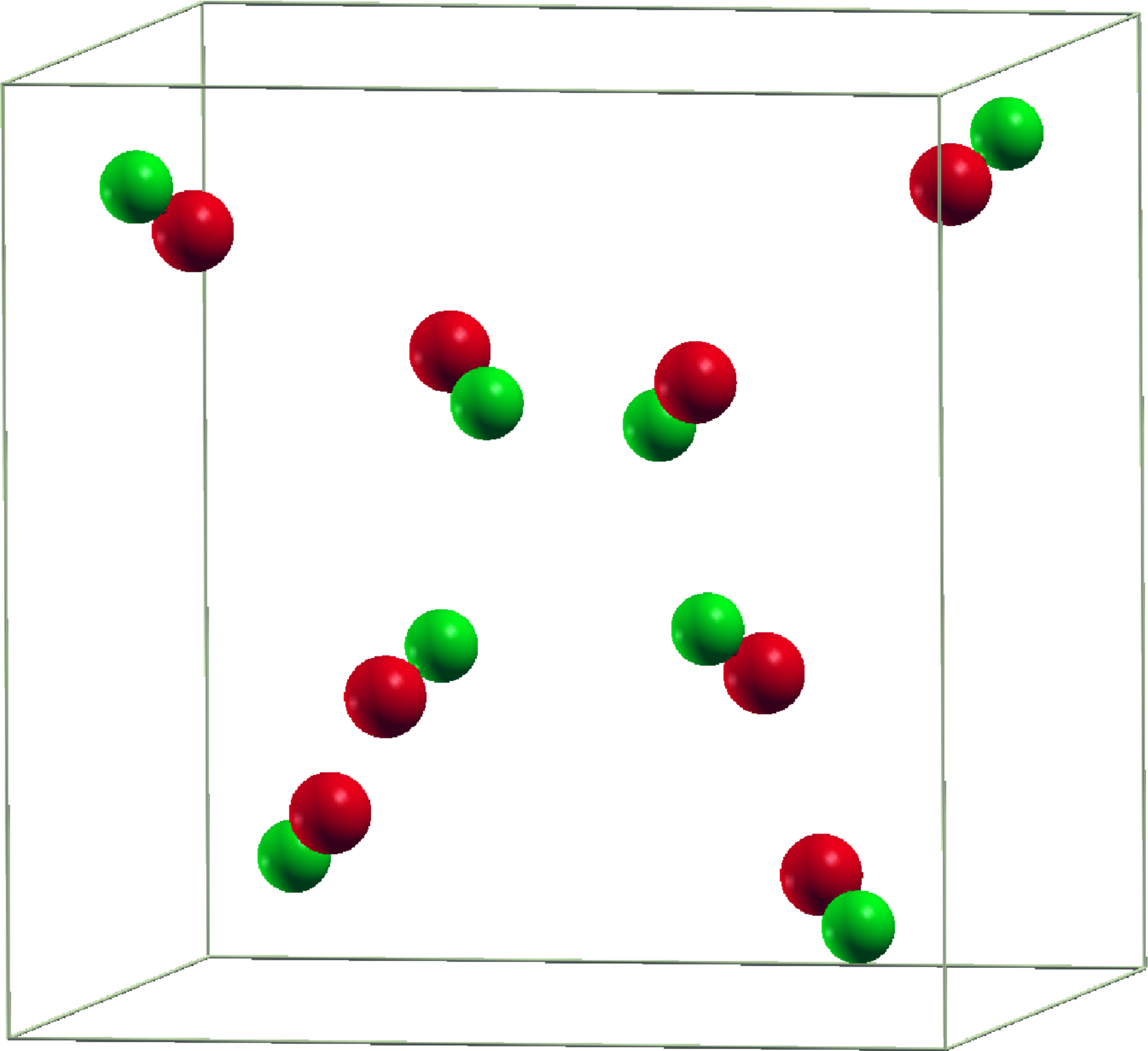}}
        \hspace{1.5cm}
        \subfigure[OT(Ag)]{\includegraphics[width=6.0cm]{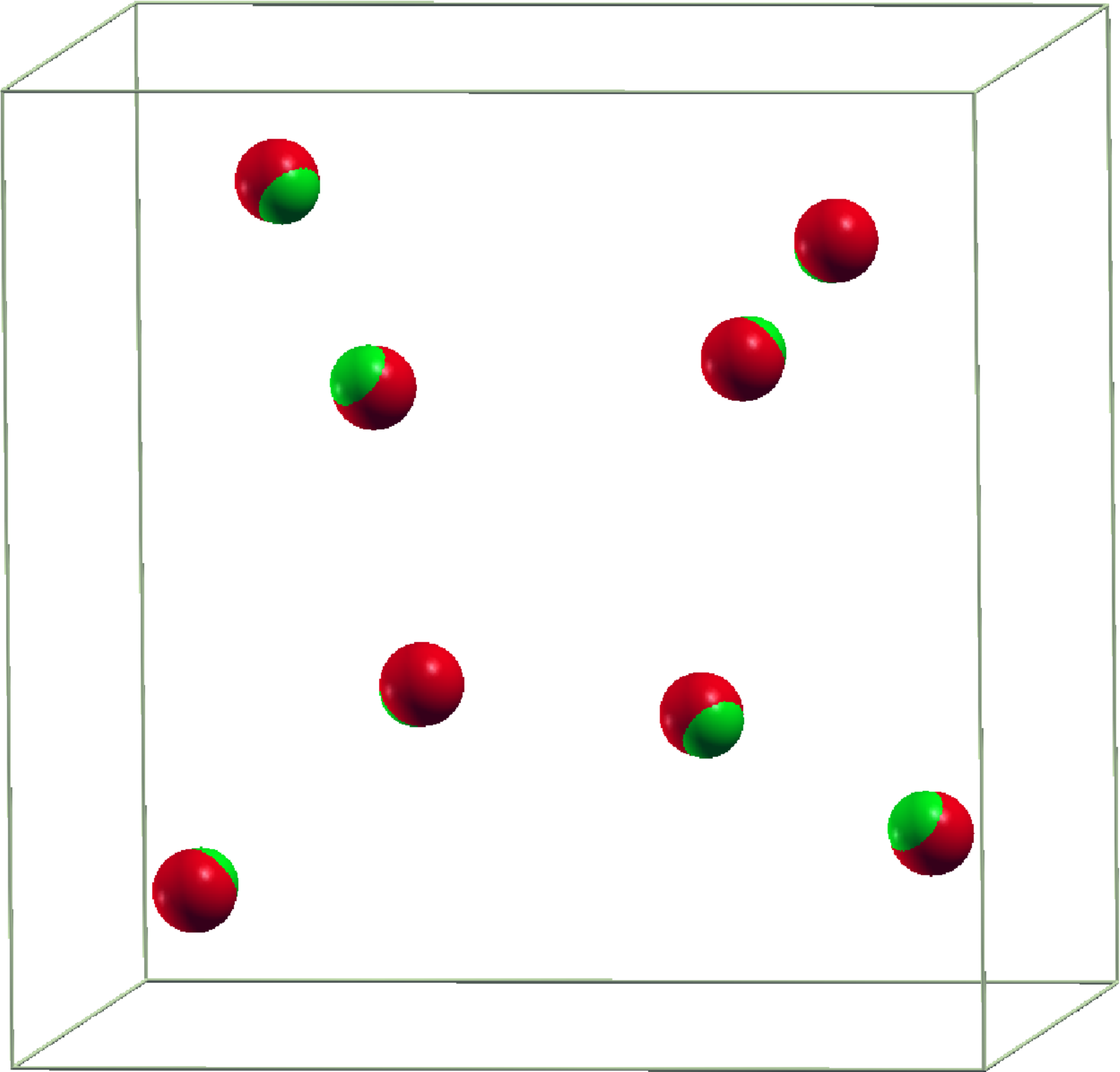}}

        \subfigure[OH(Ag$^\prime$)]{\includegraphics[width=6.0cm]{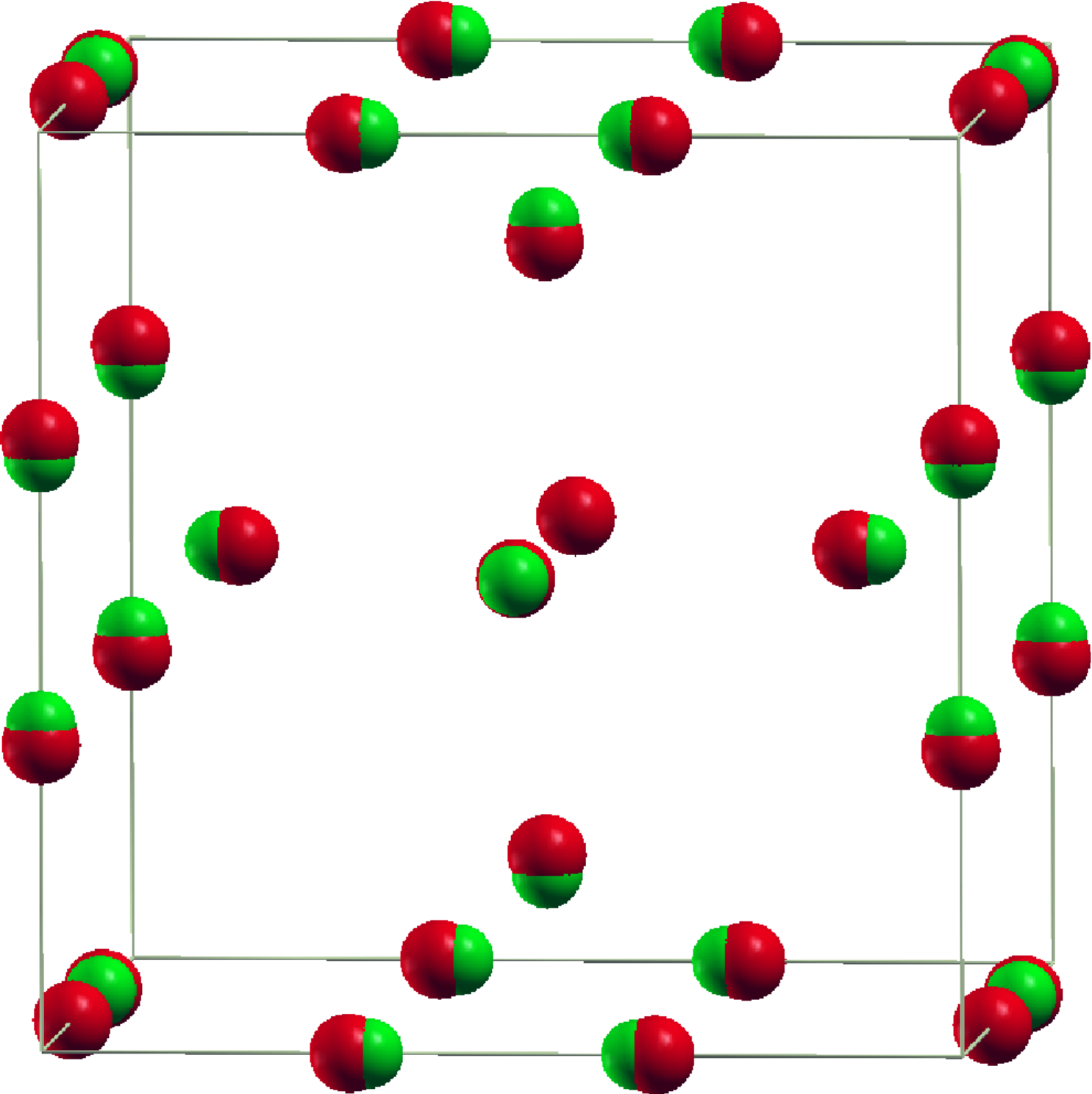}}
        \hspace{1.5cm}
        \subfigure[CO(Li$^\prime$)]{\includegraphics[width=6.0cm]{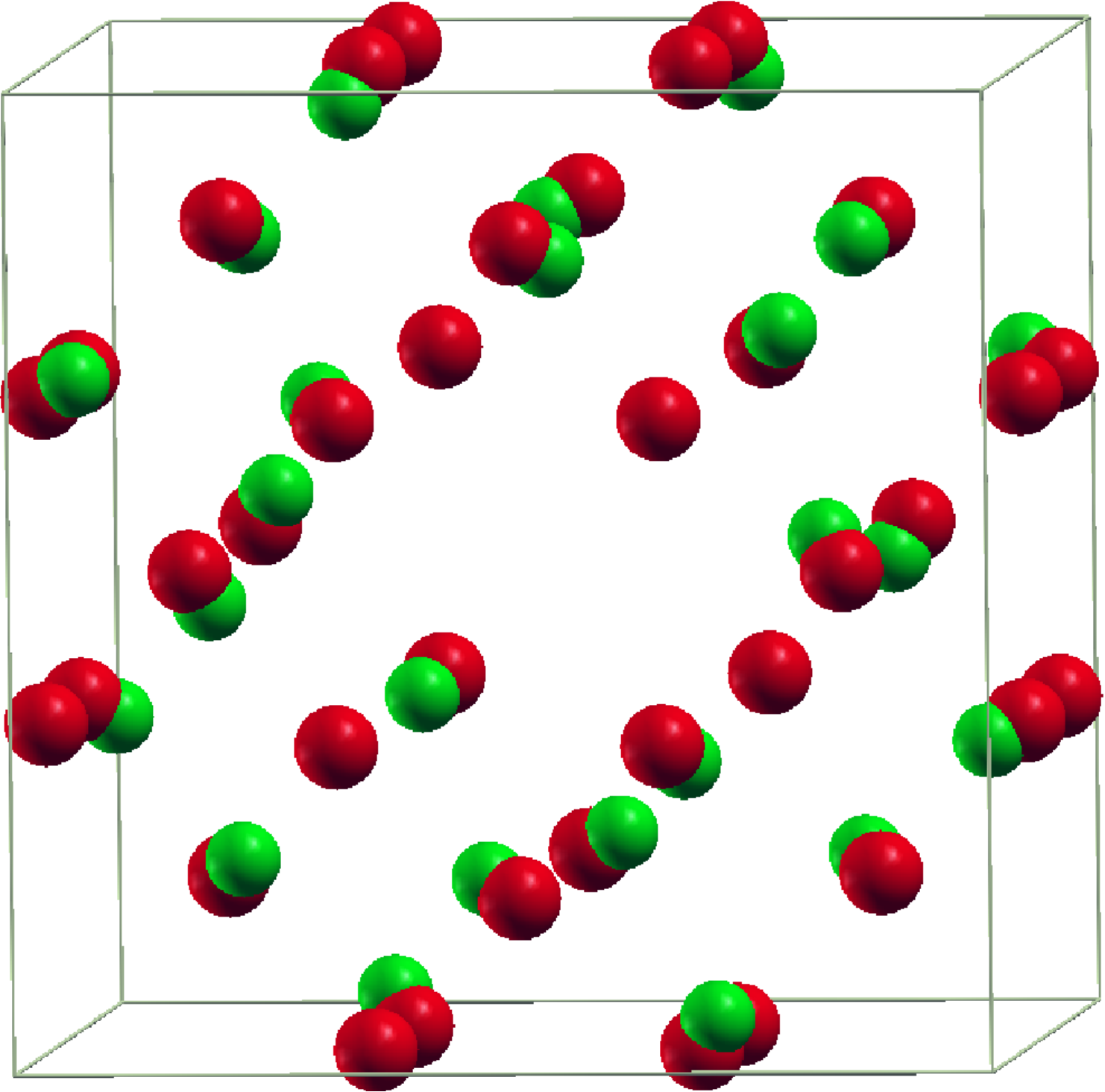}}
\caption{$\gamma$-$\rm Ag_5Li_8$ \cite{Noritake07a} (green) and the 3$\times$3$\times$3-bcc (dark red) supercell,
comparison of the subclusters IT(Li), OT(Ag), OH(Ag$^\prime$), and CO(Li$^\prime$).
Note the obvious compression of the IT(Li) whereas the other subclusters are less modified.}
\label{fig:fig69s6}
\twocolumn
\end{figure}

\newpage

\hspace{10cm}

\newpage

\noindent
along $\vec{r}$.
Outside the MT spheres, the regular basic solutions and the singular outgoing ones
are chosen as $J_{sL}(\vec{r})=\exp(-i\eta_{sl}) \; [j_L(\vec{r}) \, +
\, ik f_{sl} \,  h^{(1)}_L(\vec{r})]$
respectively $H^+_{sL}(\vec{r})=\exp(i\eta_{sl}) \; h^{(1)}_L(\vec{r})$.
The $j_L$ and $h^{(1)}_L$ (asymptotic radial part  $h^{(1)}_l(kr) \, \simeq \, i^{-l}\exp(ikr)/(ikr)$)
are vacuum solutions.
The complex scattering amplitudes of the MT potentials, $f_{sl}=\exp(i\eta_{sl}) \; \sin(\eta_{sl})/k$,
depend on the real scattering phase shifts, $\eta_{sl}(\epsilon)$.

In this framework, the representation (\ref{DELD}) of the partial DOS $n_{sl}(\epsilon)$ of the ASs is available which derives $n_{sl}$ 
from the DOS $n^0_{sl}$ of the isolated frozen ASs without environment,
 which depends only on the $\eta_{sl}(\epsilon)$, and the diagonal elements,
 $T_{sL,sL}(\epsilon)$, of the scattering-path matrix (\ref{SPMAT}).
$T_{sL,sL}(\epsilon)$ collects all scattering paths through the system which start and terminate inside 
the ASs at the valence orbital with the angular momentum $L=(l,m)$ (Figure \ref{fig:fig67s6}).
The exponentials with the 
scattering phase shifts $\eta_{sl}$ result from the transitions between muffin-tin orbitals inside and
vacuum orbitals outside the AS$s$.
The elements $P_{sL,s'L'}(\epsilon)$ of the propagation matrix $P$ describe the propagation of vacuum waves from ($s'$$L'$) to ($s$$L$)
and the elements $F_{sL,sL}(\epsilon) = (exp(2i\eta_{sl}(\epsilon)-1)/2$ of the diagonal matrix $F$
the scatterings by the spherically symmetric AS$s$.

We obtain the envisaged single-scattering approximation to the partial DOS of the ASs upon terminating the multiple scattering
in (\ref{SPMAT}) after one scattering in the environment,

\begin{equation}
      T_{sL,sL} \approx  e^{2i\eta_{sl}}  \langle sL|PFP|sL\rangle . 
\label{1SC}
\end{equation}

\noindent
This way a single-scattering approximation to the AS-DOS, $n_{sl}(\epsilon)$, is obtained for the central AS of a big cluster. 
The real phase shifts, $\eta_{sl}(\epsilon)$, at energies above the muffin-tin zero 
are calculated from the self-consistent LMTO-ASMT potentials.

For dia-C, Figure \ref{fig:fig68s6} compares this single-scattering DOS with the LMTO-ASMT calculation 
and with the bare-AS DOS (frozen ASs without environment). 
The energy scale refers to the MT0, $\epsilon_{mt0}$.

Active interference in the interatomic free propagation can be expected at those energies $\epsilon$ 
where interference conditions (\ref{BRAGG}), (\ref{RADI}) are fulfilled with
the momentum transfer $2\sqrt{\epsilon - \epsilon_{mt0}}$ on the left-hand side.
They are indicated by vertical lines with labels [(220),$R$1] and [(111),$R$2].
The interferences [(220),$R$1] remove spectral weight from the bare-AS DOS starting at 
the upper edge of the gap.
The bonding component of the b/ab-splitting appears at the upper edge of the hat-shaped feature.
Including higher terms of the multiple-scattering expansion of $T_{sL,sL}$ (\ref{SPMAT})
causes the suppression of spectral weight
in the gap to moves downwards towards the Fermi energy.

In the ES the interferences [(111),$R$2] act at the upper edge of the spectral range which has been attributed to
the stabilization of the arrangement of the centered tetrahedra.
Zinc blende phases open an additional gap in this energy range.

\subsection{Diamond- and zinc-blende phases}   

Measured bulk moduli and Pauling electronegativities of the examined alloys are shown in Table \ref{tab:tab10s6}. 
The sequence of the systems follows growing size of the SC unit cell.
EN(A) and EN(B) refer to the original respectively the corrected electronegativities. 

\begin{table}
\small
\centering
\caption{Bulk moduli (BM, \cite{BoXu13}) and Pauling electronegativities (EN,\cite{Pauling32,Alfred61}).}
\vspace{0.2cm}
        \begin{tabular}{ l | l | l | l | l }
        \hline
        phase AB\hspace*{.0cm}	& BM/GPa \hspace*{0.0cm} & EN(A) \hspace*{0.0cm} & EN(B) \hspace*{0.0cm} & avEN  \\
        \hline
	C			& 442		        & 2.55		        &  		        & 2.55  \\
	BN			& 369		        & 2.04		        & 3.04		        & 2.54  \\
	BP			& 173		        & 2.04		        & 2.19		        & 2.12  \\
	Si			& 100		        & 1.9		        & 		        & 1.9   \\
	AlP			& 86		        & 1.61		        & 2.19		        & 1.90  \\
	GaP			& 89		        & 1.81		        & 2.19		        & 2.00  \\
	Ge			& 78		        & 2.01		        & 		        & 2.01  \\
	GaAs			& 76		        & 1.81		        & 2.18		        & 2.0   \\
	AlAs			& 77		        & 1.61		        & 2.18		        & 1.9   \\
	CdS			& 62    	        & 1.69		        & 2.58		        & 2.14  \\
	InP			& 73		        & 1.78		        & 2.19		        & 1.99  \\
	CdSe			& 56    	        & 1.69		        & 2.55		        & 2.12  \\
	InAs			& 60		        & 1.78		        & 2.18		        & 1.98  \\
	GaSb			& 57		        & 1.81		        & 2.05		        & 1.93  \\
	AlSb			& 59		        & 1.61		        & 2.05		        & 1.83  \\
	InSb			& 47		        & 1.78		        & 2.05 		        & 1.92  \\
	CdTe			& 42		        & 1.69		        & 2.1		        & 1.9   \\
	Sn			& 55		        & 1.96		        & 		        & 1.96  \\
        \hline
        \end{tabular}
\label{tab:tab10s6}
\end{table}

\subsection{The $\gamma$-brasses and 3$\times$3$\times$3-bcc}   

Green bullets in Figure \ref{fig:fig69s6} denote the atomic sites of $\gamma$-$\rm Ag_5Li_8$ 
whereas the red bullets belong to 3$\times$3$\times$3-bcc.
The following shifts of the atomic sites occur in the transition 
from 3$\times$3$\times$3-bcc to $\gamma$-$\rm Ag_5Li_8$:
Li(IT) and Ag(OT) move 
along the space diagonals, Ag$^\prime$(OH) along the cubic axes,
and Li$^\prime$(CO) towards the space diagonals.

Most remarkable, the compression of the IT whereas the OT stay very close
to the sublattice sites of 3$\times$3$\times$3-bcc.
As the bcc-sites on the scale $a$ of the SC unit cell are empty, the OT preserve the memory
of bcc-structure on the scale $a$.

Inspecting the relaxation step more in detail (here not shown) reveals
that pairs of atoms on strongly reflecting
lattice planes of 3$\times$3$\times$3-bcc are moved
to symmetric positions with respect to the former planes.
We conclude that the planar interferences must be less affected
by the step from 3$\times$3$\times$3-bcc to the $\gamma$-phase.

The Tables \ref{tab:tab11s6} to \ref{tab:tab13s6} show the shortest interatomic distances $R$n 
in and between the subclusters of the 

\newpage

\begin{table}
\small
\centering
\caption{$\gamma$-$\rm Ag_5Li_8$ \cite{Noritake07a} cubic side length $a$ = 9.6066 \AA.}
\vspace{0.2cm}
\begin{tabular}{ l | l | l | l }
        \hline
         subcluster & atoms   & $\gamma$-$\rm Ag_5Li_8$ & 3$\times$3$\times$3-bcc  \\
         links      &         & \AA               & \AA                \\
        \hline
         OH-CO      & Ag$^\prime$-Li$^\prime$ & R1   = 2.737             & Q1 = 2.773              \\
         OT-CO      & Ag-Li$^\prime$          & R2   = 2.766             & Q1 = 2.773              \\
         IT-OH      & Li-Ag$^\prime$          & R3   = 2.778             & Q1 = 2.773              \\
         OH-OH      & Ag$^\prime$-Ag$^\prime$ & R4   = 2.803             & Q2 = 3.202              \\
         IT-CO      & Li-Li$^\prime$          & R5   = 2.808             & Q1 = 2.773              \\
         CO-CO      & Li$^\prime$-Li$^\prime$ & R6   = 2.842             & Q1 = 2.773              \\
         IT-OT      & Li-Ag                   & R7   = 2.879             & Q1 = 2.773              \\
         OT-OH      & Ag-Ag$^\prime$          & R8   = 2.921             & Q1 = 2.773              \\
         IT-IT      & Li-Li                   & R9   = 3.179             & Q3 = 4.529              \\
         OT-OT      & Ag-Ag                   & R10  = 4.682             & Q3 = 4.529              \\
         OH-CO      & Ag$^\prime$-Li$^\prime$ & $\tilde{R}$ = 3.045      &                         \\
        \end{tabular}
\label{tab:tab11s6}
\end{table}
\begin{table}
\small
\centering
\caption{$\gamma$-$\rm V_5Al_8$ \cite{Brandon77}, cubic side length $a$ = 9.223 \AA.}
\vspace{0.2cm}
\begin{tabular}{ l | l | l | l }
        \hline
         subcluster & atoms   & $\gamma$-$\rm V_5Al_8$ & 3$\times$3$\times$3-bcc  \\
         links      &         & \AA               & \AA                \\
        \hline
         OH-CO      & V$^\prime$-Al$^\prime$  & R4   = 2.687             & Q1 = 2.662              \\
         OT-CO      & V -Al$^\prime$          & R1   = 2.647             & Q1 = 2.662              \\
         IT-OH      & Al-V$^\prime$           & R2   = 2.659             & Q1 = 2.662              \\
         OH-OH      & V$^\prime$-V$^\prime$   & R6   = 2.726             & Q2 = 3.074              \\
         IT-CO      & Al-Al$^\prime$          & R5   = 2.712             & Q1 = 2.662              \\
         CO-CO      & Al$^\prime$-Al$^\prime$ & R7   = 2.768             & Q1 = 2.662              \\
         IT-OT      & Al-V                    & R3   = 2.680             & Q1 = 2.662              \\
         OT-OH      & V-V$^\prime$            & R9   = 2.782             & Q1 = 2.662              \\
         IT-IT      & Al-Al                   & R8   = 2.771             & Q3 = 4.348              \\
         OT-OT      & V-V                     & R10  = 4.434             & Q3 = 4.348              \\
         OH-CO      & V$^\prime$-Al$^\prime$  & $\tilde{R}$ = 3.034      &                         \\
        \end{tabular}
\label{tab:tab12s6}
\end{table}

\begin{table}
\small
\centering
\caption{$\gamma$-$\rm Cu_5Zn_8$ \cite{Gourdon07}, cubic side length $a$ = 8.866 \AA.}
\vspace{0.2cm}
\begin{tabular}{ l | l | l | l }
        \hline
         subcluster & atoms    & $\gamma$-$\rm Cu_5Zn_8$ & 3$\times$3$\times$3-bcc  \\
         links      &          & \AA               & \AA                \\
        \hline
         OH-CO      & Cu$^\prime$-Zn$^\prime$  & R2   = 2.548             & Q1 = 2.559             \\
         OT-CO      & Cu -Zn$^\prime$          & R1   = 2.547             & Q1 = 2.559             \\
         IT-OH      & Zn-Cu$^\prime$           & R4   = 2.581             & Q1 = 2.559             \\
         OH-OH      & Cu$^\prime$-Cu$^\prime$  & R3   = 2.557             & Q2 = 2.955             \\
         IT-CO      & Zn-Zn$^\prime$           & R6   = 2.631             & Q1 = 2.559             \\
         CO-CO      & Zn$^\prime$-Zn$^\prime$  & R7   = 2.643             & Q1 = 2.559             \\
         IT-OT      & Zn-Cu                    & R5   = 2.611             & Q1 = 2.559             \\
         OT-OH      & Cu-Cu$^\prime$           & R9   = 2.7044            & Q1 = 2.559             \\
         IT-IT      & Zn-Zn                    & R8   = 2.7038            & Q3 = 4.179             \\
         OT-OT      & Cu-Cu                    & R10  = 4.320             & Q3 = 4.179             \\
         OH-CO      & Cu$^\prime$-Zn$^\prime$  & $\tilde{R}$ = 2.856      &                        \\
        \end{tabular}
\label{tab:tab13s6}
\end{table}

\begin{table}
\small
\centering
\caption{Virtual valences of planar interferences in $\gamma$-brasses,
$Z(d(G^2))$ (\ref{PLZ}), $N_{act}= 52$, $G^2=h^2+k^2+l^2$.
(7),(15),\dots indicate gaps in the sequence of SC reflexes.}
\vspace{0.2cm}
\begin{tabular}{ l | l | l | l | l | l }
        \hline
         $G^2$  &$Z(d(G^2))$ & 	$G^2$ 	&$Z(d(G^2))$ 	& $G^2$ & $Z(d(G^2))$ 	\\
        \hline		
	1	& 0.0201&       11      & 0.7347&   	 21      & 1.9380\\	
	2	& 0.0570&       12      & 0.8371&	 22      & 2.0781\\
	3	& 0.1046&       13      & 0.9439&	 (23)    & 2.2214\\
	4	& 0.1611&       14      & 1.0549&	 24      & 2.3678\\
	5	& 0.2252&       (15)    & 1.1699&	 25      & 2.5173\\
	6	& 0.2960&       16      & 1.2889&	 26      & 2.6698\\
	(7)	& 0.3730&       17      & 1.4116&	 27      & 2.8253\\  
	8       & 0.4557&	18      & 1.5379&        (28)    & 2.9838\\
       	9       & 0.5437&  	19      & 1.6678&        29      & 3.1450\\
       10       & 0.6368&	20      & 1.8012&	 30	 & 3.3091\\
\end{tabular}
\label{tab:tab14s6}
\end{table}

\newpage

\noindent
$\gamma$-phases
respectively the $Q$n in the reference structure 3$\times$3$\times$3-bcc.
The order of the interatomic links between and inside the subclusters IT, OT, OH, and
CO is taken from the sequence of growing interatomic distances $R$n of $\gamma$-$\rm Ag_5Li_8$.
The Tables \ref{tab:tab12s6} and \ref{tab:tab13s6} 
use the same order despite of different sequences of growing interatomic distances $R$n.

Table \ref{tab:tab14s6} is a list of the required virtual valences for excitation of SC Bragg reflections (\ref{PLZ}).
All the 52 sites in the SC unit cell of $\gamma$-brasses are supposed to participate.
If only $N_{\rm act} < 52$ sites participate, the required virtual valences increase by factors of $52/N_{\rm act}$.
$G^2$ enclosed in braces indicates a gap in the sequence of SC Bragg reflections.

\subsection{The $d$-resonances of the $\gamma$-phases}   

The spectral positions of $d$-resonances prove important to the
electronic stabilization of $\gamma$-brasses.
The following compilation of materials-related data is derived from 
muffin-tin scattered-wave calculations on the basis of the calculated LMTO-ASMT
potentials. 
Small spectral shifts arise from different environments of the atoms (AS) in the respective structure model.
So, up to four different resonance energies can arise. 
Note that the assignment of atoms with subclusters IT, OT, OH, and CO applies only to the $\gamma$-phases.
For the properly decorated 3$\times$3$\times$3-bcc systems we take the atoms prior to the relaxation step. 
The following symbols are used: Fermi energy (F), muffin-tin zero (MT0), and the bottom of the valence band (B).

\begin{table}
\small
\centering
\caption{{\bf\bm $\gamma$-$\rm Ag_5Li_8$} \cite{Noritake07a} cubic side length $a$ = 9.6066 \AA. 
Reference energies (eV) and the $d$-resonances (eV).}
\vspace{0.2cm}
\begin{tabular}{ l | l | l | l }
        \hline
         model                 		    & from F & from F & from MT0  \\
                               		    & to MT0 & to B   & to B      \\
        \hline
	3$\times$3$\times$3-bcc        	    & -6.028 & -5.811 & +0.217    \\
	$\gamma$-$\rm Ag_5Li_8$        	    & -6.044 & -5.624 & +0.420    \\
	$\gamma$-$\rm Ag_5Li_8$-ITOT   	    & -6.202 & -5.962 & +0.240    \\
        \hline
					    &	     &	      &           \\ 
        \hline
         model                      	    & from MT0 & from F & from B    \\
        \hline
	{\bf\bm3$\times$3$\times$3-bcc}     &          &        &           \\
	Ag                                  & +1.406   & -4.622 & +1.189    \\
	Ag$^\prime$                         & +1.291   & -4.737 & +1.074    \\
	{\bf\bm$\gamma$-$\rm Ag_5Li_8$}     &   &  &     \\
	Ag(OT)                              & +1.399   & -4.645 & +0.978    \\
	Ag$^\prime$(OH)                     & +1.485   & -4.559 & +1.065    \\
	{\bf\bm$\gamma$-$\rm Ag_5Li_8$-ITOT}&   &  &     \\
	Ag(IT)                              & +1.393   & -4.809 & +1.153    \\
	Ag$^\prime$(OH)                     & +1.579   & -4.623 & +1.339    \\
        \hline
	\end{tabular}
\label{tab:tab15s6}
\end{table}

\begin{table}
\small
\centering
\caption{{\bf\bm $\gamma$-$\rm V_5Al_8$} \cite{Brandon77} cubic side length $a$ = 9.223 \AA.
Reference energies (eV) and the $d$-resonances (eV).}
\vspace{0.2cm}
\begin{tabular}{ l | l | l | l }
        \hline
         model                 		    & from F & from F  & from MT0   \\
                                            & to MT0 & to B    & to B       \\
        \hline
	3$\times$3$\times$3-bcc             & -9.642 & -10.288 & -0.646     \\
	$\gamma$-$\rm V_5Al_8$              & -9.534 & -10.010 & -0.476     \\
        \hline
                                            &        &        &             \\
        \hline
         model                              & from MT0 & from F & from B    \\
        \hline
	{\bf\bm3$\times$3$\times$3-bcc}     &          &        &           \\
	V                                   & +10.127  & +0.485 & +10.773   \\
	V$^\prime$                          & +10.009  & +0.367 & +10.655   \\
	{\bf\bm$\gamma$-$\rm V_5Al_8$}      &   &  &     \\
	V(OT)                               & +9.852   & +0.318 & +10.328   \\
	V$^\prime$(OH)                      & +9.905   & +0.370 & +10.381   \\
        \hline
\end{tabular}
\label{tab:tab16s6}
\end{table}

\begin{table}
\small
\centering
\caption{{\bf\bm $\gamma$-$\rm Cu_5Zn_8$} \cite{Gourdon07} cubic side length $a$ = 8.866 \AA.
Reference energies (eV) and the $d$-resonances (eV).}
\vspace{0.2cm}
\begin{tabular}{ l | l | l | l }
        \hline
         model                              & from F & from F  & from MT0  \\
                                            & to MT0 & to B    & to B      \\
        \hline
	3$\times$3$\times$3-bcc             & -9.310 & -11.001 & -1.691    \\
	$\gamma$-$\rm Cu_5Zn_8$             & -9.183 & -10.742 & -1.559    \\
        \hline
                                            &        &        &           \\
        \hline
         model                      	    & from MT0 & from F & from B    \\
        \hline
	{\bf\bm3$\times$3$\times$3-bcc}     &          &        &           \\
	Cu                                  & +6.073   & -3.237 & +7.764    \\
	Cu$^\prime$                         & +5.821   & -3.489 & +7.512    \\
	Zn                                  & +2.432   & -6.878 & +4.123    \\
	Zn$^\prime$                         & +1.815   & -7.495 & +3.506    \\
	{\bf\bm$\gamma$-$\rm Cu_5Zn_8$}     &   &  &     \\
	Cu(OT)                              & +5.701   & -3.482 & +7.261    \\
	Cu$^\prime$(OH)                     & +5.840   & -3.343 & +7.400    \\
	Zn(IT)                              & +2.040   & -7.143 & +3.600    \\
	Zn$^\prime$(CO)                     & +2.211   & -6.973 & +3.771    \\
        \hline
\end{tabular}
\label{tab:tab17s6}
\end{table}

\subsection{Visualization of the composition of band states}   

\begin{figure}
\centering
\includegraphics[width=5.5cm]{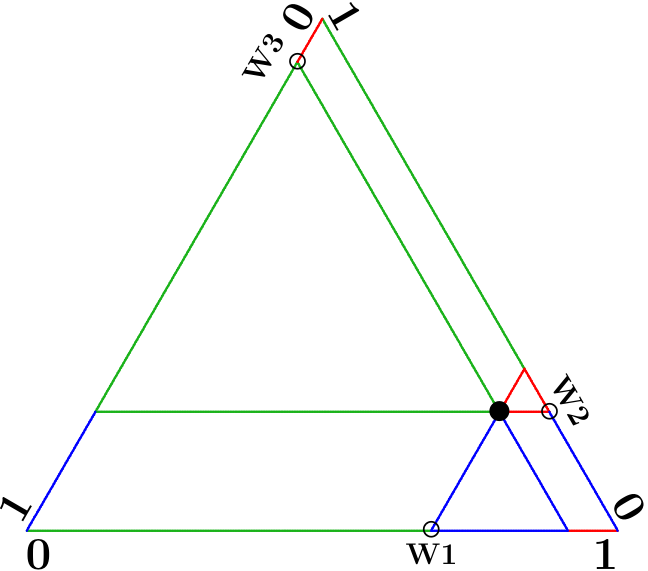}
\caption{For any decomposition of the total EQS in three subspaces, the representation of an electron band state (bullet)
  by its partial weights (W1,W2,W3) in these subspaces.}
\label{fig:fig70s6}
\end{figure}

Each band state distributes the norm One throughout the EQS.
We decompose the total EQS into any three subspaces with non-negative 
partial weights $W1, W2, W3$ where $W3$ = 1 - $(W1 + W2)$.
e.g. the decomposition into the $spd$-subspaces. 
Phase diagrams of three-component systems demonstrate how such normalized decompositions are visualized
by means of equilateral triangles (Figure \ref{fig:fig70s6}). 
All the marked triangles are equilateral triangles. 

A Cartesian coordinate system supports the navigation inside the big triangle.
We put the origin in Figure \ref{fig:fig70s6} to the intersection of the $W3$-axis with the $W1$-axis
and obtain the position of the bullet, 
\begin{eqnarray}
         x &=& W1 + \frac{W2}{2} , \\
         y &=& \frac{W2 \; \sqrt{3}}{2} .
\label{DRECK}
\end{eqnarray}

Suppose a small energy interval centered at a given energy $\epsilon$.
All the band states inside generate a distribution of bullets which is poorly represented by one bullet at the average position.
To utilize the full information content we plot the tracking curve of the average together with all bullets of the actual energy interval. 
The resulting animation reveals preferences of the fluctuation patterns which 
indicate the favored as well as the avoided
interplay between subspaces of the total EQS.

Let us now replace the $Wn$ along the sides of the big triangle by $n_{\rm sub}(\epsilon)/n_{\rm tot}(\epsilon)$ 
where $n_{\rm sub}(\epsilon)$ and $n_{\rm tot}(\epsilon)$ are the subspace-projected respectively the total densities of states.
The result will be a position very close to the average bullet above. 
Deviations are due to the special smearing of the energy eigenvalues in the LMTO-ASMT calculation.
Hence, with rising energy, the average bullet follows the DOS-based curve.

The Cartesian coordinates are very useful in order to analyze the fluctuations.
If, e.g., the fluctuation patterns point perpendicular to the W1-axis we have
$\Delta$x = 0 which provides $\Delta W$1 = -$\Delta W$2/2.
Furthermore, the norm One provides 
$\Delta W$3 = -($\Delta W$1 + $\Delta W$2) = -$\Delta W$2/2.  
This mens that equal losses of $W$1 and $W$3 result in a gain of $W$2 twice as much.
The opposite applies, too.  

This technique has the advantage that the distribution of quantum weight throughout
the total EQS is visualized by the geometry of equilateral triangles. 
\begin{footnotesize}
\begin{frenchspacing}
\vspace{3ex}
\itemsep 0.2ex
\bibliographystyle{jncs} \bibliography{lit}
\end{frenchspacing}
\end{footnotesize}

\end{document}